\documentclass[%
 reprint,
 superscriptaddress,
nofootinbib,
 amsmath,amssymb,
 aps,
 prd,
]{revtex4-2}

\usepackage{graphicx}
\usepackage{dcolumn}
\usepackage{bm}
\usepackage[T1]{fontenc}
\usepackage[utf8]{inputenc}
\usepackage[varg]{newtx}
\usepackage{comment}
\usepackage{amsmath}
\usepackage{graphicx}
\usepackage{gensymb}
\usepackage{upgreek}
\usepackage{xspace}
\DeclareRobustCommand{\ion}[2]{\textup{#1\,\textsc{\lowercase{#2}}}}
\newcommand{\hei}{\ion{He}{i}\xspace}
\newcommand{\heii}{\ion{He}{ii}\xspace}
\newcommand{\heiii}{\ion{He}{iii}\xspace}
\newcommand{\sriii}{\ion{Sr}{iii}\xspace}
\newcommand{\srii}{\ion{Sr}{ii}\xspace}
\newcommand{\sri}{\ion{Sr}{i}\xspace}
\newcommand{\Laiii}{\ion{La}{iii}\xspace}
\newcommand{\Ceiii}{\ion{Ce}{iii}\xspace}
\newcommand{\yii}{\ion{Y}{ii}\xspace}
\newcommand{\teiii}{\ion{Te}{iii}\xspace}
\newcommand{\rprocess}{\textit{r}-process\xspace}
\newcommand{\threeS}{1s2s\,$^3$S\xspace}
\newcommand{\threeP}{1s2p\,$^3$P\xspace}
\newcommand{\micron}{$\upmu$m\xspace}
\newcommand{\taubh}{$\tau_{\rm BH}$\xspace}

\newcommand{\aap}{A\&A}
\newcommand{\apjl}{Astrophys. J. Lett.}
\newcommand{\apjs}{Astrophys. J. Suppl.}
\newcommand{\mnras}{Mon. Not. R. Astron. Soc.}

\newcommand{\physrep}{Phys. Rep.}

\newcommand{\na}{New Ast.}
\newcommand{\pasa}{PASA}
\newcommand{\pasj}{PASJ}
\newcommand{\jcap}{JCAP}

\usepackage[colorlinks,linkcolor=blue,citecolor=blue,bookmarks]{hyperref}

\begin{document}

\preprint{APS/123-QED}



\title{Helium as an Indicator of the Neutron-Star Merger Remnant Lifetime \\ and its Potential for Equation of State Constraints}

\author{Albert Sneppen}
\affiliation{Niels Bohr Institute, University of Copenhagen, Jagtvej 128, Copenhagen N, Denmark}
\affiliation{Cosmic Dawn Center, Denmark}
\author{Oliver Just}
\affiliation{GSI Helmholtzzentrum für Schwerionenforschung, Planckstraße 1, D-64291 Darmstadt, Germany}
\affiliation{Astrophysical Big Bang Laboratory, RIKEN Cluster for Pioneering Research, 2-1 Hirosawa, Wako, Saitama 351-0198, Japan}
\author{Andreas Bauswein}
\affiliation{GSI Helmholtzzentrum für Schwerionenforschung, Planckstraße 1, D-64291 Darmstadt, Germany}
\affiliation{Helmholtz Forschungsakademie Hessen f\"ur FAIR (HFHF), GSI Helmholtzzentrum f\"ur Schwerionenforschung, Planckstra{\ss}e~1, 64291 Darmstadt, Germany}
\author{Rasmus Damgaard}
\affiliation{Niels Bohr Institute, University of Copenhagen, Jagtvej 128, Copenhagen N, Denmark}
\affiliation{Cosmic Dawn Center, Denmark}
\author{Darach Watson}
\affiliation{Niels Bohr Institute, University of Copenhagen, Jagtvej 128, Copenhagen N, Denmark}
\affiliation{Cosmic Dawn Center, Denmark}
\author{Luke J. Shingles}
\affiliation{GSI Helmholtzzentrum für Schwerionenforschung, Planckstraße 1, D-64291 Darmstadt, Germany}
\author{Christine E. Collins}
\affiliation{School of Physics, Trinity College Dublin, College Green, Dublin 2,
Ireland}
\author{Stuart A. Sim}
\affiliation{Astrophysics Research Centre, School of Mathematics and Physics, Queen's University Belfast, Belfast BT7 1NN, Northern Ireland, UK}
\affiliation{Niels Bohr Institute, University of Copenhagen, Jagtvej 128, Copenhagen N, Denmark}
\affiliation{Cosmic Dawn Center, Denmark}
\author{Zewei Xiong}
\affiliation{GSI Helmholtzzentrum für Schwerionenforschung, Planckstraße 1, D-64291 Darmstadt, Germany}
\author{Gabriel Martínez-Pinedo}
\affiliation{GSI Helmholtzzentrum für Schwerionenforschung, Planckstraße 1, D-64291 Darmstadt, Germany}
\affiliation{Institut {f\"ur} Kernphysik (Theoriezentrum),  Fachbereich Physik, Technische Universit{\"a}t Darmstadt, Schlossgartenstra{\ss}e 2, D-64289 Darmstadt, Germany}
\affiliation{Helmholtz Forschungsakademie Hessen f\"ur FAIR (HFHF), GSI Helmholtzzentrum f\"ur Schwerionenforschung, Planckstra{\ss}e~1, 64291 Darmstadt, Germany}
\author{Theodoros Soultanis}
\affiliation{GSI Helmholtzzentrum für Schwerionenforschung, Planckstraße 1, D-64291 Darmstadt, Germany}
\author{Vimal Vijayan}
\affiliation{GSI Helmholtzzentrum für Schwerionenforschung, Planckstraße 1, D-64291 Darmstadt, Germany}

\date{\today}

\begin{abstract}
  The time until black hole formation in a binary neutron-star (NS) merger contains invaluable information about the nuclear equation of state (EoS) but has thus far been difficult to measure. We propose a new way to constrain the merger remnant's NS lifetime, which is based on the tendency of the NS remnant neutrino-driven winds to enrich the ejected material with helium. 
  Based on the He\,{\sc i}\,$\lambda 1083.3$\,nm line, we show that the feature around 800--1200\,nm in AT2017gfo at 4.4\,days seems inconsistent with a helium mass fraction of $X_{\mathrm{He}} \gtrsim 0.05$ in the polar ejecta. Our recent neutrino-hydrodynamic simulations of merger remnants are only compatible with this limit if the NS remnant collapses within 20--30\,ms. Such a short lifetime implies that the total binary mass of GW170817, $M_\mathrm{\rm tot}$, lay close to the threshold binary mass for direct gravitational collapse, $M_\mathrm{thres}$, for which we estimate $M_{\mathrm{thres}}\lesssim 2.93\,M_\odot$. This upper bound on $M_\mathrm{thres}$ yields upper limits on the radii and maximum mass of cold, non-rotating NSs, which rule out simultaneously large values for both quantities. In combination with causality arguments, this result implies a maximum NS mass of $M_\mathrm{max}\lesssim2.3\,M_\odot$. We include an updated constraint yielding lower limits on NS radii from a previous argument that the remnant did not promptly collapse, which is independent of the consideration of the helium content.
  The combination of all limits constrains the radii of 1.6\,M$_\odot$ NSs to about 12$\pm$1\,km for $M_\mathrm{max}$ = 2.0\,M$_\odot$ and 11.5$\pm$1\,km for $M_\mathrm{max}$ = 2.15\,M$_\odot$. This $\sim2$\,km allowable range tightens significantly for $M_\mathrm{max}$ above $\approx2.15$\,M$_\odot$.
  This rules out a significant number of current EoS models. The short NS lifetime also implies that a black-hole torus, not a highly magnetized NS, was the central engine powering the relativistic jet of GRB170817A. 
  Our work motivates future developments to further corroborate and improve uncertainties in our chain of arguments, regarding NLTE spectral modeling, helium production in merger outflows, and the dependence of the remnant lifetime on the binary mass, with the potential to tighten our constraints from existing data and in particular from future events. This novel method may provide a powerful tool to get a handle on the poorly constrained remnant lifetime, the still debated central engine of short GRBs, and the high-density EoS.
\end{abstract}

\maketitle

\section{\label{sec:introduction} Introduction}

Neutron-star mergers (NSM) provide natural laboratories for studying the incompletely known properties of high-density matter quantified by the equation of state (EoS). The EoS relates the pressure and density of neutron-star matter, and is uniquely linked to the stellar parameters of neutron stars such as the mass-radius relation or the tidal deformability,  measurements of which, in turn, constrain the EoS~\cite{Tolman:1939jz,Oppenheimer:1939ne,Hinderer:2009ca,Lattimer2016,Baym:2017whm,Oertel:2016bki,Raduta:2022elz,Fan2024}. 
From the first gravitational-wave detected NS-merger, GW170817, several binary parameters like the observer distance, the total binary mass, the binary mass ratio and the dimensionless tidal deformability, $\Lambda$, could be constrained~\cite{Abbott2017c,Abbott2019}. Since $\Lambda$ scales tightly with the NS radius, this has been constrained to be \(\lesssim\)\,13.5\,km in the mass range around 1.4\,M\(_\odot\), which rules out very stiff models of high-density matter~\cite{Abbott2018b,De2018}. 

GW170817 was accompanied by the kilonova (KN) AT2017gfo~\citep{Coulter2017, Andreoni2017, Buckley2018, Cowperthwaite2017, Drout2017, Kasen2017, Kasliwal2017, Kilpatrick2017, Nicholl2017, Pian2017, Shappee2017, Smartt2017, Soares-Santos2017, Tanaka2017t, Tanvir2017}, i.e.\ optical/infrared emission resulting from radioactive decays connected to the rapid neutron-capture process (\rprocess \citep{Horowitz2018a, Arnould2020f, Cowan2021}) in the matter outflows during and after the merger \citep{Li1998,Kulkarni2005,Metzger2010}. A large number of studies have since employed the properties of the kilonova to derive additional EoS constraints all relying on the fact that the merger dynamics and thus the matter ejection are sensitive to the EoS, and therefore the electromagnetic emission should carry an imprint of the properties of high-density matter.

Early on, the argument was made that GW170817 did not result in a prompt gravitational collapse to a black hole, mainly because the high brightness of AT2017gfo disfavors such a scenario, which would be accompanied with reduced ejecta mass and therefore relatively dim kilonova luminosity~\cite{Bauswein2017,Radice2018c,Koeppel2019,Bauswein2019,Capano2020}, but see~\cite{Kiuchi2019}. This implies that the measured total binary mass of GW170817 was below the threshold binary mass for prompt collapse, 
which is an EoS-dependent quantity. Following this reasoning NS radii cannot be too small ($R\gtrsim10.5$~km) and the EoS cannot be too soft, as this would have resulted in direct black hole (BH) formation \citep{Bauswein2017,Bauswein2021}. 

Several studies also presented arguments that a black hole did eventually form during the subsequent evolution of the rotating merger remnant. First, a NS remnant surviving longer than a few seconds or more should likely have injected a large fraction of its rotational energy through magnetic spindown~\citep{Spitkovsky2006a} into the ejecta, which seems incompatible with the absence of a corresponding late-time signal in the electromagnetic emission~\cite{Margalit2017,Shibata2017,Shibata2019,Margalit2022}. Second, the detection of a short gamma-ray burst (GRB) about 1.74\,s after the GW signal provides an upper limit on the lifetime, if the GRB was launched by a black-hole torus system~\cite[e.g.][]{Fan2013,Rezzolla2018,Ruiz2018,Shibata2019}. Assuming these arguments, the aforementioned studies find upper limits for the maximum mass of cold, non-rotating neutron stars (also known as the Tolman-Oppenheimer-Volkoff (TOV) limit) of about 2.17--2.3\,M\(_\odot\). Without considering implications for the EoS, Ref.~\cite{Perego2022} speculated that strontium could be overproduced compared to AT2017gfo if a spiral wind was operating for more than about 100\,ms (also see Appendix~\ref{app:strontium}). While an influence of the remnant lifetime on the nucleosynthesis pattern was pointed out already by Refs.~\cite{Metzger2014, Lippuner2017}, the detailed evolution of NS remnants is still not well understood due to the challenges in capturing all of the relevant physics (e.g.,\ neutrino transport, numerical resolution, small-scale turbulence, general relativity). As a consequence, the NS remnant lifetime, i.e.,\ the black-hole formation time, $\tau_{\mathrm{BH}}$, in GW170817 is largely unconstrained to date, and it may even be \(\tau_{\mathrm{BH}}>\)\,1.7\,s if the GRB signal originated from a highly magnetized NS remnant (``magnetar'' scenario~\citep{Duncan1992a, Metzger2007c, Kiuchi2024a}).

A number of other studies, e.g.,~\cite{Coughlin2018,Radice2019,Dietrich2020,Breschi2021,Raaijmakers2021,Huth2022,Lund2024a}, employed the lightcurve and color evolution of the kilonova to directly link those features to NS parameters. Such an approach allows for efficient exploration of the large parameter space, but it suffers from significant systematic uncertainties, for instance, connected to the nuclear physics input, atomic data, radiative-transfer modeling, thermalization physics, or observation-angle dependence. The link between ejecta properties and the EoS is based entirely on predictions from numerical simulations, which again carry uncertainties that are difficult to quantify (numerical resolution dependence, neutrino transport physics, small-scale turbulence, approximate inclusions of long-term matter ejection; see also~\cite{Henkel2023,Janka2022,Holmbeck2021} illustrating some of the ambiguities). 

In this paper we propose a conceptually new method based on the abundance of a specific element, namely helium, to constrain the high-density EoS via the lifetime of NS merger remnants. We present evidence that the remnant in GW170817 was short-lived ($\lesssim 20$--$30$\,ms), because a longer lifetime would have resulted in a larger amount of helium with pronounced spectral features which are not found in the spectra. We use this argument to derive EoS constraints, specifically upper limits on NS radii and the maximum NS mass. In contrast to previous methodologies we employ for the first time information from the kilonova spectra to place constraints on NS parameters (and therefore the nuclear EoS). This highlights the value of the spectral information not only for nucleosynthesis but also for the merger dynamics and EoS. This work is a first exploratory study highlighting the idea and the potential of the method. Future work is needed to further corroborate and develop each of the steps taken in our study. We combine the new EoS constraint with the previously made argument that the brightness of AT2017gfo excludes prompt black-hole formation yielding lower limits on NS radii~\cite{Bauswein2017}, which we further improve in this paper.

The paper is structured as follows. In Sect.~\ref{sec:observed_helium_constraints} we describe the upper limit on the amount of helium in the outflow of AT2017gfo inferred from the spectral energy distribution. We review helium production in the outflows from neutron-star mergers based on theoretical models in Sect.~\ref{sec:HMNS_lifetime} to deduce an upper limit of the remnant lifetime. The resulting constraints on NS parameters are presented in Sect.~\ref{sec:EOS_constraints}. In Sect.~\ref{sec:discussion} we discuss additional implications of our analysis. We conclude in Sect.~\ref{sec:summary}. In this paper we use the total binary mass $M_\mathrm{tot}=M_1+M_2$ with $M_{1,2}$ being the gravitational mass of NSs in a binary with infinite orbital separation. We define the binary mass ratio as $q=M_1/M_2$ with $M_1\leq M_2$, hence $q\leq1$.

\section{\label{sec:observed_helium_constraints} The observational limits on helium production in AT2017gfo}

Spectral modeling of kilonovae has seen major progress in recent years, largely motivated by the spectroscopic data series from the kilonovae AT2017gfo \citep[e.g.,][]{Shappee2017,Pian2017,Smartt2017} and AT2023vfi \citep{Levan2024}. These spectra have already provided the first direct identification of known lines from \rprocess elements, most notably the prominent spectral feature around  1\,\micron as an absorption-emission `P~Cygni' feature of the Sr\,{\sc ii} triplet of strong lines (4p\(^6\)4d-4p\(^6\)5p) \citep{Watson2019,Domoto2021,Gillanders2022,Shingles2023,Sneppen2023c,Sneppen2024}. Compelling cases for other line identifications have also been made, \emph{viz.} Y\,{\sc ii} \citep{Sneppen2023b}, \teiii \citep{Hotokezaka2023}, La\,{\sc iii} and Ce\,{\sc iii} \citep{Domoto2022}. Given the limited atomic data, most kilonova modeling of spectral features has been done under the assumption of local thermodynamic equilibrium (LTE) \citep[although, see][]{Hotokezaka2023,Pognan2023,Pognan2024,Mulholland2024}; however, non-LTE (NLTE) effects are expected to be important for estimating level populations and ionization states.

Coincidentally, as first examined by \citet{Perego2022}, He\,{\sc i}\,$\lambda 1083.3$\,nm (1s2s\,$^3$S--1s2p\,$^3$P) may also contribute to the 1\,\micron feature under NLTE conditions. Their analysis concluded that the low helium abundance (\(\lesssim 10^{-5}\)\,$M_\odot$) produced in the dynamical ejecta of their models could not explain the observed prominence of the feature at 2-3\,days, but proposed that an order-of-magnitude increase in helium abundance (or an order-of-magnitude decrease in the luminosity) could lead to an observable effect. Indeed, \citet{Tarumi2023} corroborated these results, concluding that relatively minor quantities of helium (\(\sim 10^{-4}\)\,M\(_{\odot}\)) in their modeling could tentatively explain the observed feature at all epochs given sufficient UV line blanketing. However, by accounting for higher energy levels and including UV flux detections from the \emph{Swift} satellite, \citet{Sneppen2024_helium} showed that a helium interpretation is inconsistent with the 1\,\micron feature in the first days post-merger, both in terms of its evolution and amplitude. 

\begin{figure}
    \includegraphics[width=\linewidth,viewport=22 15 420 370 ,clip=]{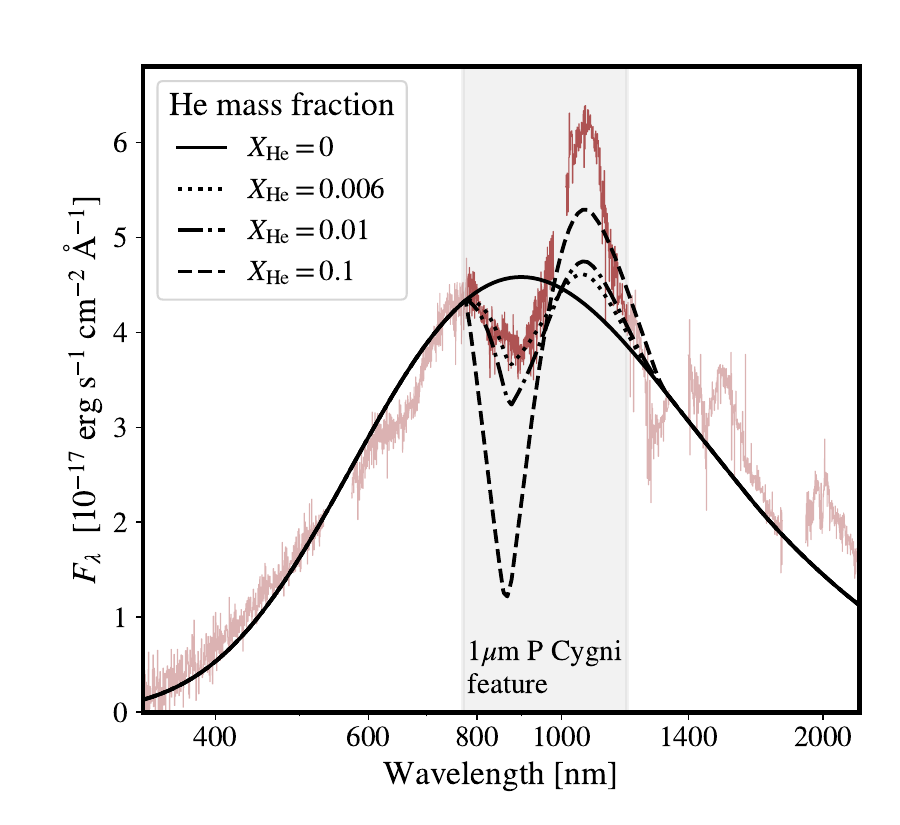}
    \caption{VLT/X-shooter spectrum of AT2017gfo 4.4\,days post-merger with a blackbody continuum overlaid ($T_{\rm BB}=3200$\,K from the best-fit blackbody compilation in \citet{Sneppen2024}) and P~Cygni features for various helium abundances computed using the model described in Sect.~\ref{sec:observed_helium_constraints}. Given a sufficient helium abundance, $X_{\rm He} \sim 0.01$, a sizeable absorption feature will be produced in the region 800--1000\,nm. The other spectral features in the spectrum have been tentatively linked to \yii (600--800\,nm) and \Laiii, \Ceiii, \teiii (1200--1600\,nm, 2000\,nm). 
    } 
    \label{fig:Xshooter_helium}
\end{figure}

Nevertheless, at around 4--5\,days post-merger optimal conditions do arise for helium to produce a detectable spectral feature at this wavelength (see App.~\ref{sec:observing_time}). In Fig.~\ref{fig:Xshooter_helium}, we show the 4.4\,day spectrum of AT2017gfo taken with the X-shooter spectrograph on the Very Large Telescope (VLT), where there may be a contribution to the absorption feature at 800--1000\,nm from He\,{\sc i}. The absorption feature's strength can be used to place a limit on the He\,{\sc i} 1s2s\,$^3$S population in the line-forming region, and hence on the total He\,{\sc i} mass via non-LTE modeling. A priori we do not know the relative contributions of Sr\,{\sc ii} and He\,{\sc i}, so the helium mass required to reproduce the observed feature may be considered an upper bound on the helium mass in that velocity range in AT2017gfo (see Sect.~\ref{sec:strontium}).

In the following subsections, we derive constraints on the maximum allowed helium mass fraction, $X_{\mathrm{He}}$, in the line-forming region, following the modeling presented in \citep{Sneppen2024_helium}. First, we summarize the radiative-collisional model used to estimate the population of helium in the relevant lower level of the 1083\,nm line (i.e.,\ \threeS). Next, we outline the P~Cygni modeling used to compare \threeS densities with the observed feature, from which our $X_{\mathrm{He}}$ constraint is derived. 

\subsection{Modeling of NLTE helium level populations}\label{sec:NLTE_helium_population}

To accurately model the NLTE population of \threeS, we compute a full collisional-radiative model, analogous to previous studies of supernovae \citep[e.g.,][]{Lucy1991} and kilonovae \citep{Tarumi2023}. Details of the computational approach can be found in \citet{Sneppen2024_helium}, but we note that the atomic data for helium is reliable (particularly in comparison to \rprocess elements) due to supporting experimental measurements, its widespread use in prior astrophysical applications and the simplicity of the few-electron system from a computational standpoint. The atomic data employed includes Einstein A coefficients \citep{Kramida2023}, thermally averaged electron-collision transition rates \citep{Ralchenko2008}, recombination rates, and photoionization cross-sections \citep{Nahar2010}. 

\begin{figure}[t]
    \includegraphics[width=0.9\columnwidth,viewport=25 25 431 663,clip=]{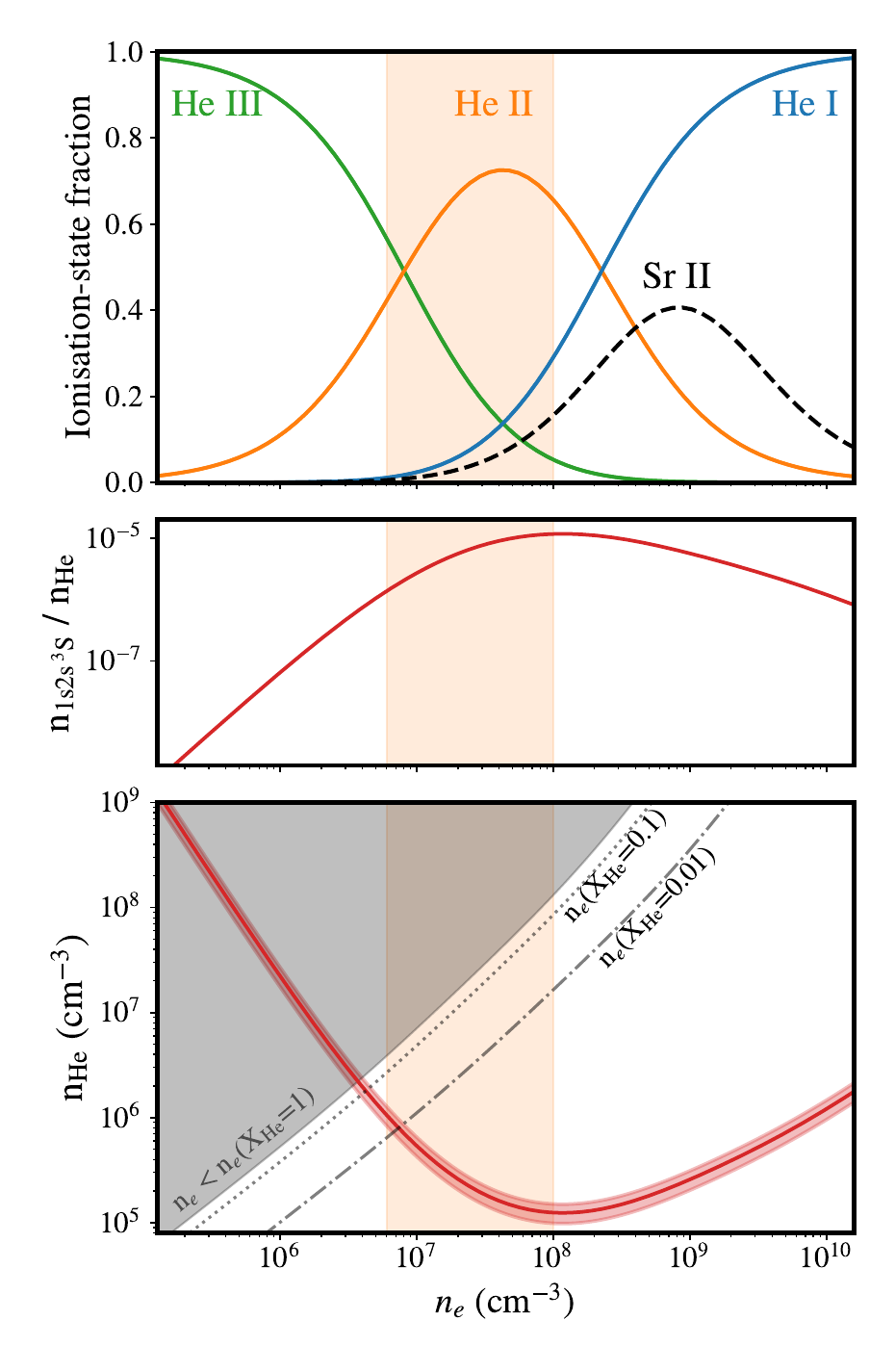}
    \caption{The fraction of helium in each ionization state (top panel), the fraction of helium in the \threeS state (middle panel) and the helium density required to produce the observed feature (red line, bottom panel) as a function of photospheric electron density. All other parameters are set to their standard values, as described in App.~\ref{app:ionisation-eq}. In the lower panel, dotted and dash-dotted lines show $n_e$ as a function of $n_{\mathrm{He}}$ given various assumed $X_{\mathrm{He}}$ and adopting a mean mass number, $A=100$, and the same mean charge as helium for all other species.  While the unphysical regime where $n_e$ is smaller than the electron density solely contributed by helium is shown with the gray shading. The electron densities expected near the photosphere (indicated with shaded orange region, $n_e \approx 6 \times 10^{6}$--$10^{8}\ {\rm cm}^{-3}$, see App.~\ref{sec:electron-density}) predict \heii should constitute a major ionization state and thus a sizeable population will be in \threeS. This implies i) a small density of helium, $n_{\mathrm{He}} \sim 10^5$--$10^{6}$\,cm$^{-3}$, would be sufficient to produce the observed feature, and ii) such electron density cannot solely be explained from the electrons contributed by helium ions, but require contributions from other ions. In the top panel, we also show the \srii fraction computed using recombination rates from \citep[][]{Banerjee2025,Singh2025} with a dashed black line, which suggests \srii may be a subdominant species at $n_e\approx10^{7} {\rm cm}^{-3}$. 
    } 
    \label{fig:Helium_ionisation}
\end{figure}

In our models, we assume homologously expanding ejecta with a power-law density dependence in velocity, $\rho = \rho_0 v^{\alpha}$. The normalization constant, $\rho_0$, is chosen such that the ejecta mass in the velocity range 0.1--0.5$c$ is 0.04\,M$_{\odot}$, around the estimated ejecta mass for AT2017gfo \citep[e.g.,][]{Tanaka2017t,Smartt2017}. We considered a wide range of power-law slopes from constant density ($\alpha=0$) to steep declines ($\alpha=-7$) but adopt $\alpha=-5$ for our fiducial model (see App.~\ref{app:RT}).
This choice yields a mass for the high-velocity ejecta ($\gtrsim0.2c$) of $\sim$0.01\,M$_\odot$, which is consistent with observational constraints from AT2017gfo \citep[e.g.,\,][]{Cowperthwaite2017,Villar2017,Siegel2019}. 
The model assumes over the line-forming region a uniform mass fraction of helium, $X_{\mathrm{He}}$, which is treated as a free parameter. The electron number density, $n_{e}$, is also assumed to follow the same velocity profile (i.e., $n_e \propto v^\alpha$) but with free normalization. Fiducially, in Appendix~\ref{sec:electron-density}, we argue that $n_e \approx 1.5\times10^7\,{\rm cm}^{-3}$ near the photosphere, but to account for potential uncertainties we explore a broader range with $n_e \approx 6 \times 10^{6}$--$10^{8}\ {\rm cm}^{-3}$.
In all cases considered here, we will adopt a photospheric velocity at 4.4\,days of $v_{\rm ph} = 0.19c$ (a slightly lower value $\sim0.15c$ can also be consistent with observations and would provide even stronger limits, see App.~\ref{sec:observing_time}), and place the outer boundary of the calculation at $v_{\rm max} = 0.5c$.
For the electron temperature, we assume, for the baseline model, the relativistically Doppler corrected blackbody temperature, i.e.,\ $T_e = 2800$\,K at 4.4\,days (but explore a broader temperature range in App.~\ref{app:sensetivity}). We note the relativistic Doppler correction leads to a slight decrease from the observed blackbody temperature of $T_{\rm BB} = 3200$\,K. We also assume that the radiation field in the model is a dilute blackbody at the same temperature (again see \cite{Sneppen2024_helium} for details). 

Each model involves solving the NLTE statistical equilibrium equations (see App.~\ref{app:ionisation-eq}) at all velocity points as input for the spectrum synthesis calculation. However, it is instructive to consider the behavior near the photosphere to illustrate when the line becomes visible. To this end, Fig.~\ref{fig:Helium_ionisation} shows, as a function of the adopted photospheric electron density (with all other parameters held at their standard values; see App.~\ref{app:ionisation-eq}), the helium ionization state, occupation fraction of the \threeS level, and the helium density required to achieve a photospheric optical depth $\tau=0.86 \pm 0.18$ (see below). As discussed by \citet{Sneppen2024_helium}, the \threeS population is governed by the balance between the outgoing pathways (i.e.,\ decay to the \hei ground state) versus the incoming recombination rate (which itself depends on the ionization state of helium in the ejecta). The recombination rate is maximized, and thus the highest population in the \threeS state is achieved, when the majority of helium is singly ionized (see Fig.~\ref{fig:Helium_ionisation}, top and middle panel). Conversely, the triplet population is minimized for either i) weak non-thermal ionization (as fulfilled in the LTE limit where the \hei ground state is the only significantly populated level) or ii) low electron densities and/or high non-thermal ionization rates (i.e.,\ the regime of \heiii dominance).

Our estimate of the helium ionization state in the ejecta for various electron densities is shown in Fig.~\ref{fig:Helium_ionisation} (top panel). Across a broad range of electron densities, $n_e \approx 10^{6}-10^{9}\,{\rm cm}^{-3}$ (which contains the likely $n_e$ of AT2017gfo at 4.4\,days, see App.~\ref{sec:electron-density}), \heii will be a significant fraction of the total helium population. For this fiducial calculation we have adopted the deposition rate of non-thermal particles, electrons and gamma-rays, from \cite{Hotokezaka2020} that is determined assuming the $\beta$-decay of an abundance pattern that reproduce the solar system \rprocess abundances. Significant suppression of \heii would require the electron density and/or the deposition rate to deviate from our fiducial expectation by orders of magnitude. For comparison, we show the corresponding \srii fraction assuming i) a work per ion of 300\,eV and ii) a recombination rate for \sriii to \srii, following \citet{Banerjee2025, Singh2025}, which are broadly consistent with independent calculations by C. Ballance (priv. comm.).  

\subsection{P~Cygni modeling}\label{sec:PCygni}
To compare the \threeS density with the observed feature, we will use the P~Cygni implementation in the Elementary Supernova model \citep{Jeffery1990}\footnote{We adapt Ulrich Noebauer's \texttt{pcygni\_profile.py} in \url{https://github.com/unoebauer/public-astro-tools} with generalisations to account for computational speed-up and special relativistic corrections \cite{Sneppen2023A&A} see \url{https://github.com/Sneppen/Kilonova-analysis}}. P~Cygni profiles are characteristic of expanding envelopes in which a spectral line yields both an emission peak near the rest wavelength and a blueshifted absorption feature. This absorption component arises from scattering of photospheric photons out of the line of sight, which means a polar observer can constrain polar ejecta from the depth of the absorption feature. In particular, the optical depth of the $\lambda_0=1083.3\,$nm line is simply related to the density in \threeS and the time post-merger by \citep{Sobolev1960,Tarumi2023}: 
\begin{equation}
    \tau_{\rm \hei\,\lambda1083\,nm}(v) = \left(\frac{ n_{\mathrm{1s2s\,^3S}}(v) }{\rm 7.4\ cm^{-3}}\right) \left(\frac{t}{\rm 1\,day}\right)
    \label{eq:he_tau}
\end{equation}
with a trivial scaling with density and a dependence on time coming from the Sobolev opacity being dependent on the inverse of the velocity gradient (e.g., for homologous expansion $|dv/dr|^{-1} \propto t$). Thus, given NLTE population densities one can compute the optical depth as a function of velocity from which the P\,Cygni line profile is calculated.
%

Synthetic line profiles calculated in this way can then be compared to the observed 4.4~day spectrum to determine which models yield compatible optical depths, $\tau_{\rm \hei\,\lambda1083\,nm} (v)$. Since this derives from $n_{\mathrm{1s2s\,^3S}}$, it in turn leads to a constraint on the \heii and total helium density via our modeling of the ionization and excitation state of helium in the ejecta. A more detailed discussion of KN P~Cygni modeling can be found in~\cite{Sneppen2023}.

\subsection{Resulting Helium mass limits}\label{sec:helium_bounds}

In Fig.~\ref{fig:Xshooter_helium}, we show the resulting spectra from calculations using our standard value of $n_e \approx 1.5\times10^7\,{\rm cm}^{-3}$ (at the photosphere, see App.~\ref{sec:electron-density}) and various helium mass fractions. If we assume helium is the cause of the feature, our best-fit P~Cygni model yields a Sobolev optical depth at the photosphere, $\tau_{\rm \hei\,\lambda1083\,nm, obs} = 0.86 \pm 0.18$ (ie. $n_{\mathrm{He}}\sim3\cdot10^{5} \,\mathrm{cm^{-3}}$) and a velocity range of the feature from approximately $0.19c$ to $\sim0.3c$ (note the ambiguities in the photospheric velocity is discussed in App.~\ref{sec:observing_time}). This corresponds to a helium mass fraction $X_{\rm He} \approx 0.006$ for our adopted total mass density, $\rho$ (recall, as discussed above, that $\rho$ is normalized to correspond with the observationally inferred ejecta mass for AT2017gfo).
As illustrated in Fig.~\ref{fig:Helium_ionisation}, assuming a lower electron density would permit a higher helium abundance. As argued in App.~\ref{sec:electron-density}, the lower limit on the electron density is $6\times10^{6}\,\mathrm{cm}^{-3}$, for which the 2$\sigma$ upper limit on the allowed helium mass fraction would be $X_{\rm He} \lesssim 0.05$. We stress that, within the framework of our modeling, this is a conservative limit as it assumes: i) the lowest likely electron density, ii) a relatively high photospheric velocity, iii) that helium dominates the feature, and iv) the 2$\sigma$ statistical upper limit.

Instead of relying solely on normalization of the density to the inferred ejecta mass, we note that $X_{\rm He}$ can also be constrained by considering the source of the free electrons. In particular, our favored combination of $n_e$ and $n_{\rm He}$ lies to the right of the gray unphysical region in the lower panel of Fig.~\ref{fig:Helium_ionisation}, indicating that charge conservation requires there be additional contributions to the free electron population beyond helium. 
In particular, for the low electron density limit $n_{\rm e} = 6\times10^{6}\,{\rm cm^{-3}}$ combined with $n_{\rm He} = (1.3\pm0.3)\times10^{6}\,{\rm cm^{-3}}$ (at the photosphere) only around one third of the free electrons can be from helium. Assuming the rest of the material is $r$-process rich with typical $A \simeq 100$ and is also weakly ionized, $Z_{\mathrm{ion}} \simeq 2$, the relative helium density to electron density implies $X_{\rm He} < 0.05$ for the 2$\sigma$ upper limit in $n_{\rm He}$. This result is consistent with the mass-based argument presented above, lending credence to our constraints. 


Lastly, we note that the derived limit corresponds to 60\% of the atoms being helium by number. Thus, these calculations support that a NSM ejecta with $X_{\rm He} \gtrsim 0.05$, i.e.,\ predominantly composed of helium by number, should yield a distinct signature in the kilonova spectra. 

\subsection{Ionization uncertainties and limitations in the current scope of modeling}\label{sec:helium_uncertainties}

In the following, we will explore the sensitivities of the model to the various assumptions used. 

The helium feature could be made invisible if \heiii completely dominates, which would be the case for an electron density significantly below $n_e \sim 10^{6} {\rm cm}^{-3}$ (see Fig.~\ref{fig:Helium_ionisation}). However, such low photospheric electron densities require very small ejecta masses (particularly for doubly ionized helium where a free electron exists for every two nucleons). An electron density of $n_e < 10^{6} {\rm cm}^{-3}$ implies $n_{\mathrm{He}} < 5\times 10^{5} {\rm cm}^{-3}$ and $m_{\mathrm{ejecta}} \lesssim 10^{-4} M_{\odot}$, which contradicts the observationally inferred ejecta mass in AT2017gfo in \citep{Smartt2017,Cowperthwaite2017,Villar2017,Siegel2019}. 
One could invoke a higher/lower deposition rate than the fiducial value of \citet{Hotokezaka2020}. However, across the broad range of ejecta conditions necessary to produce Sr, i.e., $0.25 < Y_e < 0.45$, the deposition rate per nucleon varies less than an order of magnitude up to timescales of 10 days \cite[e.g., Fig.~1 of the supplemental material in ref.][]{Wu2019b} with larger variations only seen at timescales longer than 10~days. Furthermore, the electron density is proportional to the number of ions, which depends only mildly on $Y_e$ for the range considered above. Ultimately, given the \heii ionization energy, $\mathcal{X} = 54.4$\,eV, it is difficult to see how helium would be predominantly in the \heiii state, while the typical singly ionized \rprocess element with $\mathcal{X} \approx$\,11\,eV is not much more highly ionized. Different density distributions or electron temperatures give broadly similar results over a large range in these parameters, as explored further in the Appendix (see particularly Fig.~\ref{fig:Xshooter_helium_3panel}). 

Conversely, the He 1\,\micron feature could be removed by suppressing the non-thermal particle flux. However, as the non-thermal flux is a direct product of the radioactive isotopes of the \rprocess nucleosynthesis, this requires a spatially distinct helium component insulated from the decays originating elsewhere in the ejecta. Such insulation would however have to be highly efficient as can be inferred from the lower panel of Fig.~\ref{fig:Helium_ionisation}, where a substantial weakening of the constraints require several order of magnitude shift to higher $n_e$ or equivalently weakening of deposition. Thus, the insulation would need to be nearly perfect, decreasing the ionizing flux by more than a factor of several hundred from expected values (see Fig.~\ref{fig:Xhe_verus}, middle panel). However, the ejecta are unlikely to reach this regime for two reasons. First, although highly tangled magnetic fields could partially trap charged particles, the reduced thermalisation efficiency at around 5\,days may allow non-thermal electrons to travel across the ejecta \citep{Barnes2016}. While $\gamma$-rays from nuclear decays interact less efficiently than electrons, in the context of avoiding a \hei dominated regime they are sufficient. For instance, in the models of \cite{Shingles2023} the energy deposited by $\gamma$-rays is typically 20\,\% of that by non-thermal electrons (ranging from 5\,\% to 100\,\% across the cell-to-cell variation) at a time when the model spectra are most similar to AT2017gfo -- thus consistently above the per~mille level required for a \hei dominated regime. Second, ignoring the non-thermal energy from radioactive decays in other parts of the ejecta, even the light \rprocess elements co-produced with helium in the high-$Y_e$ ejecta may release sufficient energy from nuclear decays to ensure \heii. For instance, the energy released in polar ejecta of the hydrodynamic model of a long-lived NS remnant (model sym-n1-a6; cf. Sect.~\ref{sec:HMNS_lifetime}), is characteristically around 10\,\% of the energy released in the equatorial ejecta. 

Lastly, we note there are several limitations to the current scope of modeling including i) the assumed sphericity in the P~Cygni model, ii) the photospheric approximation and an analytical prescription of density assuming smooth ejecta, iii) the steady-state approximation within the level-population modeling, iv) assuming the observed spectral continuum is indicative of the local radiation field, and v) the unknown contribution of other potential lines. These assumptions should be further tested with detailed radiative transfer calculations built on the output of realistic hydrodynamical simulations. Nonetheless, we consider several of these assumptions below and their limitations.

First, the P~Cygni implementation assumes a spherical ejecta structure (as potentially motivated by KN spectral features, see \citep{Sneppen2023,Sneppen2024}), but we here solely focus on the constraints of the absorption feature, which is most sensitive to the line-of-sight ejecta (which given the viewing angle is the polar ejecta, \citep{Mooley2018,Mooley2022}). In contrast, material in the equatorial plane should not drastically impact the line-of-sight absorption of polar ejecta. 

Second, regardless of whether a soft or sharp photosphere exists or the ejecta distribution, the outer ejecta will be highly susceptible to the helium absorption line as long as sufficient helium is present in regions with the required electron densities and radioactive heating rates (following the arguments connected to Fig.~\ref{fig:Helium_ionisation}). 
While the ejecta structure could potentially be clumped (as motivated in \citep{Gillanders2023} from the non-detection of the \srii forbidden line doublet), such clumping would need to imply several orders of magnitude increase in the dominant $n_e$ regime to affect our constraints, as the first order effect of increasing $n_e$ is to move further away from the \heiii-dominated regime. 

Third, the steady-state approximation is sensitive to the recombination timescale (i.e.,\ the slowest/bottleneck rate), which at electron densities $n_e \sim 3 \times 10^{6}$\,cm$^{-3}$ become comparable to the timescale post-merger (ie. $\sim$4.4\,days). Thus, if $n_e \lesssim 3 \times 10^{6}$\,cm$^{-3}$ the steady-state approximation would likely break down and more detailed time-dependent modeling would be needed.

Fourth, changes to the assumed radiation field within the range consistent with observed emission yield limited effect, as the pathways that dominate into (recombination rate) and away from (natural decay) the triplet \hei states are not sensitive to the radiation field. However, future radiation transport simulations can help constrain the likely properties of the local radiation field within the line-forming region, which determines photoionization/photoexcitation in the NLTE solution. 

Fifth, other lines can contribute and potentially bias inferred properties of the 1\micron feature, but such a bias would require i) near-perfectly filling in the distinct signature of strong absorption in the limited timespan where a helium feature is relevant (i.e.,\ the late-photospheric epochs), and ii) yielding no distinct evidence in preceding or subsequent epochs. Thus such a bias requires fine-tuning. The various other observed spectral features of AT2017gfo highlight how distinctly an individual element (under the correct conditions) can yield interpretable and relatively isolated spectral signatures. 

\subsection{Strontium contribution}\label{sec:strontium}

The bound on the helium abundance must be considered an upper limit, as it remains unclear whether other species (most notably \srii) contribute to the observed feature. Indeed if the majority of the feature is not due to helium (but rather another element), the upper limit on $X_{\mathrm{He}}$ could decrease and potentially fall below the range covered by our hydrodynamical simulations of Sect.~\ref{sec:HMNS_lifetime} (e.g., $X_{\mathrm{He}}\lesssim0.02$). \srii lines are known to be a significant and dominant contributor to the feature in early epochs \citep{Watson2019,Gillanders2022, Domoto2022,Vieira2023,Sneppen2023c,Sneppen2024}. 

Strontium could potentially offer similar lifetime constraints as helium if reliable NLTE models were available, particularly at earlier times when the Sr contribution is likely to be dominant. This could in principle i) provide a complementary constraint on remnant lifetime from another element and ii) decrease the helium spectral contribution, thereby tightening the upper bound on $X_{\mathrm{He}}$. Ref.~\citep{Perego2022} proposed that the abundance of Sr, specifically the observationally inferred Sr masses under LTE \citep[crudely lower-bounded at around $\sim10^{-5} M_{\odot}$,][]{Watson2019} may be inconsistent with NSM models harbouring long-lived NS remnants. However, recent constraints in the AT2017gfo literature on the strontium mass fraction, $X_{\mathrm{Sr}}$, vary by orders of magnitude, ranging from the permille level \citep{Watson2019,Gillanders2022} to the several-percent regime \citep{Tarumi2023} depending on ionization-state modeling and particularly the degree of non-thermal ionization effects. This wide range of allowed $X_{\mathrm{Sr}}$ encompasses a broad domain of allowed nucleosynthesis conditions in entropy and $Y_e$ and is therefore not particularly restrictive for hydrodynamic models of NSMs, as we detail further in App.\:\ref{app:strontium}. In this appendix, we also show that Sr is produced over a broad range of nucleosynthetic conditions and is not strongly linked to the NS remnant outflows.

A significant \srii contribution could further strengthen the helium limit in proportion to its share of the feature, so using the theoretical results for recombination rates recently reported \citep{Banerjee2025,Singh2025}, we estimate the relative abundance of \srii (cf. Fig.~\ref{fig:Helium_ionisation}, top panel). These results indicate that \srii is a prominent ion state at electron densities similar to those where \hei dominates (e.g., Fig.~\ref{fig:Helium_ionisation}, top panel). Thus, for $n_e\approx 10^{8}-10^{9}\,{\rm cm}^{-3}$ (as applicable in early epochs) \srii is a significant ionization state and dominant contributor to the feature's optical depth. If such high densities were present at 4-5 days, the inferred upper limit on $X_{\mathrm{He}}$ would decrease and might be in tension with hydrodynamic NSM models (see Sect.~\ref{sec:HMNS_lifetime}). However, a lower $n_e$ is expected at these times for generic ejecta masses (see App.~\ref{sec:electron-density}). For $n_e\approx 10^{7}\,{\rm cm}^{-3}$ higher strontium ionization states can dominate, in which case the contribution of \srii lines to the absorption feature would become negligible for a strontium mass fraction limited to a few percent, and the limit of $X_{\mathrm{He}}<0.05$ derived assuming pure helium would apply. More details on Sr constraints for NLTE modelling can be found in \cite{Tarumi2023}, Arya et al (in prep.) and Chiba et al (in prep.). Ultimately, with regard to the 4.4 day absorption feature, \srii may contribute or it may not. Given the large uncertainties in strontium ionization, we adopt the most conservative constraint, $X_{\mathrm{He}}<0.05$, leaving further tightening of this constraint and any constraint on $X_{Sr}$ to future studies.


\section{\label{sec:HMNS_lifetime} Helium production in neutron-star merger models}

\begin{figure}
  \centering
    \includegraphics[width=0.95\linewidth,viewport=22 20 350 345 ,clip=]{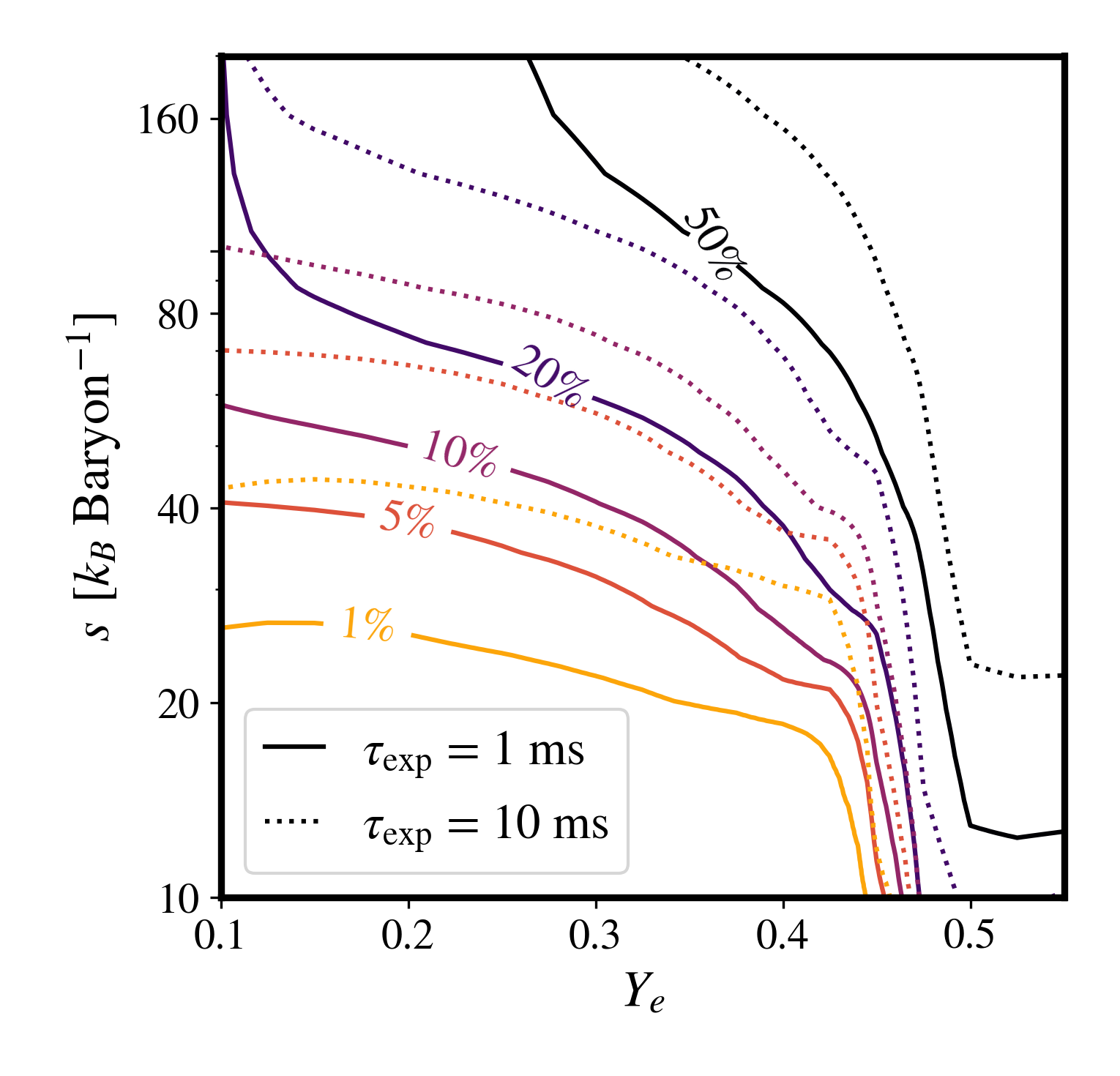}
  \caption{Selected contours of the helium mass fraction as a function of electron fraction, $Y_e$, and entropy, $s$, for parametrized outflow conditions (as in \citep{Lippuner2015}) and expansion timescales typical of dynamical ejecta ($\tau_{\mathrm{exp}}=$1\,ms; solid lines) and post-merger ejecta ($\tau_{\mathrm{exp}}=$10\,ms; dotted lines). The nucleosynthesis calculations are from \citep{Gross2023} except that the nuclear network is started at 8\,GK instead of 6\,GK to account for quasi-statistical equilibrium corrections \citep{Meyer1998,Hix.Thielemann:1999a}. Starting at 6\,GK would result in an artificially steep transition to large $X_{\mathrm{He}}$ (see Fig.~3 of \citep{Perego2022} for comparison). The absence of a helium feature in AT2017gfo rules out a significant fraction of the ejecta to have $Y_e \gtrsim 0.45$ and $s \gtrsim40 \,k_B \, {\rm baryon^{-1}}$.} 
    \label{fig:He_nucleosynthesis}
\end{figure}

The contours of helium yields shown in Fig.~\ref{fig:He_nucleosynthesis} for exemplary, parametrized trajectories highlight what is known already from numerous previous nucleosynthesis studies \citep[e.g.,][]{Woosley1992a, Takahashi1994, Hoffman1997, Goriely2015, Lippuner2017, Perego2022, Kawaguchi2022, Jacobi2025arXiv}, namely that helium is efficiently produced in nucleosynthesis conditions with high electron fractions, $Y_e \gtrsim 0.45$, and/or high entropies, $s \gtrsim 40 \,k_B\,{\rm baryon^{-1}}$. The dynamical outflows expelled during the first $\sim 10\,$ms of NS mergers \citep[][]{Wanajo2014a, Goriely2015, Radice2018b, Foucart2023a, Kullmann2021a}, as well as the viscous post-merger ejecta \citep{Fernandez2013b, Just2015a, Siegel2018c, Miller2019a, Fujibayashi2020b}, typically show a broad pattern of conditions for $Y_e$ and $s$ with only a small fraction of material, if any, exhibiting sufficiently high values of either quantity to enable helium production. These ejecta components are therefore relatively inefficient sources of helium.

In contrast, high values of $Y_e$ and $s$ are characteristic of outflows powered by, or even just irradiated by, neutrinos. Neutrino-driven winds are well known from the field of core-collapse supernovae (CCSNe) \cite[e.g.,][]{Witti1994, Qian1996, Thompson2001a, Arcones2007, Hudepohl2010a, Fischer2010, Roberts.Woosley.Hoffman:2010, Wanajo2013a, Fischer.Guo.ea:2024} and by now have also been studied in a significant number of works in the context of NSMs \citep[e.g.,][]{Metzger2008b, Dessart2009, Surman2011a, Wanajo2012, Metzger2014, Perego2014a, Just2015a, Fujibayashi2018a, Just2023, Bernuzzi2024a, Cheong2024b}. The subsequent sections will elaborate on the basic physics of and expected electron fraction in neutrino-driven winds and the implication of the observational $X_{\mathrm{He}}$ limit from Sect.~\ref{sec:observed_helium_constraints} for the NS remnant lifetime, followed by a discussion of the modeling uncertainties.

\subsection{Expected electron fraction of neutrino-driven winds}\label{sec:expect-electr-fract}

\begin{figure*}
  \includegraphics[width=\linewidth]{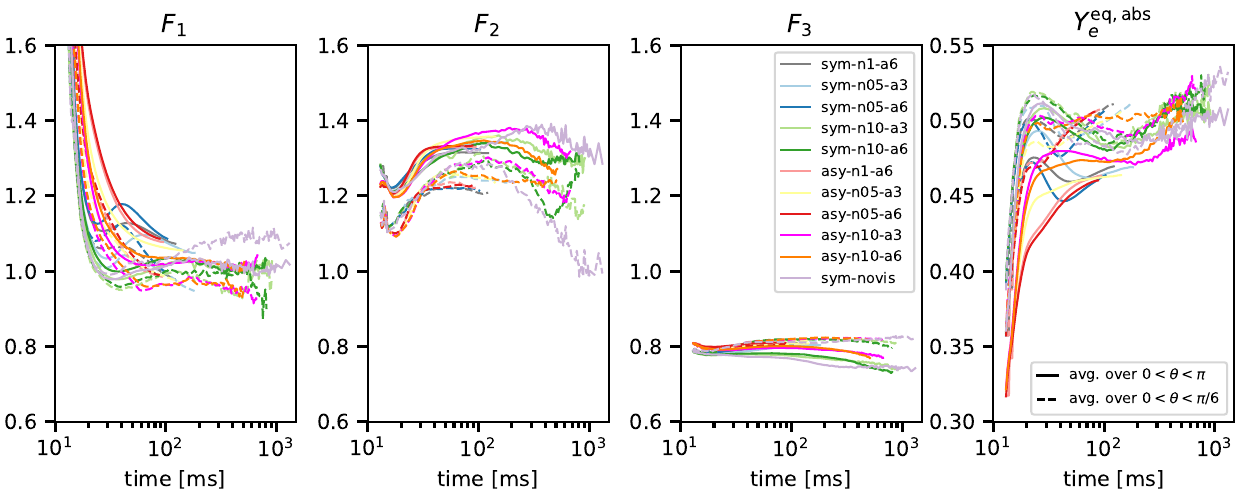}
  \caption{Factors $F_{1/2/3}$ of Eqs.~(\ref{eq:F1})--(\ref{eq:F3}) (three panels from left) entering the estimate of the electron fraction resulting in neutrino winds, $Y_e^{\rm eq,abs}$ (cf. Eq.~(\ref{eq:yeeq})), as well as $Y_e^{\rm eq,abs}$ itself (right panel) for all hydrodynamic simulation models in which the NS remnant survives longer than 10\,ms. All quantities are measured at a radius of 500\,km. Solid lines are obtained using quantites averaged across the entire sphere, while dashed lines use quantities averaged only within $30^\circ$ from the north pole, thus more appropriately probing the conditions in the polar wind. Lines are plotted only until BH formation, because neutrino winds launched from the BH-torus systems are significantly less massive.}
  \label{fig:yeeq}
\end{figure*}

In neutrino-driven winds, material surrounding hot and dense regions of intense neutrino emission is ejected as a consequence of continuous deposition of energy due mainly to absorption of electron-type neutrinos on free nucleons through the following reactions:
\begin{align}\label{eq:beta}
   \nu_e + n &\rightarrow p + e^- \, , \nonumber \\
   \bar\nu_e + p &\rightarrow n + e^+ \, .
\end{align}
These neutrinos were originally emitted from the cooling layer via the inverse of these reactions, with energy emission rates per unit mass that are strongly temperature dependent \citep[e.g., $\propto T^6$ ][]{Bruenn1985}. In the region surrounding the cooling layer, the so-called gain (or heating) layer, neutrino emission is inefficient, owing to the lower temperatures, and dominated by neutrino absorption. In this region, material gains enough thermal energy (ranging up to $\sim$100\,MeV per baryon) for the ensuing $pdV$ expansion to unbind the material from the gravitational potential of the central object. The heating rate per unit of mass associated with the reactions of Eq.~(\ref{eq:beta}) is approximately proportional to $L_{N,\nu} \langle\epsilon^3\rangle_{\nu}$ \citep[e.g.,][]{Bruenn1985}, which means that the mass-outflow rates in neutrino-driven winds grow with the number-loss rates of electron-type neutrinos, $L_{N,\nu}$, and with some higher moment of the neutrino energy distribution\footnote{The number-loss rate $L_{N,\nu}$ used here is related to the otherwise often quoted energy luminosity $L_{\nu}$ by $L_{N,\nu}\langle\epsilon\rangle_{\nu}=L_{\nu}$. The energy moments are defined as $\langle\epsilon^n\rangle_\nu=(\int \epsilon^{2+n}f_{\nu}d\epsilon)/(\int \epsilon^{2}f_{\nu}d\epsilon)$ for $n=1,2,\ldots$ with $f_{\nu}$ being the distribution function of neutrino species $\nu$.}. Therefore, strong neutrino-driven winds are naturally expected to result from hot NSM remnants that release neutrinos with high luminosities and high average energies.

The absorption reactions, Eq.~(\ref{eq:beta}), not only heat the material and raise its entropy, but also increase the electron fraction of the ejecta. An estimate of $Y_e$ resulting in neutrino-driven winds can be obtained by neglecting neutrino emission and considering the equilibrium state achieved when both rates of Eq.~(\ref{eq:beta}) become equal. This defines the value that $Y_e$ would relax to for given, fixed thermodynamic conditions and neutrino distributions. The resulting value for this absorption equilibrium is given by \citep{Qian1996, Horowitz1999a}:
\begin{equation}\label{eq:yeeq}
  Y_e^{\rm eq,abs} \approx  \frac{1}{1+ F_1 F_2 F_3}\, ,
\end{equation}
where the three factors $F_{1/2/3}$ are given by the ratio of number-luminosities of both electron-type neutrinos,
\begin{equation}\label{eq:F1}
  F_1 = \frac{L_{N,\bar\nu_e}}{L_{N,\nu_e}} \, ,
\end{equation}
the ratio of mean-squared energies (modulo corrections including the rest-mass difference between neutrons and protons, $Q_{np}\approx 1.29$\,MeV),
\begin{equation}\label{eq:F2}
  F_2 = \frac{\langle\epsilon^2\rangle_{\bar\nu_e}-2 Q_{np} \langle\epsilon\rangle_{\bar\nu_e}} {\langle\epsilon^2\rangle_{\nu_e} + 2 Q_{np} \langle\epsilon\rangle_{\nu_e}} \, ,
\end{equation}
and a factor representing weak-magnetism and recoil corrections \citep{Horowitz2002a},
\begin{equation}\label{eq:F3}
  F_3 = \frac{1-7.22\langle\epsilon^3\rangle_{\bar\nu_e}/\langle\epsilon^2\rangle_{\bar\nu_e}}{1+1.02\langle\epsilon^3\rangle_{\nu_e}/\langle\epsilon^2\rangle_{\nu_e}} \, .
\end{equation}
Once the dynamical merger phase is over and the remnant has relaxed into a nearly axisymmetric configuration ($t\gtrsim 10\,$ms), the hydrodynamic evolution of the remnant is mainly determined by viscous heating and angular-momentum transport, and by cooling of the initially hot gas via neutrino emission. These processes operate on relatively long timescales of $\gtrsim 0.1-1\,$s compared to the typical weak-interaction timescale of $\lesssim 0.001\,$s \citep[e.g.,][]{Perego2014a, De2021k, Just2022a, Radice2023c}. Thus, $F_1=L_{N,\bar\nu_e}/L_{N,\nu_e}\rightarrow 1$, because the average $Y_e$ of the cooling layer converges towards an emission equilibrium \cite{Just2022a} characterized by equal rates of the inverse reactions of Eqs.~(\ref{eq:beta}). The spectral ratio $F_2$ is slightly above unity, because the neutron-rich conditions in the cooling layer cause absorption reactions of $\bar\nu_e$ on protons to occur less frequently than $\nu_e$ absorption on neutrons, allowing $\bar\nu_e$ neutrinos to escape from deeper, and therefore hotter, regions of the remnant compared to $\nu_e$. Lastly, $F_3$ is slightly smaller than unity, because weak-magnetism and recoil corrections reduce the opacity of $\bar\nu_e$ in the gain layer. Altogether, the product $F_1\, F_2\, F_3$ is expected to reach values close to unity and, therefore, $Y_e^{\rm eq,abs}$ generically lies close to 0.5 in neutrino-driven winds.

Figure~\ref{fig:yeeq} illustrates the values of $F_1, F_2, F_3,$~and~$Y_e^{\rm eq,abs}$ resulting in our recently developed ``end-to-end'' neutrino-hydrodynamics simulations, which follow all three evolutionary phases of matter ejection, namely the dynamical merger, the NS-torus phase, and the final BH-torus evolution (see \citet{Just2023} and Appendix~\ref{sec:hydrodynamic-models} for more details). These models adopt sophisticated neutrino treatments and describe angular-momentum transport due to small-scale turbulence using a recently developed two-parameter viscosity prescription inspired by the $\alpha$-viscosity scheme of Ref.~\citep{Shakura1973}. In contrast to CCSNe, remnants of NSMs are characterized by large angular variations with respect to the rotational axis, leading to a much stronger impact of neutrino irradiation along the poles compared to the equator \citep[e.g.,][]{Dessart2009, Perego2014a, Just2015a, Richers2015a, Miller2019a, Gizzi2021b, Foucart2024b}. The dashed lines in Fig.~\ref{fig:yeeq}, which only refer to the conditions at the poles, confirm the aforementioned expectation that $Y_e^{\rm eq,abs}\rightarrow 0.5$ in polar neutrino winds.

Figure~\ref{fig:yeeq} suggests neutrino-driven winds to be suitable sites of helium production, however, Eq.~\ref{eq:yeeq} only provides a rough estimate of the actual $Y_e$, because it adopts simplifications regarding the neutrino field and assumes the absorption reactions to occur fast enough for $Y_e$ to reach the weak-equilibrium value. As detailed in the next section, our models indeed exhibit favorable conditions for helium production in the polar winds from the NS remnant.

\subsection{Lifetime constraint from observational helium abundance limit}\label{sec:lifetime-constraint}

\begin{figure*}
  \includegraphics[width=\linewidth]{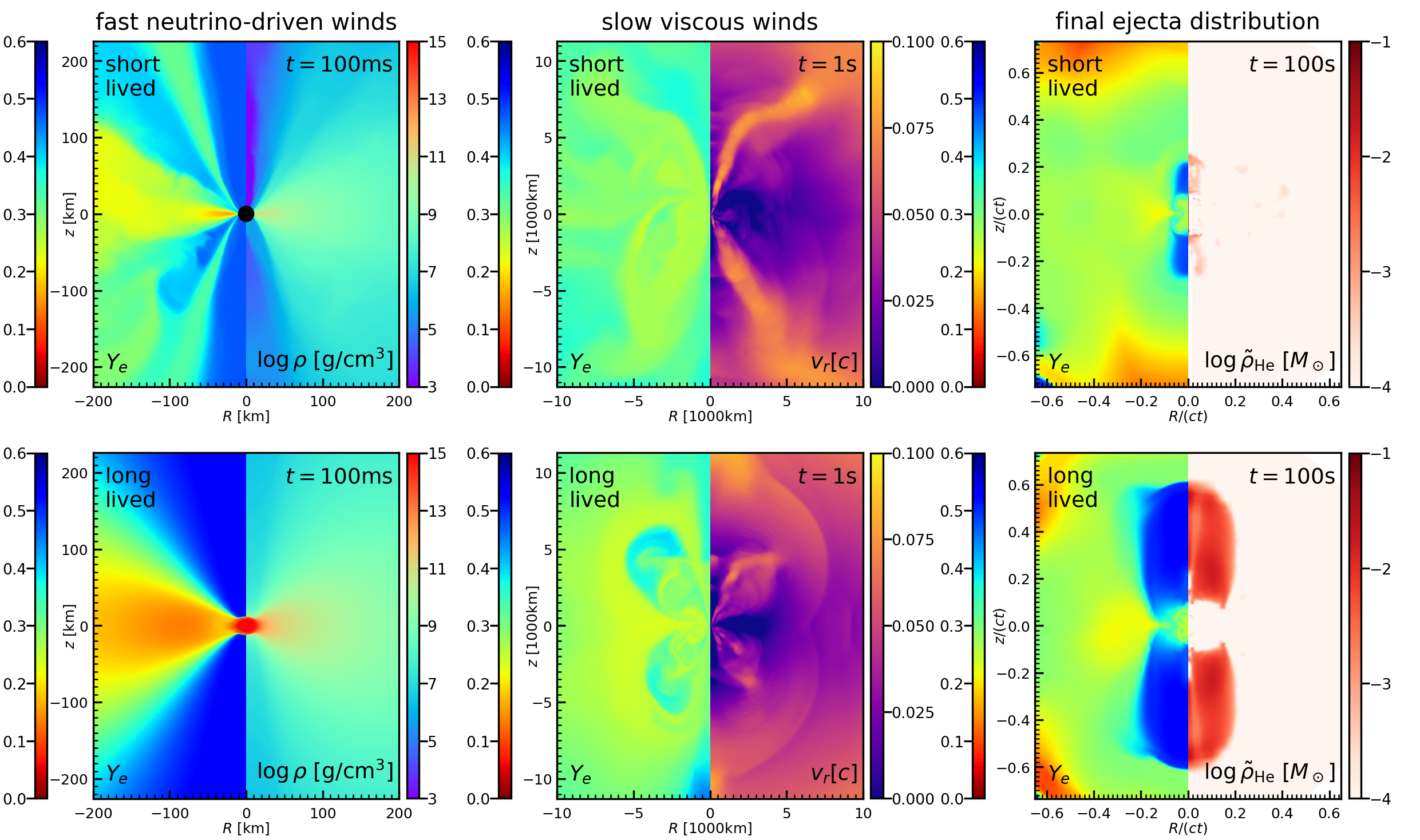}
  \caption{Snapshots from two numerical simulations in which the NS remnants are short lived (model ``sym-n1-a6-short'' with $\tau_{\mathrm{BH}}=10$\,ms; top row) and long lived (``sym-n1-a6'' with $\tau_{\mathrm{BH}}=122$\,ms; bottom row) at three characteristic times after merger illustrating the launch of early, fast outflows (left column) and late, slow outflows (middle column) as well as the final ejecta configuration in velocity space (right column), where $Y_e, \rho$, and $v_r$ are the electron fraction (measured before onset of the \rprocess), mass density, and radial velocity, respectively, and $\tilde\rho_{\mathrm{He}}=\mathrm{d}M_{\mathrm{He}}/\mathrm{d}\beta_r/\mathrm{d}\Omega = \rho X_{\mathrm{He}}\beta_r^2(c t)^3$ is the final helium density in dimensionless velocity space (with $\beta_r=r/(ct)$ and solid-angle element $\mathrm{d}\Omega$) rescaled by $\beta_r^2$ to enhance visibility for outflows with strong density decline. Being a stronger source of neutrinos than a BH-torus remnant, the NS remnant produces a far more massive and extended helium-rich, high-$Y_e$ ejecta component at early times, while the late BH-torus outflows are inefficient helium sources in both cases.}
  \label{fig:sim_snapshots}
\end{figure*}

Neutrino-driven, or neutrino-irradiated, winds can in principle be launched both before and after collapse of the NS remnant (which we denote as HMNS hereafter). However, since BH-tori are considerably weaker sources of neutrino emission than HMNSs, the neutrino-wind masses in BH-torus remnants are predicted to be much smaller than those of other ejecta components \citep{Fernandez2013b, Just2015a, Fujibayashi2020a, Kawaguchi2025a}. Conversely, neutrino winds from the HMNS can be as massive as -- or even exceed -- the dynamical and BH-torus ejecta \citep{Metzger2014, Just2023, Fujibayashi2020b, Jacobi2025arXiv}, particularly when being enhanced due to magneto-hydrodynamic effects \citep{Combi2023a, Curtis2024a, Kiuchi2024a}. Thus, the ejecta launched during the HMNS phase are likely distinguished from the dynamical and BH-torus ejecta in that they produce a substantially greater amount of helium.

This notion is supported by our hydrodynamic models. The snapshots shown in Fig.~\ref{fig:sim_snapshots} illustrate the crucial difference between a model with a short-lived HMNS remnant (with BH-formation time at $\tau_{\mathrm{BH}}\approx 10\,$ms, top row) and a long-lived one (with $\tau_{\mathrm{BH}}\approx 120\,$ms, bottom row): In the latter case the HMNS, due to its longer lifetime, gives rise to a massive and extended high-$Y_e$, high-entropy neutrino-driven wind along both polar directions, leading to substantial helium enrichment at final ejecta velocities of $0.15\lesssim v/c\lesssim 0.6$ within a cone of $\sim 20-40^\circ$ half-opening angle (see right panels of Fig.~\ref{fig:sim_snapshots} for the spatial distribution of helium in the final ejecta configuration as well as Fig.~\ref{fig:Y-s-histograms} for the mass distribution in $Y_e$ and $s$).

\begin{figure}
    \includegraphics[width=\linewidth]{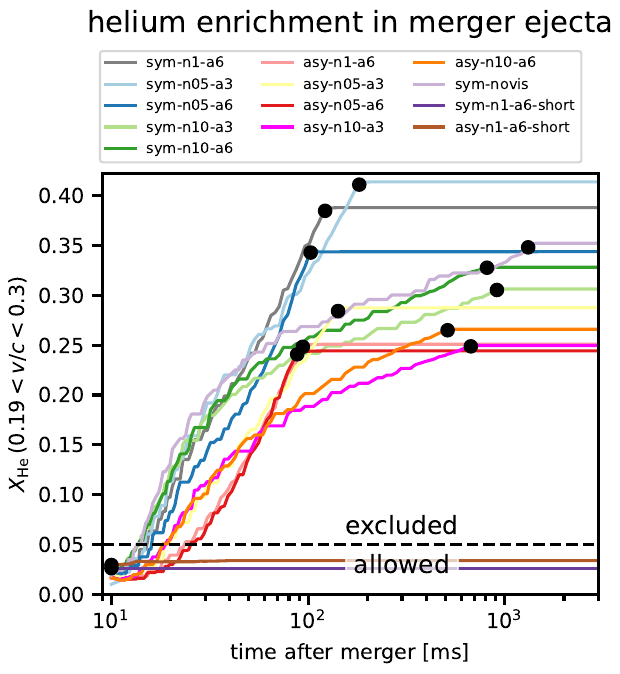}
    \caption{
    Mass of material ejected until the given time that will end up as helium relative to the mass of all material ejected until that time in our hydrodynamic simulation models (cf. Appendix~\ref{sec:hydrodynamic-models}). Only material in the observationally relevant velocity range, $0.19<v/c<0.3$, is considered. We count as ejecta all material that at a given time is radially expanding faster than 0.1\,$c$. The helium abundance plateaus following BH formation at $t=\tau_{\mathrm{BH}}$, indicated for each model with black circles. The observational constraint $X_{\mathrm{He}}=0.05$ is shown by the dashed line. Neutrino winds continuously inject helium into the ejecta but only as long as the HMNS is present, leading to a strong correlation between the helium abundance and the HMNS lifetime. In this set of models only the two short-lived models (denoted by ``short'' suffixes) with $\tau_{\mathrm{BH}}=10\,$ms satisfy the observational constraint.}
    \label{fig:He_HMNS_time}
\end{figure}

Since the HMNS injects the wind with mass fluxes that vary only slowly with time, the relative fraction $X_{\mathrm{He}}(t)$ of mass ending up as helium in the observationally relevant (cf. Sect.~\ref{sec:observed_helium_constraints}) velocity band $0.19\lesssim v/c\lesssim 0.3$ keeps growing continuously for the long-lived ($\tau_{\mathrm{BH}}> 10\,$ms) models, reaching values of 20--40\,$\%$ until abruptly saturating when the HMNS undergoes BH formation; see Fig.~\ref{fig:He_HMNS_time}. Given a sufficiently long HMNS lifetime, helium becomes the most abundant element (by mass) in the entire outflow. Clearly, all the long-lived models shown in Fig.~\ref{fig:He_HMNS_time} are immediately ruled out by the observational constraint $X_{\mathrm{He}}<0.05$ (cf.\ Sect.~\ref{sec:observed_helium_constraints}), while the short-lived models are compatible. Assuming that our set of models is representative concerning the behavior of $X_{\mathrm{He}}(t)$ (see discussion below), Fig.~\ref{fig:He_HMNS_time} implies that the observational constraint can only be fulfilled for relatively short lifetimes of
\begin{equation}\label{eq:tauconstraint}
  \tau_{\mathrm{BH}}\,(\,\mathrm{AT2017gfo}\,)\,\lesssim 20-30\,\mathrm{ms} \, .
\end{equation}
In other words, the HMNS remnant in AT2017gfo must have collapsed within a few tens of milliseconds after the merger, because otherwise it would have blown out enough helium to be clearly observable in the kilonova spectra according to the P~Cygni analysis of Sect.~\ref{sec:observed_helium_constraints}. Importantly, given the rapid growth of $X_{\mathrm{He}}(t)$, even a less constraining bound, of say $X_{\mathrm{He}}<0.1$, would result in a strong lifetime constraint. Additionally, we note the helium-enriched models also display typical electron densities in the range $2.5\times10^{7}$\,cm$^{-3}-2\times 10^{8}$\,cm$^{-3}$ at 4.4 days, across grid cells in the polar ejecta at the velocities of interest (see App.\,\ref{sec:electron-density}), which motivates the observability of helium under the specific radiative transfer conditions of the hydrodynamical simulations.

We note in passing that the large angular anisotropy created by the polar neutrino winds would also be at odds with the quasi-spherical geometry suggested by the observed spectral features in AT2017gfo (cf. for instance the P~Cygni features and discussions in \citep{Sneppen2023,Sneppen2024}, but see also \citep{Collins2023}).

\subsection{Merger-modeling uncertainties}\label{sec:discussion-1}

A few comments are in order regarding the lifetime constraint, Eq.~\ref{eq:tauconstraint}. While the idea of HMNS remnants producing high-$Y_e$ winds is not new \citep[e.g.,][]{Dessart2009, Metzger2014, Perego2014a, Lippuner2017, Foucart2016a, Fujibayashi2018, Just2023, Curtis2024a}, the question of how fast these winds enrich the ejecta with helium has not been addressed so far.
The set of models is still relatively small and therefore probably not exhaustive regarding the impact of different progenitor masses, mass ratios, and EoSs. However, the model-by-model variation of the time corresponding to $X_{\mathrm{He}}(t)=0.05$ is not more than about a factor of two, even for cases where the lifetimes \taubh differ by one order of magnitude, suggesting a robustness of the helium-enrichment mechanism and only a relatively mild sensitivity of the lifetime constraint, Eq.~\ref{eq:tauconstraint}, with respect to these uncertainties.

We now briefly discuss additional sources of uncertainty from the physics approximations adopted to make the long-term simulations computationally feasible. The mapping from 3D to 2D simulations at $t=10\,$ms onto an axisymmetric post-merger simulation code using a different (more approximate) treatment of general relativity creates an artificial perturbation that results in the ejection of a wave of spurious ejecta right after the mapping. However, this initial transient-like outflow carries away at most a few per cent of the total amount of ejecta and it does not exhibit a significant helium mass fraction. Thus, it should have little to no impact on the helium production in the neutrino-driven wind at later times. In order to test the sensitivity of our 2D simulation results, we continued two 3D merger simulations (which adopt a general relativistic gravity treatment; see Appendix~\ref{sec:hydrodynamic-models}) to later times and analyzed the helium yields produced during just the 3D evolution phase. The $X_{\mathrm{He}}(t)$ curves resulting from these 3D models, shown in Fig.~\ref{fig:vimal_xhe}, resemble the behaviour of the 2D models, which suggests that the steep increase of $X_{\mathrm{He}}(t)$ due to polar HMNS winds is not particularly sensitive to the mapping, axisymmetry, or approximate treatment of gravity in our models.

Next, our viscosity treatment approximates the effects of small-scale turbulence due to, e.g., the magneto-rotational \citep{Balbus1991} or other instabilities \citep[e.g.,][]{Margalit2022}. While this treatment has been used extensively in the literature of accretion disks, in the context of NSMs it has been rigorously compared with more realistic 3D magneto-hydrodynamic simulations only in rare cases \citep[e.g.,][]{Fernandez2019b, De2021k, Just2022a, Kiuchi2024b}. Nevertheless, our set of models covers substantial variations in the chosen viscosity parameters, suggesting only a mild sensitivity of the $X_{\mathrm{He}}(t)$ curves to the viscosity. In fact, the fastest rise in helium abundance is obtained for the non-viscous model (sym-novis), proving that the powerful enrichment of helium (relative to the total mass) is not artificially enhanced by the viscosity treatment. Besides triggering turbulence, magnetic fields may also develop large-scale structures that contribute in driving outflows in the polar direction \citep[e.g.,][]{Combi2023a, Curtis2024a, Aguilera-Miret2023b, Kiuchi2024a, Musolino2024c, Kalinani2025a}. No obvious reason exists for $Y_e^{\rm eq,abs}$ to be significantly reduced in such magnetically-driven outflows. However, the actual $Y_e$ values may freeze out at lower values than $Y_e^{\rm eq,abs}$ if the expansion proceeds very fast in these outflows (namely faster than absorption reactions drive $Y_e$ towards $Y_e^{\rm eq,abs}$). Since in our set of viscous models the polar outflows are launched mainly via neutrino heating, their $Y_e$ may turn out to be artificially close to $Y_e^{\rm eq,abs}$ and thereby possibly overestimate the resulting helium abundance compared to the case in which the winds are (additionally) powered by MHD effects. Future studies incorporating MHD models with reliable neutrino transport (see next paragraph) will need to investigate the impact of MHD on the lifetime suggested by our viscous models, Eq.~\ref{eq:tauconstraint}.


Another potential source of uncertainty is the neutrino transport. Since six-dimensional Boltzmann schemes are very expensive \citep[e.g.,][]{Miller2019a,Foucart2024b, Kawaguchi2025a}, all existing long-term simulations of merger remnants, including ours, adopt approximations to deal with the direction dependence of the neutrino field in momentum space. While we can only speculate without Boltzmann solutions at hand, we expect some quantitative impact of the M1 scheme, but likely no qualitative change of the steep increase of $X_{\mathrm{He}}(t)$. This would require considerable errors in the predictions of $F_{1/2/3}$ and $Y_e^{\rm eq,abs}$ in Sect.~\ref{sec:expect-electr-fract} and imply a significantly different behavior of neutrino-driven winds in NSMs than in CCSNe. Among long-term HMNS simulations available in the literature, the energy-dependent M1 transport scheme used here is relatively advanced. It resolves the energy spectrum of neutrinos unlike grey neutrino-transport schemes \citep[e.g.,][]{Andresen2024a,Cheong2024c}, and it takes into account weak-magnetism corrections -- both important for reliable estimates of $Y_e^{\rm eq,abs}$ (cf. Eqs.~(\ref{eq:yeeq}){--}(\ref{eq:F3})).

Regardless of the uncertainties discussed above, the dichotomy between long-lived and short-lived models seen in Fig.~\ref{fig:He_HMNS_time} is striking, and we leave it to future work to explore in more detail the uncertainties of the $X_{\mathrm{He}}(t)$ dependence and of the implied lifetime constraint, Eq.~\ref{eq:tauconstraint}. A meaningful comparison with other literature results for the $X_{\mathrm{He}}(t)$ evolution is difficult, if not impossible, at this point, because so far only a small number of long-term merger-remnant simulations exist that are capable of describing neutrino winds\footnote{Pure neutrino-leakage schemes (based on Ref.~\citep{Ruffert1996a} without additional treatment of neutrino absorption), which are often adopted in the merger literature, only describe (net) neutrino cooling, i.e., no heating, and are therefore unable to capture neutrino winds.} combined with a (full or approximate) treatment of general relativistic gravity. Only a fraction of those report helium abundances and none of them report helium abundances in just the relevant velocity band $0.19\lesssim v/c\lesssim 0.3$. We remark, however, that the results reported by Refs.~\citep{Lippuner2017, Fujibayashi2020a, Jacobi2025arXiv} for the helium production in polar neutrino winds of long-lived HMNSs appear to be in broad agreement with our models, while models with short-lived remnants do appear to be systematically deficient of high $Y_e$ material \citep[e.g.,][]{Kiuchi2023}. This work motivates more in-depths analyses of the conditions for helium nucleosynthesis in merger remnants, particularly for the so far sparsely sampled cases with NS-remnant lifetimes of a few hundred milliseconds.

Our radiative transfer results (cf. Fig.~\ref{fig:Xshooter_helium}) indicate that the upper limit of $X_{\mathrm{He}}$ may even be as low as 0.01 (or lower). Such low values are not compatible with our current set of hydrodynamic models (cf. Fig.~\ref{fig:He_HMNS_time}) in which the smallest value of $X_{\mathrm{He}}$ is about 0.03. This means that either systems with lower $X_{\mathrm{He}}$ exist that are not covered by our set of models (presumably with shorter lifetimes \taubh$<10$\,ms), or our radiative transfer modeling is too approximate to reliably limit $X_{\mathrm{He}}$ in the sub-percent range. At this point it is unclear if a relation between the $X_{\mathrm{He}}$ limit and the lifetime holds in the regime of very low helium mass fractions. However, for our purposes this regime of low helium abundances is not critical provided that such low He mass fraction is incompatible with long-lived merger remnants. Under this condition it is reasonable to adopt a lifetime limit \taubh$<20$\,ms.

We finally point out that the polar nature of the helium-rich wind is particularly well suited for our observational constraint because of the near-polar viewing angle in AT2017gfo and the circumstance that absorption features in KN spectra are formed mainly along the line-of-sight. The broad velocity distribution of the wind, ranging from 0.1\,$c$ to 0.6\,$c$ is further auspicious for observability, because it safely encompasses the velocities of the line-forming region, $0.19c\lesssim v\lesssim 0.3c$, at around 4-5\,days post-merger. In earlier spectra, the observed line-forming region could constrain ejecta at larger velocities - even reaching out to $\sim$0.45$c$ at 1.17\,days \cite[][]{Sneppen2024}. However, as deliberated in App.~\ref{sec:observing_time}, at earlier times radiative transitions will suppress the \hei feature and imply weaker abundance constraints for such outer layers.

\section{\label{sec:EOS_constraints} Equation of state constraints }
\subsection{Lifetime and the proximity to the threshold mass of prompt collapse}
The upper limit on the remnant lifetime can be turned into an EoS constraint by recognizing that the lifetime indicates the proximity of the measured total binary mass, $M_\mathrm{tot}$, to the threshold mass for prompt black hole formation, $M_\mathrm{thres}$. The lifetime is expected to steeply decrease with higher total binary mass to reach roughly zero at $M_\mathrm{thres}$. In essence, therefore, the absence of significant amounts of He implies an upper limit on $M_\mathrm{thres}$. $M_\mathrm{thres}$ scales well with stellar parameters (e.g.,\ radii, maximum mass, tidal deformability) of non-rotating NSs~\citep{Bauswein2013b,Bauswein2017b,Koelsch2022,Koeppel2019,Agathos2020,Bauswein2020,Bauswein2021,Tootle2021,Kashyap2022,Ecker2024,Perego2022prl}, which can thus be constrained. This line of argument has already been presented in~\citep{Bauswein2017} (see Fig.~5 in that paper for a hypothetical case).

\begin{figure}
    \includegraphics[width=\linewidth]{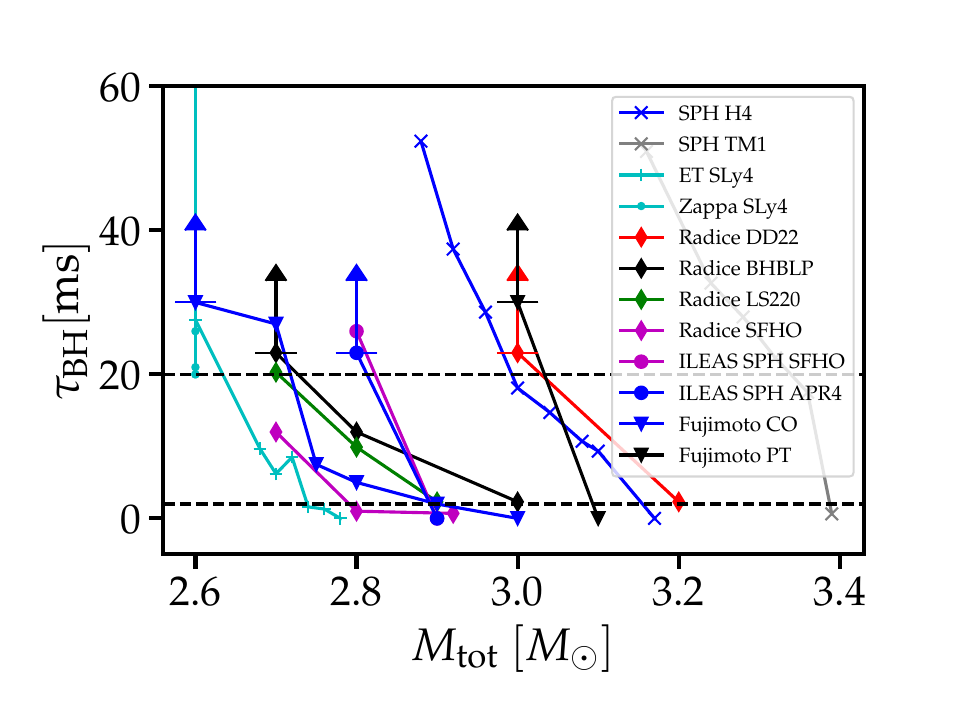}\\
    \includegraphics[width=\linewidth]{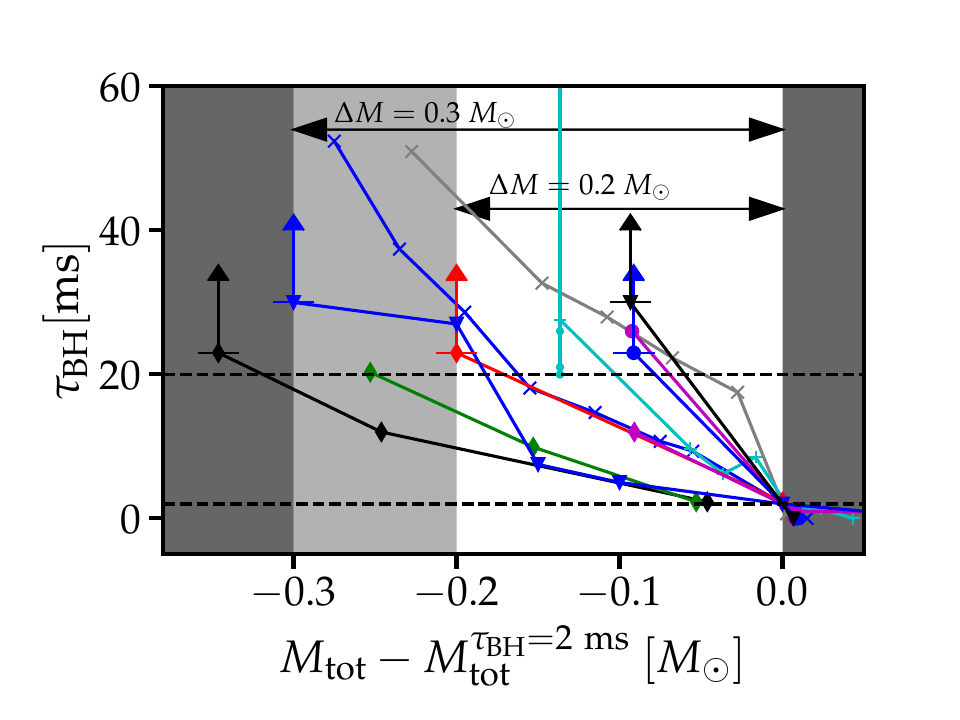}
    \caption{Upper panel: Remnant lifetimes in sequences of simulations of equal-mass mergers varying only the total binary mass $M_\mathrm{tot}$. Colors and symbols refer to different EoS and simulation tools, respectively. From the literature we adopt data from~\citep{Radice2018c,Zappa2023,Fujimoto2024} (from~\cite{Zappa2023} we only include calculations at a fixed binary mass for different simulation settings to indicate uncertainties). Dashed lines indicating lifetimes of 2\,ms and 20\,ms are drawn to estimate $\Delta M=M_\mathrm{thres}-M_\mathrm{tot}^{\tau_\mathrm{BH}=20\,\mathrm{ms}}$ for the various EoS models. Arrows display lower limits on \taubh, where \taubh is given by the end of the simulation time, until which no gravitational collapse took place. Lower panel: Same sequences as in the upper panel but shifted by $-M_\mathrm{tot}^{\tau_\mathrm{BH}=2\,\mathrm{ms}}$ to read off an estimate of $\Delta M$.}
    \label{fig:lifetime}
\end{figure}
In a first step, we thus consider the dependence $\tau_\mathrm{BH}(M_\mathrm{tot})$, where one expects that the lifetime decreases with $M_\mathrm{tot}$ since more mass destabilizes the remnant (e.g.,~\cite{Hotokezaka2013b}). The loss and redistribution of energy and angular momentum of the central object are governed by the magneto-hydrodynamical evolution, gravitational-wave emission and neutrino emission. For higher binary masses these processes take less time to drive the remnant to a more compact and ultimately unstable configuration. The exact dependence of the lifetime on the binary masses is notoriously difficult to determine because the remnant lifetime in numerical simulations is strongly affected by numerics, e.g.,\ the numerical resolution, discretization schemes or the choice of the initial orbital separation, and the physics included (e.g.,\ neutrinos, magnetic fields or effective viscosity)~\citep{Zappa2023,Koelsch2022}. In addition, $\tau_\mathrm{BH}(M_\mathrm{tot})$ may not even be unique for a given numerical scheme and input physics but be to a certain extent subject to stochastic simulation-to-simulation variations. One should also expect a dependence on the EoS and the binary mass ratio as well. Surveying the literature there are hardly any studies available that explicitly determine $\tau_\mathrm{BH}(M_\mathrm{tot})$ by running sequences of models with fine spacing in $M_\mathrm{tot}$ within the range of interest (0\,ms~$\lesssim\tau_\mathrm{BH}\lesssim 40$\,ms) based on a set of consistent simulations (i.e.,\ with calculations varying \emph{only} $M_\mathrm{tot}$ but otherwise using the same numerical and physical setup), which is essential to obtain values that can be meaningfully compared (but see~\citet{Lucca2020,Holmbeck2021} for a meta-study, which however does not include sufficiently fine spaced model setups within a consistent treatment; and see discussions and Fig.~6 in~\citet{Koelsch2022} as well as Fig.~11 in~\citet{Fujimoto2024}, which includes $q<1$ cases that show a dependence of $\tau_\mathrm{BH}(M_\mathrm{tot})$ very similar to that of equal-mass binaries). For the sake of our argument, however, a coarse estimate suffices taking advantage of the fact that $\tau_\mathrm{BH}(M_\mathrm{tot})$ is likely a very steep function (as suggested by simulations). We intend to estimate an upper limit on $\Delta M$ being the difference between $M_\mathrm{thres}$ and the measured binary mass of GW170817.

In Fig.~\ref{fig:lifetime} we collect data from different simulations showing the lifetime as a function of the total binary mass for equal-mass binaries. This includes different calculations with our smoothed-particle hydrodynamics (SPH) code employing the conformal flatness approximation as in~\citep{Oechslin2002,Criswell2023}\footnote{Note that the calculations in~\citep{Criswell2023} employ a different SPH kernel function as compared to earlier simulations.}. The runs labelled ``ILEAS SPH'' incorporate a account for neutrino emission effects~\citep{Ardevol-Pulpillo2019a}. We also include runs with the Einstein Toolkit~\cite{zachariah_etienne_2021_4884780} in full general relativity as in~\citep{Soultanis2022} but for the Sly4 EoS~\citep{Douchin2001} and a set of simulations from the literature in full GR partly in combination with a neutrino treatment~\citep{Radice2018c,Fujimoto2024}. For the latter, the simulation setups are coarsely spaced in $M_\mathrm{tot}$. To indicate uncertainties we add calculations for a fixed binary mass and EoS, but with different numerical resolution, neutrino treatment and partly a scheme to model turbulent viscosity from \citet{Zappa2023}. For clarity we drop some data points with a prompt collapse, which presumably have a total binary mass much in excess of $M_\mathrm{thres}$. There are different definitions of a prompt collapse and of $M_\mathrm{thres}$ discussed in the literature~\citep{Koeppel2019,Koelsch2022,Ecker2024}. To estimate $\Delta M=M_\mathrm{thres}-M_\mathrm{tot}$, we adopt the notion of \citet{Agathos2020} and \citet{Koelsch2022}, defining $M_\mathrm{thres}$ as the system with \taubh$=2$\,ms. In Fig.~\ref{fig:lifetime} the dashed horizontal lines indicate lifetimes of 2\,ms and 20\,ms. Reading off $\Delta M$ by the intersections of the respective curves for the various EoSs at 2\,ms and 20\,ms, we find values in the range between 0.048\,M$_\odot$ and 0.319\,M$_\odot$ with only two out of the eleven EoSs exceeding 0.2\,M$_\odot$ (see lower panel in Fig.~\ref{fig:lifetime}). We stress again that these values are only tentative because of the limited number of sequences, the, in places, coarse sampling in $M_\mathrm{tot}$ and our poor knowledge of underlying (numerical or physical) uncertainties.

The sparseness of the current data prevents us from estimating $\Delta M$ for a limit of \taubh$=30$\,ms or even \taubh$=40$\,ms, but $\tau_\mathrm{BH}(M_\mathrm{tot})$ typically becomes steeper in this range (as indicated in the figure) and thus $\Delta M$ should not be much affected by the exact limit on \taubh implied by the observed lack of helium.

For the following derivation of the EoS constraints we therefore adopt $\Delta M=0.2$\,M$_\odot$ as a sensible choice and we present results for $\Delta M=0.3$\,M$_\odot$ and $\Delta M=0.4$\,M$_\odot$ as a more conservative approach in App.~\ref{app:sense}. We note that Fig.~6 of~\citep{Koelsch2022} indicates a similar range of $\Delta M\approx0.2$\,M$_\odot$ (also for asymmetric binaries). A value of
$\Delta M=0.2$\,M$_\odot$ implies that for the measured binary mass of $M_\mathrm{tot}^\mathrm{GW170817}=2.73^{+0.04}_{-0.01}$\,M$_{\odot}$~\citep{Abbott2019}, the threshold mass for prompt gravitational collapse is unlikely to exceed $\approx 2.97$~M$_{\odot}$.

\subsection{Constraints on stellar parameters: radius, tidal deformability and maximum mass}
An upper limit on the threshold mass implies constraints on NS radii and the maximum mass of non-rotating NSs, $M_\mathrm{max}$, because $M_\mathrm{thres}$ scales tightly with these properties~\citep{Bauswein2013b,Bauswein2017b,Koelsch2022,Koeppel2019,Agathos2020,Bauswein2020,Bauswein2021,Tootle2021,Kashyap2022,Ecker2024,Perego2022prl}. Generally, the threshold mass increases with $M_\mathrm{max}$ and NS radius, $R$, (or equivalently the tidal deformability, $\Lambda$). Consequently, both quantities, $M_\mathrm{max}$ and $R$, cannot simultaneously become too large to accommodate a given upper limit on $M_\mathrm{thres}$. Furthermore, $M_\mathrm{thres}$ depends on the binary mass ratio $q$~\citep{Bauswein2020,Bauswein2021,Perego2022prl,Koelsch2022}. A number of fit formulae for $M_\mathrm{thres}$ have been developed describing these dependencies based on the analysis of a large set of numerical simulations determining $M_\mathrm{thres}$ for different EoS models and mass ratios~\citep{Bauswein2021,Koelsch2022}. The exact fit formulae and their tightness are affected by the number and type of considered EoS models, the binary mass ratio, and the numerical tool.
\begin{figure*}
    \centering 
    \includegraphics[width=0.49\linewidth,viewport=20 20 428 315 ,clip=]{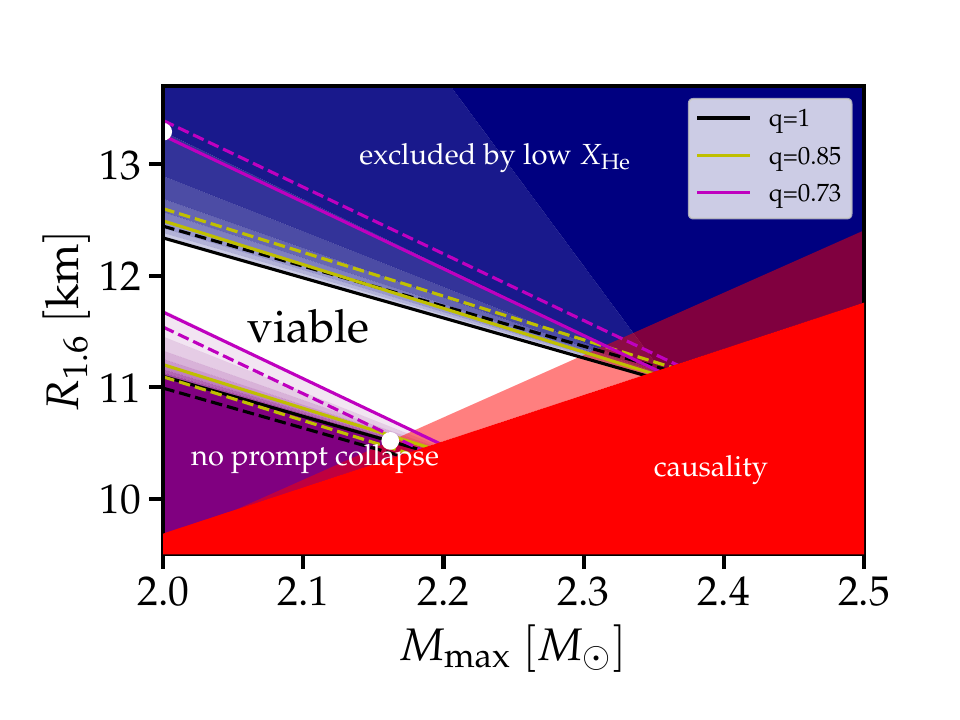}
    \includegraphics[width=0.49\linewidth,viewport=14 20 426 315,clip=]{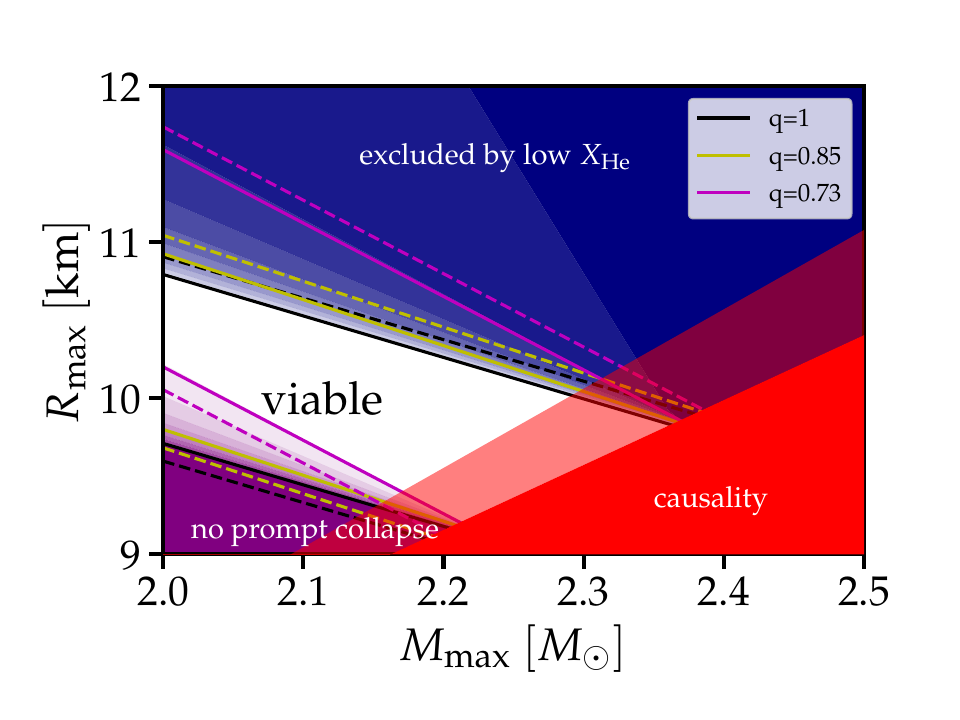}
      \caption{Constraints on the radius of a 1.6\,M$_\odot$ NS (left) and the radius, $R_\mathrm{max}$, of the maximum-mass configuration (right) as function of maximum mass $M_\mathrm{max}$. Stellar parameters in the upper right corner (blue area) are ruled out for a total binary mass $\sim0.2\,$M$_\odot$ below the threshold mass for prompt black-hole formation $M_\mathrm{thres}$.  This binary mass is derived from the maximum lifetime of $\sim$20\,ms inferred for the merger remnant from the absence of strong He features in the kilonova spectrum, as explained in the text. Small NS radii in the lower left corner are excluded if GW170817 did not undergo a prompt gravitational collapse (as argued in~\citep{Bauswein2017,Bauswein2021}) based on the high kilonova brightness (purple area). Solid lines in the respective regions display constraints for assumed binary mass ratios of $q=1$ (black), $q=0.85$ (yellow) and $q=0.73$ (magenta). Dashed lines of the same color include additionally the uncertainty from the scatter in the fit formula for $M_\mathrm{thres}(q,M_\mathrm{max},R)$. The color shading of the blue and purple area indicates confidence levels of exclusion (in 10\% steps) from considering the posterior distribution of the binary mass ratio of GW170817. Shading for the area between the 0\% and 10\% levels are not plotted. For the upper limit, note the initial steep increase with the 50\% level close to the yellow line ($q=0.85$). For the lower limit, the confidence of exclusion drops steeply between the 90\% level (close to the black solid line for $q=1$) and the 50\% level (close to the yellow solid line for $q=0.85$). Causality excludes the dark red region. Stellar parameters in the light red area are empirically not found in a large set of microphysical EoS models. White dots in the left panel show absolute limits at the 90\% confidence level given by $M_\mathrm{max}=2.0\,$M$_\odot$ for the upper limit and the intersection with the empirical exclusion region (light red area) for the lower limit. These dots show the $R_{1.6}$ constraints visualized in Fig.~\ref{fig:tovlimit}. See main text for more information.}
    \label{fig:r16limit}
\end{figure*}

\begin{figure}
    \includegraphics[width=\linewidth,viewport=10 20 430 315,clip=]{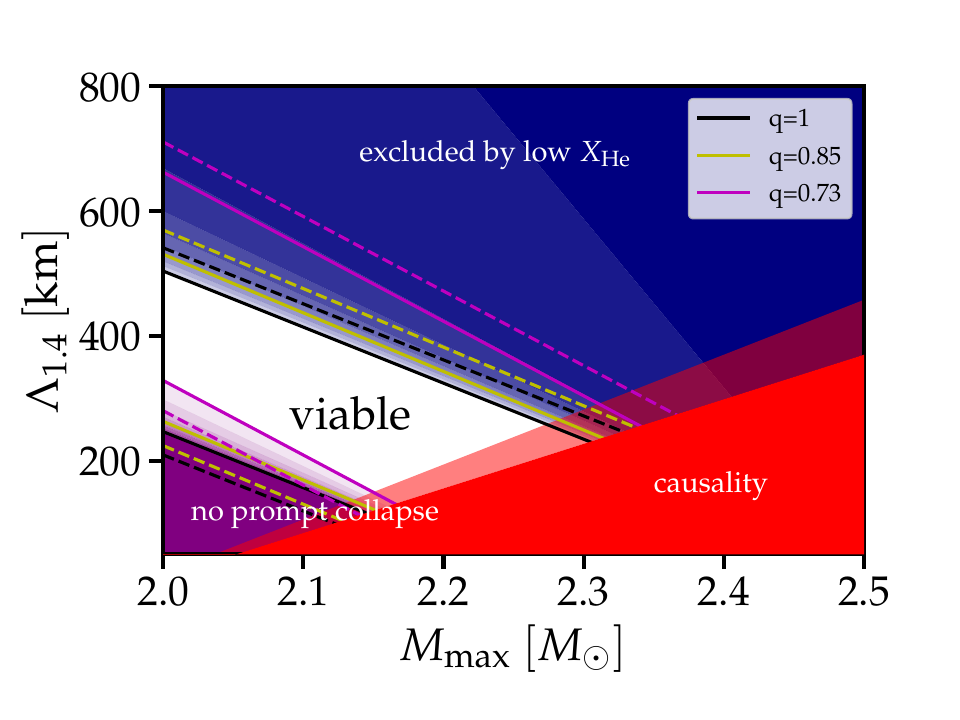}
    \caption{Same as Fig.~\ref{fig:r16limit} but for the tidal deformability $\Lambda_{1.4}$ of a 1.4~M$_\odot$ NS.}
    \label{fig:lam14limit}
\end{figure}

\begin{figure}
    \includegraphics[width=\linewidth,viewport=25 23 428 315 ,clip=]{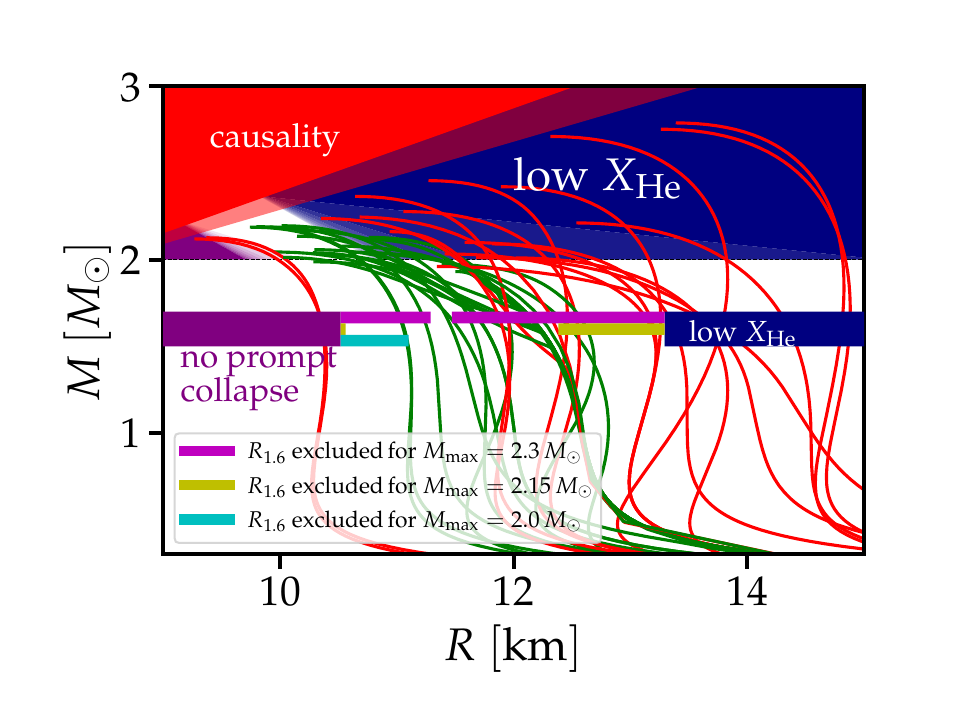}
    \caption{Constraints on NS parameters as in Fig.~\ref{fig:r16limit} overplotted with exemplary mass-radius relations for various microphysical EoSs (green and red lines)~\cite{Banik2014,Fortin2018,Marques2017,Hempel2010,Typel2010,Typel2005,Alvarez-Castillo2016,Akmal1998,Goriely2010,Wiringa1988,Lattimer1991,Shen2011,Lalazissis1997a,Douchin2001,Steiner2013,Sugahara1994a,Hempel2012,Toki1995,Hempel2012,Read2009a,Muther1987,Alford2005,Lackey2006,Kaltenborn2017,Bastian2018,Fischer2018,Bauswein2019,Bastian2020,Cierniak2018,Engvik1996,Glendenning1985,Schneider2019}. 
    If the remnant of GW170817 was short-lived, this excludes stellar parameters in the blue areas (see main text). NS radii are constrained from below (purple areas) because the brightness of the kilonova AT2017gfo points to no prompt collapse yielding a lower limit on $M_\mathrm{thres}$. The color shadings of the constraints on $\{M_\mathrm{max},R_\mathrm{max}\}$ above $M=2.0\,$M$_\odot$ (thin dashed horizontal line) resemble the exclusion probability implied by the posterior distribution of the binary mass ratio in GW170817~\citep{Abbott2019} similarly as in Figs.~\ref{fig:r16limit} and~\ref{fig:lam14limit}. Causality rules out the area in the upper left corner. Note that a mass-radius curve is only excluded if its $\{M_\mathrm{max},R_\mathrm{max}\}$ lies in one of the excluded regions. The horizontal bars around $M=1.6\,$M$_\odot$ display the upper and lower limit on $R_{1.6}$ with the purple and blue areas providing the absolute limits independent of $M_\mathrm{max}$ (at 90\% confidence based on the binary mass ratio distribution; see black dots in left panel of Fig.~\ref{fig:r16limit} and main text). The magenta, yellow and cyan bars give the bounds for specific values of $M_\mathrm{max}$. For $M_\mathrm{max}\gtrsim2.15~M_\odot$ the causality or empirical limit, respectively, become more stringent then the lower bound from the ``no prompt collapse'' argument and the allowed range of $R_{1.6}$ becomes increasingly smaller approaching zero for $M_\mathrm{max}\approx2.3~M_\odot$ and thus EoSs with $M_\mathrm{max}\gtrsim2.3~M_\odot$ are ruled out. Mass-radius relations shown in red are excluded by our constraints. See main text for more detailed explanations.}
    \label{fig:tovlimit}
\end{figure}

For our constraint we employ the fit formulae from Ref.~\citep{Bauswein2021}, which simultaneously include a dependence on the EoS and $q$ for a very large number of EoSs (more than 20 models). We note that the influence of the binary mass ratio turns out to be considerable, which is why it is important to use a fit that equally covers the EoS and mass ratio dependence. We adopt the prescription
\begin{equation}\label{eq:mthr}
    M_\mathrm{thres}(q,M_\mathrm{max},R)= c_1 M_\mathrm{max} + c_2 R + c_3 + c_4\delta q^3 M_\mathrm{max} + c_5 \delta q^3 R
\end{equation}
from~\citep{Bauswein2021} with fit parameters $c_i$ and $\delta q \equiv 1-q$. The radius $R$ may be the radius $R_{1.6}$ of a 1.6\,M$_\odot$ NS or the radius $R_\mathrm{max}$ of the maximum mass configuration. $R$ may also be replaced by the tidal deformability of a NS with fixed mass, e.g.,\ 1.4~M$_\odot$. See Tab.~VI in~\citep{Bauswein2021} for the fit parameters $c_i$ resulting from different underlying datasets; we choose the fits for the EoS sample `b', i.e.,\ the subset of purely baryonic EoS models which are compatible with pulsar observations~\citep{Antoniadis2013,Fonseca2021,Romani2022} and the tidal deformability from GW170817~\citep{Abbott2019} (see App.~\ref{app:sense} for further comments on the choice of the fit formula; for convenience Tab.~\ref{tab:fits} lists coefficients of all relations employed in this study).

For $M_\mathrm{thres}\leq M_\mathrm{tot}^\mathrm{GW170817} + \Delta M$, it immediately follows that
\begin{equation}\label{eq:upper}
    R\leq\frac{M_\mathrm{tot}^\mathrm{GW170817}+\Delta M -c_1 M_\mathrm{max} -c_3 -c_4\delta q^3 M_\mathrm{max}  }{ c_2+c_5\delta q^3},
\end{equation}
where $\Delta M$ can be further increased to include additional sources of error, e.g.,\ the tightness $\delta M$ of the fit formulae. We display the constraint on $R_{1.6}$ resulting from Eq.~\eqref{eq:upper} in Fig.~\ref{fig:r16limit} (blue area in the upper right in the left panel). In Fig.~\ref{fig:r16limit} the lines refer to fixed binary mass ratios adopted in Eq.~\eqref{eq:upper}. 
In Eq.~\eqref{eq:upper} we include the total binary mass of GW170817 using the well measured chirp mass $\mathcal{M}$ via $M_\mathrm{tot}=\mathcal{M} q^{-3/5} (1+q)^{6/5}$. Dashed lines indicate the constraints for fixed mass ratios adding the mean deviation $\delta M=0.017\,$M$_\odot$ of the fit from the underlying data as additional error to $\Delta M$ (see eighth column in Tab.~VI for $\delta M$ in~\citep{Bauswein2021}).

In GW170817, the binary mass ratio was found to be in the range $0.73\leq q\leq 1$ at the 90\% confidence level assuming a low spin prior~\citep{Abbott2019}. Figure~\ref{fig:r16limit} shows a strong dependence on the binary mass ratio with $q=0.73$ yielding the weakest constraint unless $M_\mathrm{max}\gtrapprox 2.3\,$M$_\odot$. Equation~\eqref{eq:upper} shows a complicated behavior with $q$ depending on the chosen $M_\mathrm{max}$, which for intermediate $M_\mathrm{max}$ can even be non-monotonic. For smaller $M_\mathrm{max}$, the maximum allowed radius is increasing with the binary mass asymmetry in Eq.~\eqref{eq:upper}. 
The posterior probability of $q$ from GW170817 shows a relatively flat distribution for $0.8\lesssim q\lesssim 1$ to decrease more steeply below $\sim0.8$ (see Fig.~7 in~\citep{Abbott2019}). Thus, it is statistically very unlikely that the mass ratio was in the range $q\lesssim 0.7$. Using the posterior distribution of $q$ from Ref.~\citep{Abbott2019}, we use ten different shadings to indicate the exclusion levels in Fig.~\ref{fig:r16limit} in 10\% steps resulting from the distribution of $q$, which we propagate through Eq.~\eqref{eq:upper}. 

Note that the dependence on $q$ in Eq.~\eqref{eq:upper} is such that stronger deviations from the $q=1$ case only occur for very asymmetric systems because of the $\delta q^3$ terms in Eq.~\eqref{eq:mthr}. The line for $q=0.85$ (yellow) is very close to that of the equal-mass mergers (black) in Fig.~\ref{fig:r16limit}. Thus the probability that a radius is excluded rises quickly in the region between the lines with $q=1$ and $q=0.85$ (different shadings are hardly distinguishable in this range in Fig.~\ref{fig:r16limit}). The 50\% exclusion contour is very close to the $q=0.85$ line (yellow); the 90\% limit follows closely the $q=0.73$ line (magenta) for $M_\mathrm{max}\lesssim2.4\,$M$_\odot$.

Large radii cannot be ruled out if GW170817 was very asymmetric. The reason for this lies in the behavior of $M_\mathrm{thres}(q)$, which is relatively flat for small binary mass asymmetries ($q\approx 1$) and declines stronger for larger asymmetries, i.e.,\ smaller $q$ (see e.g.,\ Fig.~4 in~\citep{Bauswein2021} or~\citep{Bauswein2020,Perego2022prl,Koelsch2022}). Even a stiff EoS could thus yield a relatively small $M_\mathrm{thres}$ if the binary was very asymmetric implying only a weak constraint on the radius. In addition, significant binary mass asymmetries imply a higher total binary mass of GW170817, for which only the chirp mass is well known, and thus weaken our upper radius limit.

We note that the fit formulae do not consider intrinsic spins of the NSs, which however affect $M_\mathrm{thres}$ only for very large and probably unrealistic values~\citep{Tootle2021,Helbich2023}. We show the impact of strong first-order phase transitions on our constraints in App.~\ref{app:sense} and find that they would weaken the radius constraints by a few hundred meters if $M_\mathrm{max}\sim 2\,$M$_\odot$ and if phase transitions are as extreme as the ones adopted in~\citep{Bauswein2021}. We also present the limits resulting from alternative fit formulae~\citep{Koelsch2022} in the appendix, which only mildly affect the quantitative results. Thus the main uncertainty still remains that of the binary mass ratio.

As already apparent from Eq.~\eqref{eq:upper}, the new upper limit on $R_{1.6}$ depends on the maximum mass and becomes stronger for larger $M_\mathrm{max}$. This is clear because $R$ and $M_\mathrm{max}$ both increase $M_\mathrm{thres}$ (cf.\ Eq.~\eqref{eq:mthr}), and a larger $M_\mathrm{max}$ is only compatible with $M_\mathrm{thres}\lesssim 2.9\,$M$_\odot$ if the radius is correspondingly smaller. The fact that our limits depend on $M_\mathrm{max}$ make them more constraining than an individual number suggests: large radii are only compatible with relatively small $M_\mathrm{max}$ -- just above the current lower bound~\citep{Antoniadis2013,Fonseca2021,Romani2022}. This is rather atypical for many EoS models, which often reach far beyond 2\,M$_\odot$ if the radius $R_{1.6}$ is larger than $\sim12$\,km. Our constraint thus rules out a significant number of current EoS models, which can be directly seen in Fig.~\ref{fig:sensedm}, where the stellar parameters of a sample of microphysical EoS models are overplotted (see also Tab~\ref{tab:eostab}).

We include two additional constraints in Fig.~\ref{fig:r16limit}, which do not depend on arguments about the presence of helium in the outflow of AT2017gfo. In the lower left part we display the excluded region (purple) derived from the argument that GW170817 was likely not a prompt collapse event, which would be incompatible with the relatively high brightness of the kilonova~\citep{Bauswein2017}. A prompt collapse is likely connected with reduced mass ejection and, thus, one concludes that $M_\mathrm{thres}>M_\mathrm{tot}^{GW170817}$. Following~\citep{Bauswein2017,Bauswein2021}, this implies a lower limit on the radius. We update and improve this constraint in comparison to~\citep{Bauswein2017,Bauswein2021} by employing for consistency the same $q$-dependent fit formula for $M_\mathrm{thres}$ (Eq.~\eqref{eq:mthr}) and by considering the posterior sample of $q$ from GW170817. Instead of only computing an absolute lower limit as in~\citep{Bauswein2021} we 
show the lower limit as function of $M_\mathrm{max}$. 
As for the upper limit, the lower limits directly result from Eq.~\eqref{eq:mthr} as
\begin{equation}\label{eq:lowerlimit}
    R>\frac{M_\mathrm{tot}^\mathrm{GW170817} -c_1 M_\mathrm{max} -c_3 -c_4\delta q^3 M_\mathrm{max}  }{ c_2+c_5\delta q^3}.
\end{equation}
Again we find a significant impact from the binary mass ratio especially for $M_\mathrm{max}\approx 2\,$M$_\odot$. This can be seen from the lines in the lower left of the figure, where in contrast to the upper limit the $q=1$ case represents the more conservative limit. The dashed lines again indicate the uncertainties of the fit formula (by shifting $M_\mathrm{tot}^\mathrm{GW170817} \rightarrow M_\mathrm{tot}^\mathrm{GW170817} - \delta M$ in Eq.~\eqref{eq:lowerlimit}). We again propagate the posterior sample of $q$ from GW170817 through Eq.~\eqref{eq:lowerlimit} and use different shadings in Fig.~\ref{fig:r16limit} to visualize the exclusion level in steps of 10\% (purple area). As for the upper limit the resulting distribution becomes very steep in the range corresponding to small binary mass asymmetries. The 90\% level is close to the black solid line ($q=1$) and the 50\% level follows closely the yellow $q=0.85$ line.

Like the upper limit, the lower limit on the radius also depends on the maximum mass and effectively the combined constraint appears like a ``sliding window'', where larger radii are favored for relatively small maximum masses $\sim2\,$M$_\odot$ and smaller radii are only compatible with larger $M_\mathrm{max}$. The sliding window essentially is a result of our main argument that $M_\mathrm{tot}^\mathrm{GW170817} < M_\mathrm{thres}\leq M_\mathrm{tot}^\mathrm{GW170817}+\Delta M$. Recall that the lower limits are independent of $\Delta M$ and the presence or absence of helium, and $\Delta M$ determines the width of the allowed range in $R_{1.6}$.

The second additional constraint we consider arises because causality limits the stiffness of any EoS. In the lower right of Fig.~\ref{fig:r16limit} (left panel) we display an area which is excluded by causality requiring that the speed of sound $v_s$ cannot exceed the speed of light $c$. This limits the maximum stiffness of the EoS and consequently rules out large $M_\mathrm{max}$ for a given $R_{1.6}$. Being less conservative, we obtain an ``empirical''  limit by considering pairs $\{R_{1.6},M_\mathrm{max}\}$ from a large set of microphysical EoSs and determining the limit such that all models lie within this phenomenological bound. See~\citep{Bauswein2017,Bauswein2019,Bauswein2021} and App.~\ref{app:sense} for the details on the ``causality limit'' and the ``empirical limit''.

One may expect that an upper limit on $M_\mathrm{thres}$ also implies a constraint on $M_\mathrm{max}$. This constraint is visible in Fig.~\ref{fig:r16limit}, where the intersection between the empirical or causal limit (red area) and the upper limit on $R_{1.6}$ (blue area) provides the highest possible $M_\mathrm{max}$. Based on the argument of a low helium mass fraction, we can rule out $M_\mathrm{max}\gtrsim2.3$\,M$_\odot$ for $\Delta M=0.2\,$M$_\odot$. We emphasize that large $M_\mathrm{max}$ far in excess of 2\,M$_\odot$ are only compatible with a relatively narrow range of radii because for $M_\mathrm{max}\gtrsim2.15\,$M$_\odot$ the phenomenological and causality constraints become stronger than the bound from the no-prompt-collapse argument. This significantly tightens the ``allowed window'' in this $M_\mathrm{max}$ range. Note that the numbers for the lower radius limit derived in~\citep{Bauswein2017,Bauswein2021} (i.e.,\ $\sim10.5$\,km) essentially correspond to the point where the lines of the lower limit (cf. Eq.~\eqref{eq:lowerlimit}) intersect with the empirical or causality constraint (red area). This represents a more conservative limit independent of $M_\mathrm{max}$ and provides an absolute lower bound. Table~\ref{tab:nopc} lists the absolute upper limits on $M_\mathrm{max}$, which range from about 2.3\,M$_\odot$ to 2.5\,M$_\odot$ depending on $\Delta M$, and the absolute lower and upper limits on NS radii.

In Fig.~\ref{fig:r16limit} (right panel) we visualize the constraint on the radius $R_\mathrm{max}$ of the maximum mass configuration employing a fit formula $M_\mathrm{thres}(q,M_\mathrm{max},R_\mathrm{max})$ but otherwise following the same line of argument and derivation (fit for set `b' in Tab.~VI in~\citep{Bauswein2021}). For the dark red exclusion region (from causality arguments) we use results from Refs.~\citep{Koranda1997,Lattimer2016}, whereas the corresponding empirical limit (light shading) is again obtained from considering a large set of microphysical EoS models (see App.~\ref{app:sense}, Tab.~\ref{tab:nopc}).

Figure~\ref{fig:lam14limit} shows the constraints on the tidal deformability $\Lambda_{1.4}=\Lambda(1.4~M_\odot)$ (fit for set `b' in Tab.~VI in~\citep{Bauswein2021}). The excluded area in the lower right (red) is obtained in a way equivalent to the corresponding constraint on $R_{1.6}$ (see App.~\ref{app:sense}). 

We summarize the various constraints on NS radii in Fig.~\ref{fig:tovlimit} in comparison to a set of different microphysical EoSs (same sample as in~\citep{Bauswein2021}; cf. Tab.~\ref{tab:eostab}). For the maximum-mass configurations, we display the constraints on $\{M_\mathrm{max},R_\mathrm{max}\}$ (in the upper part of the diagram) with the shading reproducing the likelihood of different binary mass ratios of GW170817 as in Fig.~\ref{fig:r16limit}. A mass-radius relation of an EoS can cross this exclusion region and the model remains still viable if $\{M_\mathrm{max},R_\mathrm{max}\}$ lie in the allowed region.

The exact constraints on $R_{1.6}$ are not straightforward to visualize in Fig.~\ref{fig:tovlimit} because they depend on $M_\mathrm{max}$. As discussed above the position of the ``allowed'' window shifts with $M_\mathrm{max}$ but its width is approximately constant for $M_\mathrm{max}\lesssim2.15\,$M$_\odot$. For $M_\mathrm{max}\gtrsim2.15\,$M$_\odot$ the allowed range of $R_{1.6}$ becomes increasingly smaller (see left panel of Fig.~\ref{fig:r16limit}). For $M_\mathrm{max}\approx2.3~M_\odot$ this range decreases to zero and models with $M_\mathrm{max}\gtrsim2.3\,M_\odot$ are excluded. The wide purple and blue bars at $M\approx1.6~M_\odot$ display the 90\% exclusion limits for a respective absolute limit independent of $M_\mathrm{max}$. For the upper bound the absolute limit is given by the constraint for $M_\mathrm{max}=2.0\,$M$_\odot$, whereas the absolute lower bound results from $M_\mathrm{max}=2.126\,$M$_\odot$ (see markers in left panel of Fig.~\ref{fig:r16limit}). We stress that for any other $M_\mathrm{max}$ the limits are more stringent, which is illustrated by the magenta, yellow and cyan bars showing the corresponding bounds on $R_{1.6}$ for specific values of $M_\mathrm{max}$ in Fig.~\ref{fig:tovlimit} (adopting the 90\% limits from the posterior distribution of the binary mass ratio). Thus, the absolute limits (wide blue and purple bars) do not adequately represent the constraining power of our constraints on $R_{1.6}$. For instance, a $M-R$ curve passing close to the blue region at $M=1.6$\,M$_\odot$ (``no He'' constraint) cannot reach much higher than 2\,M$_\odot$. Hence, some of the shown models compatible with $R<13.29$\,km (blue) are in fact excluded. Larger $M_\mathrm{max}$ are only compatible with smaller $R_{1.6}$. Similarly, if an $M-R$ relation passes close to the purple exclusion region (``no prompt collapse'' constraint), its maximum mass should be relatively large and an EoS model with $M_\mathrm{max}\approx 2.0\,$M$_\odot$ would be excluded. Generally, our constraints favor models with mass-radius relations which tend to strongly bend over at higher masses. This may potentially point to a stronger softening of the EoS at higher densities, which can be caused by the occurrence of non-nucleonic degrees of freedom (cf.~\cite{Bauswein2025}).

The resulting constraints on the stellar properties of NSs for adopting $\Delta M=0.3\,$M$_\odot$ or even $\Delta M=0.4\,$M$_\odot$ are provided in App.~\ref{app:sense}. We stress that a chosen $\Delta M$ is equivalent to adopting an upper bound on $M_\mathrm{thres}=M_\mathrm{tot}^\mathrm{GW170817}+\Delta M$. This implies that the upper limits discussed in our study would generally result from any upper bound on the threshold mass for prompt black hole formation, which may be inferred in the future. Thus our constraints on stellar parameters for different $\Delta M$ will be directly applicable to any upcoming upper limit on $M_\mathrm{thres}$.

Considering Eqs.~\eqref{eq:upper} and~\eqref{eq:lowerlimit} it is clear how future NSM observations may further improve the constraints presented here. As already argued in~\citep{Bauswein2017,Bauswein2021}, a bright kilonova with a higher chirp mass would imply a stronger constraint on the lower limit of $R$ (or $\Lambda$). As apparent from the figures, an event with the same chirp mass but a well measured binary mass ratio $q<1$ would, if a prompt collapse could be excluded, as in GW170817, shift the lower bound to larger radii. Similarly, a GW170817-like detection but with well-measured $q\approx1$ would imply a decrease of the upper limit on $R$ or $\Lambda$ and mean a stronger EoS constraint. Also, if a new detection with lower chirp mass showed no signs of a significant helium enrichment, the upper bounds on $R$ or $\Lambda$ would be shifted to smaller values and the upper limit on $M_\mathrm{max}$ strengthened as well. We finally remark that the (absolute) upper limit on $R$ or $\Lambda$ obviously depends on the lower bound on $M_\mathrm{max}$, which is given by the most massive NS mass measurement (for simplicity we adopt a value of 2\,M$_\odot$ consistent with~\citep{Antoniadis2013,Fonseca2021,Romani2022}). Clearly, a well-determined lower bound on $M_\mathrm{max}$ with a greater value would further improve our radius constraint.

\section{Further implications}\label{sec:discussion}

If AT2017gfo indeed harbored a short-lived HMNS remnant, as we argue here, this would have several further interesting implications.
\subsection{Magnetar-powered short GRBs?}
GW170817 was accompanied by an sGRB (GRB170817A) powered by a highly-relativistic jet outflow launched in the aftermath of the merger. A short HMNS lifetime of just a few tens of milliseconds renders the magnetar scenario (e.g., \citep{Metzger2007c, Kiuchi2024a}) for launching this jet very unlikely in this system. It strongly supports the BH-torus scenario, i.e.,\ the jet was probably powered by the Blandford-Znajek process \citep{Blandford1977} (or possibly by neutrino-pair annihilation \citep{Eichler1989} or a combination of both, though neutrino-pair annihilation is less favored given its relatively low efficiency \citep{Just2016}). Assuming further that the binary system in GW170817 is representative of the broader population of neutron star mergers -- consistent with its inferred component masses ($1.16-1.6\,M_{\odot}$) and mass ratio ($q\sim0.73-1$) overlapping with known Galactic BNS systems \citep{Ozel2012}, then most observed sGRBs would be powered by BH-torus central engines.

\subsection{GW190814 interpretation}

The nature of the secondary compact object in the gravitational-wave event GW190814 \citep{Abbott2020} with a mass of $2.5$-$2.67\, M_{\odot}$ is not safely identified so far. While a BH seems to be likely \citep[e.g.,][]{Essick2020a, Tews2021a}, a NS cannot be completely ruled out at present \citep[e.g.,][]{Biswas2021a}. The upper limit for the maximum NS mass presented here, i.e., $M_{\rm max}<2.3\,M_\odot$, would exclude a NS interpretation unless the secondary was rapidly spinning, which implies the secondary to be the lightest BH ever discovered. This conclusion even holds for the very conservative assumption of $\Delta M=0.4\,M_\odot$.

\begin{figure}
    \includegraphics[width=\linewidth]{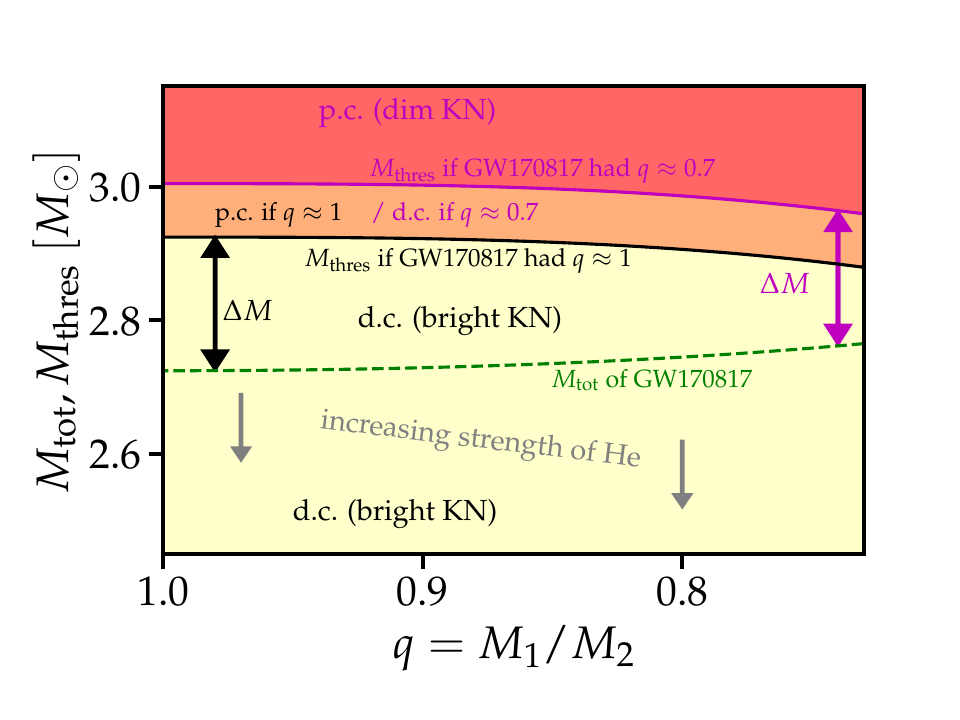}
    \caption{ Sketch of the merger outcome as a function of the total binary mass and the binary mass ratio. Green dashed line displays the measured total binary mass of GW170817. In this study, we argue that the threshold mass for prompt collapse (p.c.) lies at most $\Delta M\approx0.2~M_\odot$ above the total binary mass. Above $M_\mathrm{thres}$, one would expect a dim kilonova (KN), whereas below $M_\mathrm{thres}$ a delayed collapse (d.c.) occurs associated with a bright KN. Since $q$ was not well determined in GW170817, the exact location of the threshold mass cannot be precisely fixed (see black (magenta) curve for $q=1$ (for $q=0.7$) in GW170817). The precise value of $\Delta M$, i.e., the proximity of $M_\mathrm{tot}$ and $M_\mathrm{thres}$, is not known and could in principle be smaller than indicated in the figure because the limit on helium as discussed provides only an upper limit on $\Delta M$.}
    \label{fig:sketch}
\end{figure}

\subsection{Predictions for future observations} 
If the total mass $M_{\mathrm{tot}}^\mathrm{GW170817}$ in GW170817 was indeed relatively close to the threshold mass for prompt collapse, then NSMs with somewhat higher total masses would lead to prompt BH formation (cf. Fig.~\ref{fig:sketch}). This implies that for future observations of such events with $M_\mathrm{tot} \gtrsim M_\mathrm{tot}^\mathrm{GW170817} + \Delta M$ no gravitational-wave signal from the oscillating NS remnant will be detectable even for nearby events and the kilonova would be expected to be significantly fainter than AT2017gfo because of a smaller ejecta mass. The exact location of $M_\mathrm{thres}$ also depends somewhat on the binary mass ratio of GW170817, and if that was very asymmetric the threshold mass would be more elevated. We would also infer that any system more massive than GW170817 should also not exhibit a strong spectral signature of helium (unless GW170817 was very asymmetric, cf. Fig.~\ref{fig:sketch}, in which case a slightly more massive system with $q\approx1$ might feature a longer lifetime). If the merger outcome can be unambiguously identified, for instance through the kilonova brightness or the post-merger GW signal, our prediction of a prompt-collapse event for binary masses close to $M_\mathrm{tot}^\mathrm{GW170817}+\Delta M\approx 2.9\,$M$_\odot$, which is relatively low compared to most predicted $M_\mathrm{thres}$ (see e.g.,\ Tab.~IX in~\citep{Bauswein2021}), can be seen as a strong test of the arguments laid out in this paper. Here we assumed that the mass ratio $q$ in GW170817 was close to unity, whereas the increase in total mass would need to be more pronounced to obtain a prompt collapse if GW170817 was very asymmetric. The sketch in Fig.~\ref{fig:sketch} illustrates the different possibilities depending on the binary mass ratio of GW170817.

On the other hand, in future events with a smaller total mass than GW170817 and consequently \taubh in the several tens, hundreds or thousands of milliseconds, a helium feature should be present for a polar observer in late photospheric epochs, due to the longer lifetime and correspondingly higher neutrino-wind masses. Specifically, we predict this to be the case for binary masses below a certain critical mass $M_\mathrm{tot}^{\mathrm{He}}$ with $M_\mathrm{tot}^\mathrm{GW170817}-\Delta M < M_\mathrm{tot}^{\mathrm{He}}<M_\mathrm{tot}^\mathrm{GW170817}$, where the exact value depends on how close  $M_\mathrm{tot}^\mathrm{GW170817}$ was to $M_\mathrm{thres}$. 
A clear helium spectral feature may point to a longer lifetime than in GW170817 assuming the ejecta properties (e.g.,\ electron densities and deposition rates) to be similar to those of GW170817. In the event of a NSM detection with a nearly identical total mass to GW170817, a strong helium spectral feature may point to a longer lifetime than in GW170817. This could be explained by a more symmetric progenitor mass configuration ($q$ closer to unity) compared to GW170817 (see Fig.~\ref{fig:sketch}), as stability of the remnant typically decreases with smaller mass ratios \citep{Bauswein2020,Bauswein2021,Koelsch2022,Perego2022prl}.

Future detections of kilonovae and sGRBs could also be made without detecting the GW signal, i.e.,\ lacking information on the mass of the binary (e.g.,\ GRB\,211211A \citep{Rastinejad2022,Troja2022a} and GRB\,230307A \citep{Levan2024}). In such ``orphan-kilonova'' cases, the appearance or absence of a helium feature in the observed kilonova spectra at a constraining epoch may provide an indication of the mass relative to GW170817. 

The helium constraints proposed in this paper can be tested if future observations can reveal the remnant lifetime independently, e.g.,\ through the shut-off of the post-merger GW signal~\cite{Easter2021,Breschi2022} or the ringdown of the newly formed BH~\cite{Dhani2024} or, possibly, through extended X-ray emission found in a subset of observed sGRBs \citep[e.g.,][]{Rowlinson2013a, Gompertz2014a, Ravi2014a}. Such a direct detection through the GW signal requires a high signal-to-noise ratio, however, and will only be possible for a very close event and/or with next-generation GW detectors (such as the Einstein telescope \citep{Maggiore2020c}).



\section{Summary}\label{sec:summary}

This paper develops and applies a novel approach for using multi-messenger observations of NSMs to constrain NS properties and hence the high-density EoS based on the imprint of helium on the kilonova spectrum. Specifically, the helium abundance in the ejecta constrains the HMNS remnant lifetime, \taubh, and therefore the threshold mass for prompt collapse, $M_{\mathrm{thres}}$. Applied to AT2017gfo, our analysis yields upper limits for $\tau_{\mathrm{BH}}$ and $M_{\mathrm{thres}}$, implying constraints on stellar parameters of cold, non-rotating NSs and thus the EoS. To our knowledge this is the first study to use spectral features from observations of a kilonova for EoS constraints. It opens up a new window to probe NSMs, e.g.,\ the HMNS lifetime and threshold for BH formation, which is complementary to the information from the GW signal, kilonova lightcurve, and GRB signal. 
We stress that our constraints are based on a chain of arguments where each individual point is well motivated, but our reasoning relies on aspects which, to date, are only incompletely understood due to the complexity of the models involved. We thus emphasize the need for future work to corroborate specific aspects of our work, and our constraints should, in this sense, be considered preliminary.

The main results are:

\begin{enumerate}
\item Using a collisional-radiative NLTE model for helium, we find that in order to be compatible with the observed spectrum of AT2017gfo at 4.4\,days post-merger, the helium mass fraction in the line-forming region (i.e.,\ at velocities $0.19\lesssim v/c\lesssim 0.3$) must be limited to $X_{\rm He} \lesssim 0.05$.

\item From a recently developed set of neutrino-hydrodynamic simulations of NSMs that capture all phases of matter ejection \citep{Just2023}, we find that NS remnants enrich the ejecta with substantial amounts of helium through neutrino winds with characteristically high electron fractions, $Y_e\sim 0.5$. The steep increase of $X_{\rm He}(t)$ in our models, with $t$ being the lifetime of the NS remnant,  suggests a short lifetime of the NS remnant in GW170817 of $\tau_{\mathrm{BH}}\lesssim\,$20{--}30\,ms to satisfy the observational limit $X_{\rm He}\lesssim 0.05$.

\item A short HMNS lifetime implies a close proximity of the total binary mass in GW170817, $M_\mathrm{tot}^\mathrm{GW170817}=2.73^{+0.04}_{-0.01}$~M$_{\odot}$, to the threshold mass for prompt collapse, $M_\mathrm{thres}$. Motivated by several sequences of merger simulations and a literature survey, we argue that $M_\mathrm{thres}$ likely lies no more than $\Delta M\approx 0.2\,M_\odot$ above $M_\mathrm{tot}^\mathrm{GW170817}$ in order that $\tau_{\mathrm{BH}}\lesssim 20\,$ms.

\item Using empirical fit functions, we identify a large region in the $M_\mathrm{max}${--}$R_{1.6}$ plane that is excluded by the inferred low value of $X_{\mathrm{He}}$ (see Fig.~\ref{fig:r16limit}, left panel). This region defines upper limits on the NS radius, $R_{1.6}$, that become stronger for larger values of the maximum NS mass, $M_\mathrm{max}$, because $M_{\mathrm{thres}}$ grows with both $R_{1.6}$ and $M_\mathrm{max}$. Namely, the upper limit on $R_{1.6}$ decreases from 13.2 to 11.4\,km for $M_\mathrm{max}$ between 2.0 and $2.3\,M_\odot$. A significant number of currently viable EoS models are ruled out, mainly because our constraints simultaneously limit $R_{1.6}$ and $M_{\mathrm{max}}$. The nature of our upper, helium-based limit decreasing with $M_{\mathrm{max}}$ contrasts with the usual behavior of many EoS models, where large radii are typically accompanied by large values of $M_{\mathrm{max}}$.

\item Physical limits on the EoS stiffness (such as from causality) constrain the maximum possible $M_{\mathrm{max}}$ for given $R_{1.6}$, defining another exclusion region in the $M_\mathrm{max}${--}$R_{1.6}$ plane. The combination of this region and the helium exclusion region provides an upper limit for $M_\mathrm{max}$ of about 2.3\,M$_\odot$ (cf.\ Fig.~\ref{fig:r16limit}, left panel). Furthermore, we discuss an update of previous constraints providing lower bounds on neutron-star radii, which are $M_\mathrm{max}$ dependent, via the argument that GW170817 did not collapse promptly. This constraint is independent of the consideration of the helium abundance in the ejecta. The combination of all limits results in a narrow range of allowed stellar parameters, e.g.,~$R_{1.6}$, the radius of the maximum mass configuration $R_\mathrm{max}$, or the tidal deformability of a 1.4~M$_\odot$ NS, $\Lambda_{1.4}$. 

\item We find sensitivity of the EoS constraints with respect to the binary mass ratio, $q$, which weakens the upper limit for very asymmetric binaries.

\item With a HMNS lifetime of just $\sim$\,20\,ms, the sGRB jet responsible for GRB170817A was most likely produced by a BH-torus central engine, not by a magnetar. 

\item Future events with a total mass only somewhat higher than that of GW170817 are likely to undergo a prompt collapse and yield smaller ejecta masses and therefore fainter kilonovae, which should, like AT2017gfo, show no spectral signature of helium. Conversely, future kilonovae with smaller total binary masses (and similar mass ratio) are expected to show strong helium features. An event with a similar total mass as in GW170817 but with a distinguishable helium feature would point to a smaller mass ratio $q$ than in GW170817. In the case of an ``orphan-kilonova'' with a missing GW signal the appearance or absence of a helium feature could indicate whether the total binary mass was below or above that of GW170817, respectively, assuming a similar mass ratio.
  
\end{enumerate}

A particularly powerful characteristic of our constraint is the inverse relationship between $R_{1.6}$ and $M_{\mathrm{max}}$ for given $M_{\mathrm{thres}}$: An upper limit of $M_{\mathrm{thres}}$ therefore rules out a significant number of currently existing EoS models with simultaneously large values of $R_{1.6}$ (or $R_{\mathrm{max}}$ or $\lambda_{1.4}$) and $M_{\mathrm{max}}$.

Based on this work, finding a combination of the EoS and mass ratio that can lead to a kilonova both helium-poor and as bright as AT2017gfo may be non-trivial, considering that recent kilonova models \citep{Kawaguchi2021a, Kawaguchi2023, Just2023} seem to favor long- over short-lived scenarios to produce sufficiently high ejecta masses. 
 
Finally, a word of warning is in order. Although we adopt reasonably conservative estimates in every step of this study, due to the complexity of the techniques employed, our constraint may be affected by modeling uncertainties. This concerns, for instance, the determination of the spectral contribution of helium, which assumes idealized ejecta conditions in the line-forming region, the prediction of $X_{\mathrm{He}}(\tau_{\mathrm{BH}})$ using NSM simulations with approximate treatments of general relativity, neutrino transport, and angular-momentum transport, or insufficient numerical resolution to accurately determine the HMNS lifetime, resulting in a possible over- or under-estimation of the proximity $\Delta M$ to the threshold mass corresponding to a given HMNS lifetime $\tau_{\mathrm{BH}}$. Our study motivates future work to examine in more detail these uncertainties and to improve our understanding of the enrichment of helium in NSMs, the relationship between the total mass and the HMNS lifetime (in particular also for unequal-mass binaries), and the modeling of helium features in kilonova spectra. \newline

\section*{Acknowledgements}

AS, AB, RD, DW, CEC, SAS and VV are funded/co-funded by the European Union (ERC, HEAVYMETAL, 101071865). OJ, LJS, GMP and ZX acknowledge support by the European Research Council (ERC) under the European Union’s Horizon 2020 research and innovation programme (ERC Advanced Grant KILONOVA No. 885281).  Views and opinions expressed are, however, those of the authors only and do not necessarily reflect those of the European Union or the European Research Council. Neither the European Union nor the granting authority can be held responsible for them. AS, RD and DW are part of the Cosmic Dawn Center (DAWN), which is funded by the Danish National Research Foundation under grant DNRF140.  AB, TS, OJ, GMP, LJS and ZX acknowledge support by the Deutsche Forschungsgemeinschaft (DFG, German Research Foundation) through Project - ID 279384907 – SFB 1245 (subprojects B06, B07) and MA 4248/3-1. AB, OJ, GMP, and TS acknowledge funding by the State of Hesse within the Cluster Project ELEMENTS. CEC is funded by the European Union’s Horizon Europe research and innovation programme under the Marie Skłodowska-Curie grant agreement No. 101152610.

\bibliographystyle{apsrev4-2} 
\bibliography{refs}

\appendix

\setcounter{figure}{0}
\renewcommand{\thefigure}{A.\arabic{figure}}

\section{Notes on NLTE ionization modeling}
\label{app:RT}
In this appendix, we motivate the fiducial input parameters for the radiative-transfer calculations and explore a broad range of conditions (e.g., density structure, non-thermal deposition rates and composition) to test the robustness of our constraints against radiative-transfer uncertainties.

\subsection{The optimal timeframe for constraining helium}\label{sec:observing_time}

We discuss the optimal timeframe post-merger for observing (and constraining) a spectral feature from He\,{\sc i}\,$\lambda 1083.3$\,nm.
The population in the lower energy level of the transition (i.e.,\ 1s2s\,$^3$S, the lowest-lying level of triplet helium) determines the strength of the feature. As this state is 19.8\,eV above the \hei ground state (which should be compared with $k_B T \approx 0.2$--$0.4$\,eV suggested by the spectra of AT2017gfo several days post-merger) the triplet states are not efficiently populated by excitation due to thermal photons or thermal electrons. The triplet states are instead populated by recombination from \ion{He}{ii}, which is itself populated by ionization of \hei by radioactive decay particles (e.g., electrons and $\gamma$-rays). This means that most effective population of the \threeS-state will occur when the majority of helium is singly ionized.
Thus, modeling the level population under NLTE conditions requires i) the non-thermal ionization rate, ii) the recombination rate and iii) collisional/radiative rates. These are detailed in Sect.~\ref{sec:NLTE_helium_population}. 

Importantly, in the first days post-merger photoionization will rapidly depopulate triplet helium \citep{Sneppen2024_helium}. However, around 4-5 days post-merger, the characteristic temperature suggested by the blackbody ($T\lesssim3000$\,K) implies that photoionization and radiative transitions between higher levels are inconsequential - only the two lowest energy levels of triplet helium (\threeS and \threeP) are significantly populated with the radiative pathway connecting these two levels being the He\,{\sc i}\,$\lambda 1083.3$\,nm line. At this time, the pathway leaving triplet helium is due to the decay to the \hei ground level from 1s2s\,$^3$P \citep{Sneppen2024_helium}. Here, helium provides a well understood NLTE system, in contrast to most \rprocess elements, as the required atomic data is largely known \citep[e.g.,][]{Kramida2023,Berrington1987,Ralchenko2008,Nahar2010} and has been applied in astrophysical context for decades in the interpretation of SN Ib spectra \citep[e.g.,][]{Lucy1991}. 

Conversely, at late-times robust constraints are difficult to attain, because the ionization state of the elements in the ejecta is largely unknown. If the electron density is large (in excess of $10^{9}-10^{10} {\rm cm}^{-3}$) neutral helium is likely to dominate, whereas \heiii is likely to dominate when the electron density drops below $10^{5}-10^{6} {\rm cm}^{-3}$ (see Fig.~\ref{fig:Helium_ionisation}). Thus, as the ejecta expand and the electron density dilutes, the typical ionization state of helium is likely to increase. To estimate whether \heiii-dominated ejecta are important for constraints in this paper, one needs to estimate the ionization state of the ejecta, which depends on recombination rates and, specifically, the electron density. As we will argue in App.~\ref{sec:electron-density} in the line-forming region around 4.4\,days, $n_e  > 6 \times 10^6\, {\rm cm}^{-3}$, which is a conservative lower limit assuming Saha-Boltzmann ionization at temperature 2800\,K. Impact ionization by non-thermal particles produced in radioactive decays is expected to increase the electron density by a factor of several, but as discussed in Sect.~\ref{sec:NLTE_helium_population} the low ionization provides the weakest constraint on the helium abundance. At later times, the continued homologous expansion will decrease the electron density further, which implies an increasingly weakened net constraint on the helium abundance. 

Thus, at intermediate times around 4--5 days post-merger, a particularly opportune time window exists for the detectability of a \hei feature. Radiative transitions (particularly photoionization) have diminished, thereby blocking the strong pathways away from triplet helium, while the electron densities remain high enough for \heii to be an abundant species, thus allowing the recombination pathways into triplet helium. 

At this time, the 1083\,nm wavelength and the observed absorption suggests a photospheric velocity of $\sim0.19c$ \citep{Sneppen2024_helium}, while the inferred blackbody velocity and photospheric velocity inferred for a \srii interpretation of the feature (with a mean rest wavelength of \srii at $\sim$1045\,nm) would suggest a somewhat slower velocity ($\sim0.15c$) \citep{Sneppen2023}. Ultimately, there is some ambiguity in what inner velocity to adopt, but we fiducially follow the $0.19c$ velocity implied by the helium interpretation of the feature. Importantly assuming a lower velocity would only make the upper bound observational limit on the helium mass fraction stronger, as this additional line-forming region would add further absorption.

\subsection{Formalism for ionization state calculation\label{app:ionisation-eq}}
Detailed discussion on collisional-radiative modeling of helium in the context of kilonovae can be found in \citet{Tarumi2023,Sneppen2024_helium}, but for clarity we here quickly summarize the adopted formalism for determining the various rates between ionization-states. 

The recombination-rate from ionization state $i+1$ to $i$ is set by the density in the upper state $n_{i+1}$, the electron density, $n_e$ and the recombination-rate coefficient which depends on the electron temperature, $\alpha_{\rm r}(T_e)$ \citep{Nahar2010}: 
\begin{equation}
    R_{\mathrm{r}} = n_{i+1} \, n_e \, \alpha_{\rm r}(T_e).
\end{equation}
The temperature assumed for this and all radiative transitions is the Doppler-corrected blackbody temperature ($T_e = 2800$\,K).

For both the \hei ground state and the populated triplet levels, the photoionization-rate is negligible for the analysed timeframe (i.e., in late photospheric epochs) given the low UV-flux. The photoionization-rate from any energy-level (having number density $n_{\mathrm{PI}}$) can be computed given i) $W= \left(1-\sqrt{1-(v_{\mathrm{ph}}/v)^2}\right)/2$, the geometric dilution factor accounting for anisotropies in the photon flux, ii) the photon number density per energy interval $4\pi B(E,T)/(c E)$ where the radiation temperature $T=T_e$, and iii) the photoionization cross-section, $\sigma_{\mathrm{PI}}$, taken from~\citep{Nahar2010}: 
\begin{equation}
    R_{\mathrm{PI}} = n_{\mathrm{PI}} W \int_{E_{\mathrm{th}}}^{\infty} \sigma_{\mathrm{PI}} \frac{4\pi B(E,T)}{E} dE,
\end{equation}
with $E_{\mathrm{th}}$ the threshold energy for ionization.

Due to the high energies required, we expect that Helium ionization is dominated by non-thermal particles produced by radioactive decay of r-process nuclei.  This requires knowledge of the non-thermal spectrum that itself depends on understanding how high-energy particles thermalize in the ejecta (see e.g.,~\cite{Barnes2016}). This is a non-trivial problem that we treat following the formalism of refs.~\cite{Tarumi2023,Sneppen2023}. The ionization rates by non-thermal particles from ionization state $i$ to $i+1$ is given by the density in the lower ionization state $n_{i}$, the radioactive heat per unit time and per ion, $\Dot{q}_{\mathrm{non-thermal}}$, and the work per ion, $w_i$: 

\begin{equation}
    R_{\mathrm{non-thermal}} = n_{i} \, \frac{\Dot{q}_{\mathrm{non-thermal}}}{w_i}.
\end{equation}

For He, we use $w_{\hei} = 550\,\mathrm{eV}$ and $w_{\heii} = 3200\,\mathrm{eV}$, derived using the Spencer-Fano solver, \texttt{pynonthermal}, given a composition of $X_\mathrm{He}=0.01$ and approximating the \rprocess stopping power with Sr as a representative element \citep{Shingles2024}. We note that this is within 10\% of the values $w_{\hei} = 593\,\mathrm{eV}$ and $w_{\heii} = 3076\,\mathrm{eV}$ derived from ref.~\citet{Tarumi2023}. There are notable compositional and ionization-state uncertainties attached to these values for which we discuss the downstream implication in Sect.~\ref{app:sensetivity}, but this remains a much smaller variation than the several-orders-of-magnitude uncertainty quoted for the non-thermal deposition rate (e.g., see Sect.~\ref{sec:helium_uncertainties}). We assume that beta-decay is the main source of radioactive heating, $\Dot{q}_{\mathrm{non-thermal}} \approx  \Dot{q}_\beta$. We use $\Dot{q}_\beta = \Dot{q}_{1\,\mathrm{day}} \,t_d^{-1.3}$~eV~s$^{-1}$~ion$^{-1}$~\cite{Hotokezaka2020,Tarumi2023} with $t_d$ in days and $\Dot{q}_{1\,\mathrm{day}}\approx1$. We note that the heating distributed into photons can still provide sufficient non-thermal ionization if tangled magnetic fields trap electrons locally (see Sect.~\ref{sec:helium_uncertainties}). Additionally, we explore the impact of a larger or smaller  $\Dot{q}_{\mathrm{non-thermal}}$ in Sect.~\ref{sec:helium_uncertainties} where we show that changes of an order of magnitude in the deposited energy (or in the work per ion) does not affect our constraints (see also Sect.~\ref{app:sensetivity} and Fig.~\ref{fig:Xhe_verus}). The deposition is effectively assumed to be instantaneous and local, although we discuss a more distributed deposition from photons in Sect.~\ref{sec:helium_uncertainties}.

The resulting equilibrium between these rates (i.e., balancing $R_{\mathrm{non-thermal}} + R_{\rm PI} \approx R_{\mathrm{non-thermal}}$, $R_{\mathrm{r}}$, see Fig.~\ref{fig:Helium_ionisation}), determines the fractional population in the various ion-states: 
\begin{equation}
    n_{i+1} = n_{i} \frac{\Dot{q}_{\mathrm{non-thermal}}}{w_i \, n_{e} \, \alpha_{\rm r}(T_e)},
\end{equation}
where we assume that photoionization constitutes a negligible contribution under the expected conditions. In our fiducial setup, photoionization is two orders-of-magnitude weaker than the dominant pathways leaving triplet helium, so to be a significant contributor the radiative field in the UV would have to be significantly greater. Specifically, if the UV flux were scaled equally at all wavelength, $F(\lambda\sim340\,\mathrm{nm})$ would need approximately to be comparable to or exceeding the observed flux at $800\,\mathrm{nm}$.

\begin{figure}
  \includegraphics[width=\linewidth,viewport=10 10 465 400 ,clip=]{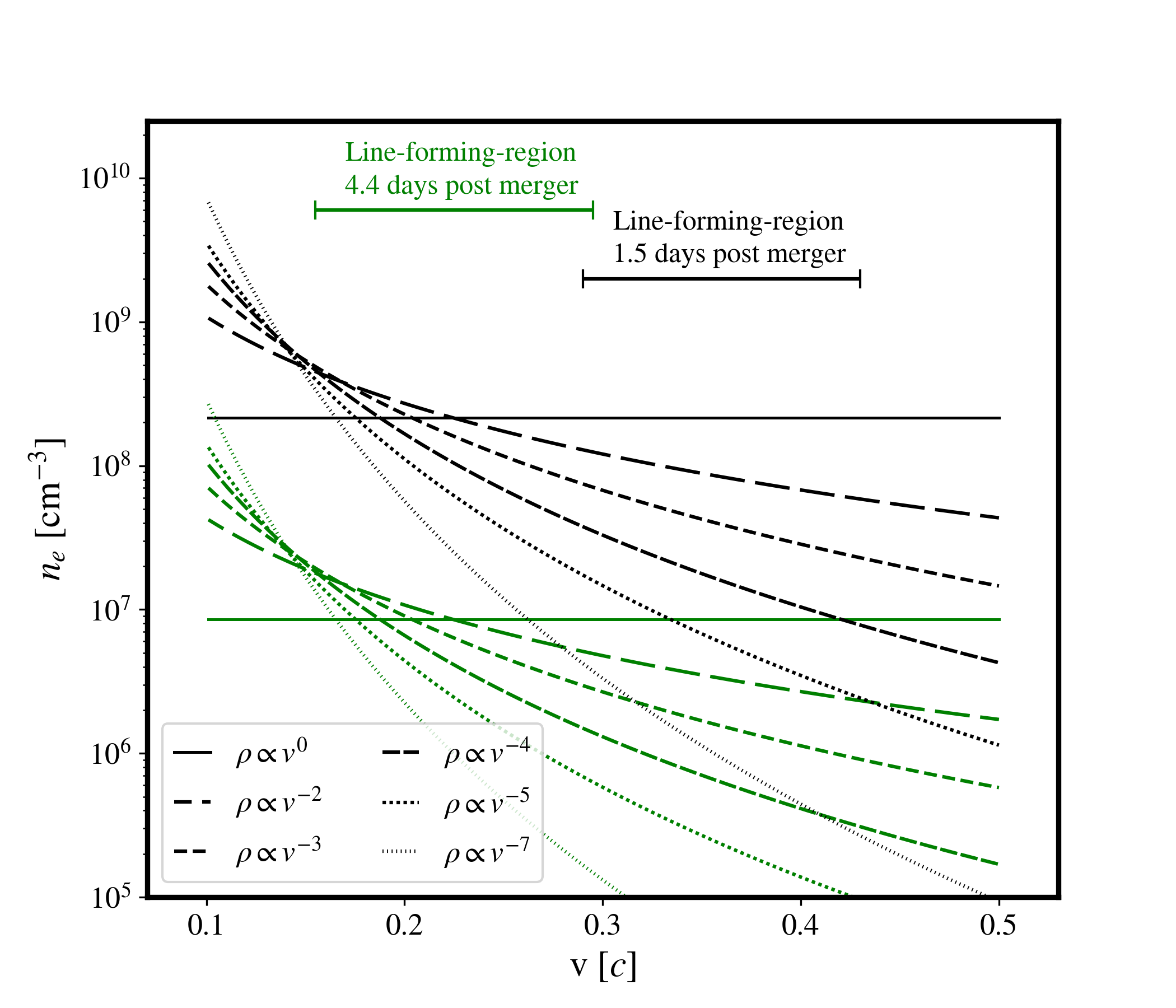}
  \caption{Electron-density versus velocity assuming ejected matter is power-law distributed, $\rho\,=\,\rho_0 \, v^{\alpha} \, t^{-3}$, with various power-law indices $\alpha$. Black and green lines indicate densities respectively 1.5 and 4.4 days post-merger. This includes a constant density model (i.e., $\alpha=0$), an equal mass at all radii (i.e., $\alpha=-2$) and various slopes with the typical mass increasingly ejected at lower velocities (i.e., $\alpha=-3$ to $\alpha=-5$). The normalization is set assuming i) the total mass ejected is similar to AT2017gfo (i.e., $\sim 0.04 M_{\odot}$), ii) the average nucleon-number is $A\sim100$ and iii) the average ionization-state is lowly ionized (i.e., $Z_{\mathrm{ion}}=2.5$).} 
    \label{fig:Density_structure}
\end{figure}

\begin{figure*}
    \includegraphics[width=\linewidth,viewport=22 15 1070 380 ,clip=]{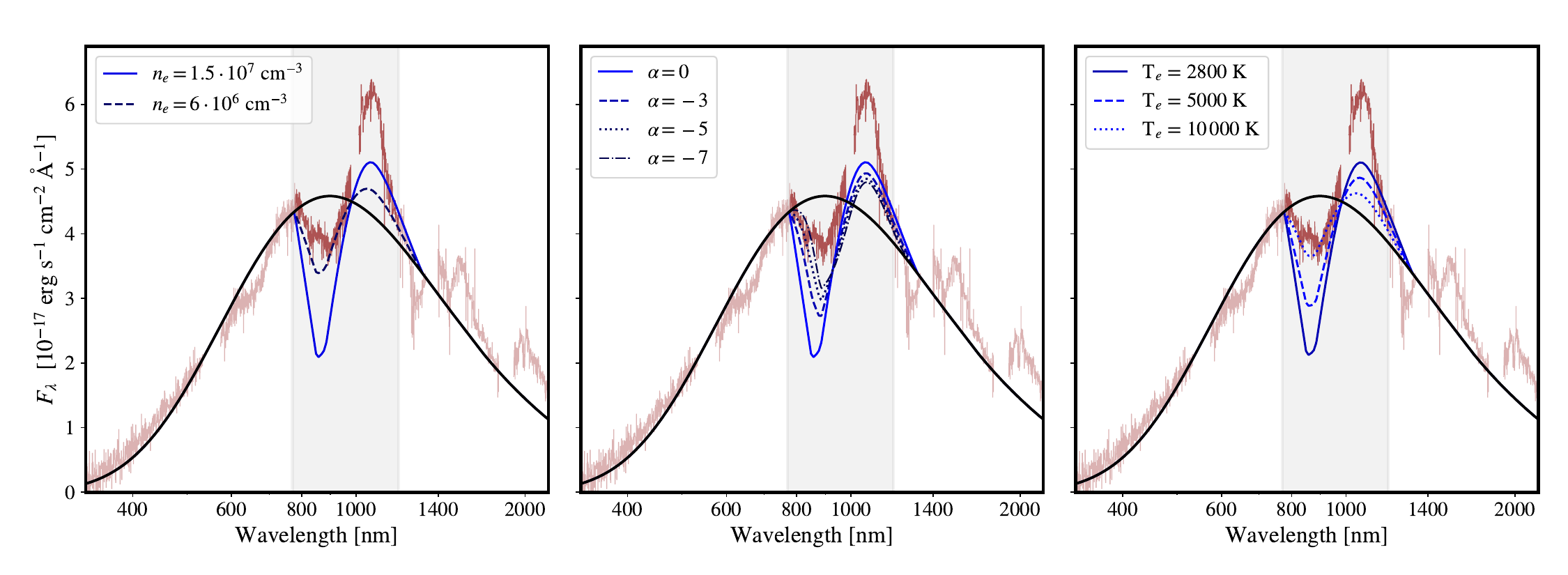}
    \caption{VLT/X-shooter spectrum of AT2017gfo 4.4 days post-merger with overlaid blackbody continuum ($T_{\rm BB}=3200$\,K from best-fit blackbody temperature compilation in \citet{Sneppen2024}) and P~Cygni feature assuming $X_{\mathrm{He}}=0.01$ -- but assuming a range of electron-densities ($n_e$, left panel), density power-law slopes ($\alpha$, center panel) and electron temperatures ($T_e$, right panel). For each panel, we assume the parameters from the other panel indicated with fully drawn lines (i.e.,\ $n_e=1.5\times10^{7}$ cm$^{-3}$, $\alpha=0$, $T_e=2800$\,K). For the central panel, the reference electron density $n_e=1.5\times10^{7}$ cm$^{-3}$ is at $0.19c$. 
    } 
    \label{fig:Xshooter_helium_3panel}
\end{figure*}

\begin{figure*}
    \includegraphics[width=\linewidth,viewport=21 24 840 270 ,clip=]{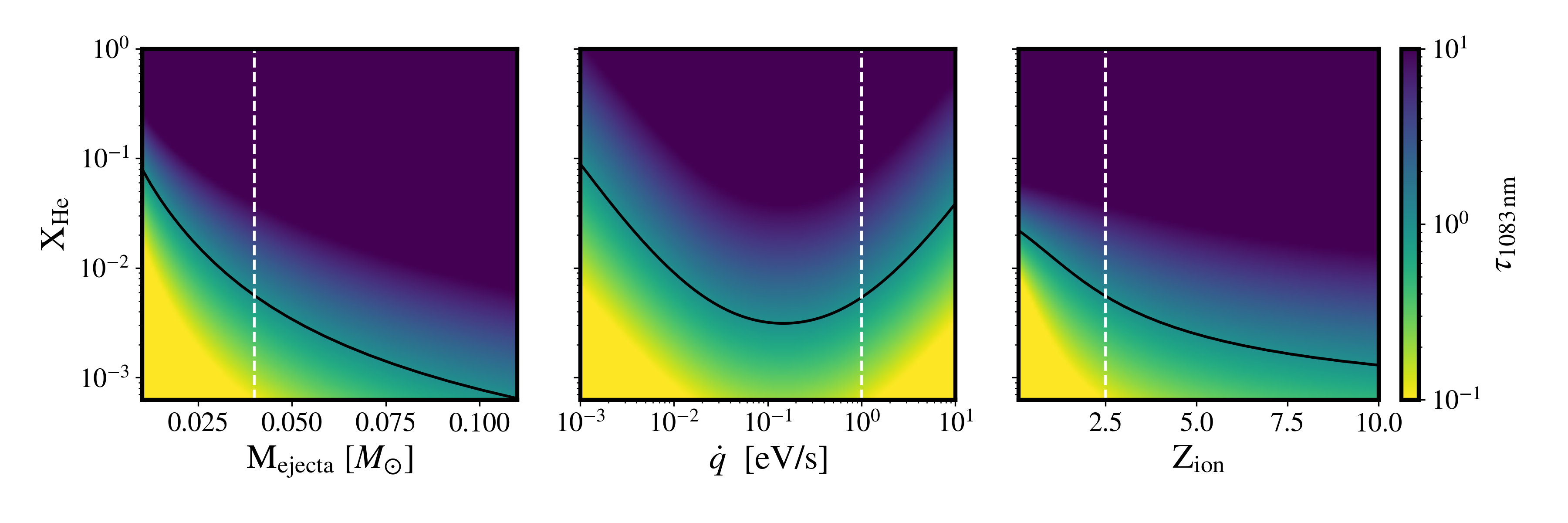}
    \caption{ Optical depth of He\,{\sc i}\,$\lambda 1083.3$\,nm line at photosphere given various ejecta properties: helium mass fractions, $X_{\mathrm{He}}$, total ejecta-mass, $M_{\mathrm{ejecta}}$, radioactive heating deposited per unit time per ion, $\dot{q}$ (expressed at the value at 1 day post-merger with the temporal decay following $t^{-1.3}$) and typical ionization degree of the \rprocess elements, $Z_{\mathrm{ion}}$. For all plots, we include the computed dependence of the work per ion on composition, $X_{\mathrm{He}}$ (see text), while we assume the power-law slope $\alpha=-5$ (from 0.1-0.5c) and typical nucleon mass number $A\sim100$. The black contour line indicates $\tau_{1083\,\mathrm{nm}}=1$, while the dotted white lines indicate the fiducial value of each parameter used for computations in the other panels. Lower ejecta masses, higher deposition-rates or lower ionization degree result in a higher helium mass upper bound. While we only vary one of the parameters $M_{\mathrm{ejecta}}$, $\dot{q}$ and $Z_{\mathrm{ion}}$ in any given panel (keeping the other two fixed), the latter two parameters are not fully independent as high deposition rates will also increase the typical ionization degree of the \rprocess elements. We note that this is a relatively subtle effect with orders-of-magnitude increase in $\dot{q}$ resulting in at most a few more ionized electrons per nuclei. 
    } 
    \label{fig:Xhe_verus}
\end{figure*}

\begin{figure}
    \includegraphics[width=\linewidth]{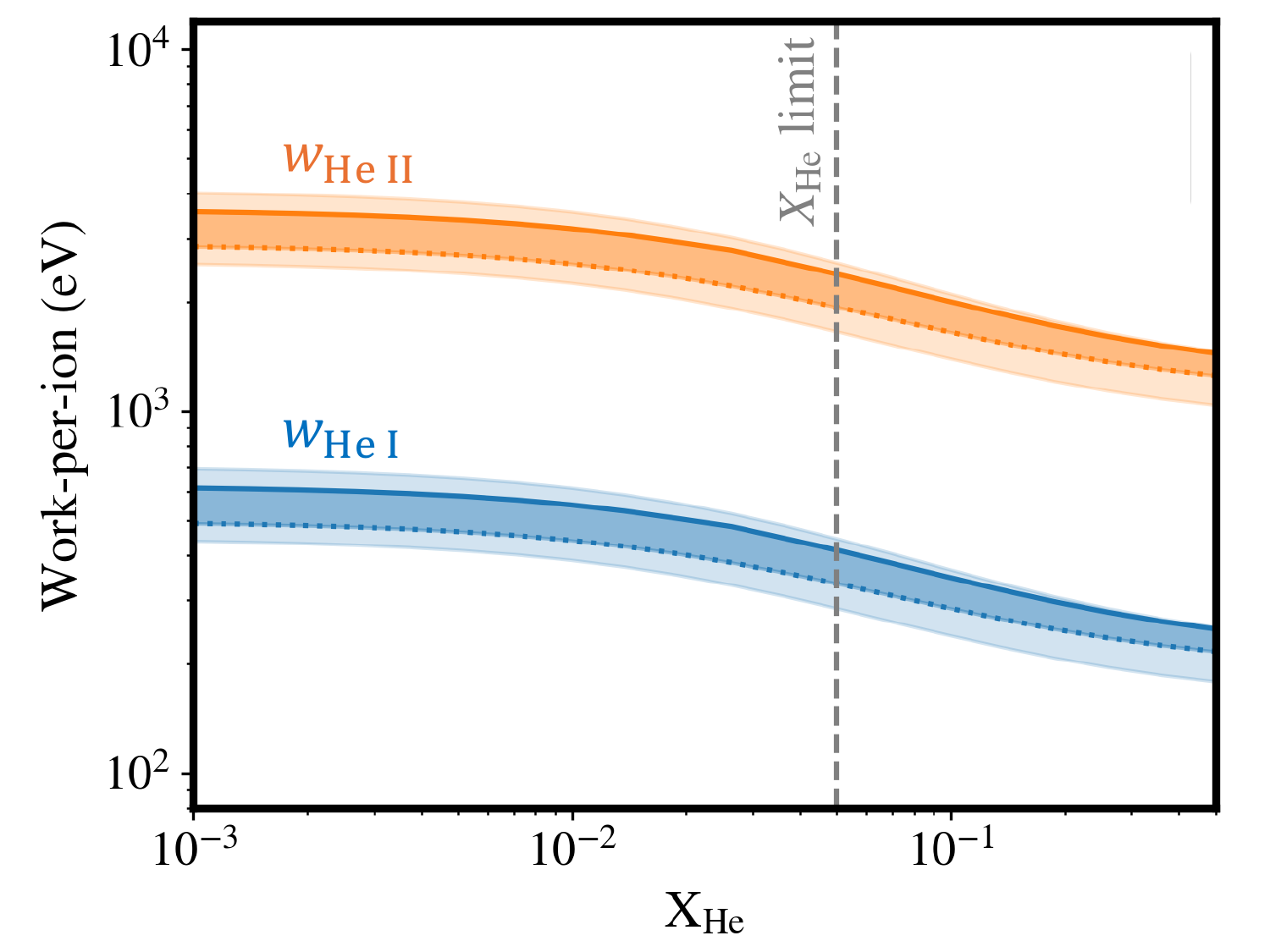}
    \caption{ Work per ion for \hei and \heii as a function of helium composition, $X_{\mathrm{He}}$. The contribution of heavy elements is approximated with Sr (see text for Fe comparison), while the ionization balance for both light and heavy species are solved given the fiducial non-thermal deposition-rate and a range of electron densities from $n_e=10^{8} \mathrm{cm}^{-3}$ (dotted line) to $n_e=6\cdot10^{6} \mathrm{cm}^{-3}$ (solid line). We also illustrate the impact of the non-thermal deposition-rate by varying this from a 0.01-10 times the fiducial value for $n_e=6\cdot10^{6} \mathrm{cm}^{-3}$ (lightly shaded region). 
    In the limit of $X_{\mathrm{He}}<0.01$ ($X_{\mathrm{He}}>0.1$), r-process elements (helium) control the number of free electrons and thus set the loss to thermal electrons. Our fiducial value for the work per ion is computed at $X_{\mathrm{He}=0.01}$ (and $n_e=6\cdot10^{6} \mathrm{cm}^{-3}$), but only a 25\% difference are expected at the observational limit $X_{\mathrm{He}}=0.05$ (vertical dashed line), which remains smaller than the order-of-magnitude uncertainty associated with the deposition-rate, $\dot{q}$. }
    \label{fig:Work_per_ion_versus_Xhe}
\end{figure}

\subsection{Estimating the electron density}\label{sec:electron-density}
To provide a sense of the main dependencies, in Fig.~\ref{fig:Density_structure}, we show the electron density as a function of velocity (at 1.5 and 4.4 days post merger) given a simplified parametric ejecta distribution. Specifically, we assume a relatively low ionization state in the ejecta (i.e., fiducially $Z_{\mathrm{ion}}=2.5$, as characteristically expected from NLTE ionization modeling \citep{Pognan2023}) in which the average mass number is $A\sim100$. We will assume that the ejecta have a density that follows a power-law distribution with radius, $\rho\,=\,\rho_0 \, v^{\alpha} \, t^{-3}$ (i.e., as in the main text), such that when integrated over volume (from 0.1c to 0.5c) the ejecta have a total mass similar to AT2017gfo ($\sim 0.04 M_{\odot}$ \citep{Cowperthwaite2017,Villar2017,Siegel2019}). The electron density is then related to the matter density as $n_e =  Z_{\text{ion}}\rho/(A m_u)$, with $m_u$ the mass number unit.

Figure~\ref{fig:Density_structure}  highlights the two strongest functional dependencies in the electron density. Firstly, the homologous expansion rapidly decreases the density in any layer ($\rho \propto t^{-3}$), which implies at late enough times the ejecta will reach low electron-densities and become inefficient at recombination. Secondly, for steep density gradients one infers relatively consistent electron-densities in the receding line-forming region (these velocity regions inferred from the wavelength-range of the P~Cygni are indicated with the black and green ranges in Fig.~\ref{fig:Density_structure}). The expectation from hydrodynamic simulations and lightcurve modeling of AT2017gfo, e.g.,~\cite{Villar2017}, is that the bulk of matter is ejected at lower velocities, so as the line-forming region recedes deeper into the ejecta the mass-density (and by extension $n_e$) should increase if solely considering this effect. For instance, $n_e(0.3c, 1.5\,d) \sim (2-6) \times \,n_e(0.19c, 4.4\,d)$ for $\alpha$ between $-3$ and $-5$. This relatively slow temporal decline in the line-forming-region's $n_e$ may be an important component behind the observed spectral features of AT2017gfo being remarkably consistent over such a large range of times and velocities \citep{Sneppen2024}. Higher order effects like variations in nucleon-number, $A$, and the typical ionization-state, $Z_{\mathrm{ion}}$, is at most a factor of a few, while the heating rate per nucleon is rather consistent over a broad range of compositions for the time-scales of interest. Based on the simple profiles in Fig A.1, we expect that the electron density at the photosphere around 1.5 days should be roughly $(4-10)\cdot10^{7} \,\mathrm{cm^{-3}}$, dropping to $\sim10^{7} \,\mathrm{cm^{-3}}$ by 4.4 days. These estimates are supported by additional quantitative arguments. First, the rapid recombination from 1.43-1.47 days that appears to be needed to explain the observed \sriii to \srii transition constrains the line-forming-region at this time and at these velocities ($v>0.3c$) to have an electron-density, $n_e\gtrsim10^8\,$cm$^{-3}$~\citep{Sneppen2024}. Specifically, this follows from the intuition that once the ejecta has entered the radiative regime where \srii is for the first time permitted (e.g., $T_{rad}\lesssim5000\,$K), the bottleneck timescale is the recombination timescale from \sriii, which is given by 
\begin{equation}
    \tau_{rec} = 1/(n_e \alpha_{\mathrm{Sr\,III}}).
\end{equation} 
For the calculated recombination rates of strontium $\alpha_{\mathrm{Sr\,III}}\approx2\cdot10^{-12} \mathrm{\,cm^3/s}$ \citep[e.g., C. Ballance (priv. comm.) and][]{Banerjee2025} and observed recombination-timescale of a few hours, this requires the aforementioned electron density.
Assuming homologous expansion, these velocity-layers will be diluted $\sim 25$ fold by 4.4\,days, while the resulting density at fixed time would be a factor of 2.5-10 (4-32) times greater at a velocity of 0.19c (0.15c) than 0.3c (from $\rho \propto v^{-2}$ to $\rho \propto v^{-5}$) -- implying $n_e \gtrsim 0.6-4 \times 10^{7}\,{\rm cm}^{-3}$ ($n_e \gtrsim 1-13 \times 10^{7}\,{\rm cm}^{-3}$) at $t=4.4$ days (if $n_e\gtrsim10^8\,$cm$^{-3}$ at $0.3c$ and $t=1.5$ days). 

Furthermore, densities from analytical parametrizations can be compared with the output of any specific hydrodynamical simulation. For instance, using the same approximations for composition and ionization as above, the $n_e$ in the helium enriched hydrodynamical simulations in Fig.~\ref{fig:sim_snapshots} is between $2.5\times10^{7}$\,cm$^{-3}$ and $2\times 10^{8}$\,cm$^{-3}$ at 4.4 days across grid cells in the polar ejecta with velocities in the interval $0.19-0.3c$. As for the analytical prescriptions, even the singly ionized domain of electron density is characterised by density conditions where the Sobolev optical depth, $\tau_{\rm \hei\,\lambda1083\,nm}>10$, would be inconsistent with the absence of a strong helium feature in observations. In contrast, helium-deficient models have conditions for an optically thin helium line.

\subsection{P~Cygni profiles sensitivity to assumed physical conditions}\label{app:sensetivity}

Ultimately, the electron density constraints presented in the paper are motivated by the observational constraints (e.g., recombination-timescales of features and existence of specific ionization-states like \srii) and the typical ejecta-mass estimates for AT2017gfo. In Fig.~\ref{fig:Xshooter_helium_3panel} (left panel), we show the strength of the P Cygni for various values of the photospheric electron-density, $n_e$. For sufficiently low electron density the strength of the feature diminishes. The feature is observed to be optically thin, which favours modest electron densities for $X_{\mathrm{He}}\gtrsim0.01$; higher densities, weaker non-thermal ionization, and generally any conditions that yields an optically thick helium feature are incompatible with this. In Fig.~\ref{fig:Xshooter_helium_3panel} central panel, we show the strength of the P Cygni for various power-law slopes, $\alpha$, characterizing the density decline in velocity space. Particularly, the blue wing of the absorption feature becomes weaker for more rapid declines - with a more negative $\alpha$ corresponding to a more rapid decline in the optical depth $\tau$ with velocity (ie. less scattering from higher velocity layers). However, the feature is not strongly sensitive to the assumption of the exact density distribution for a large range in $\alpha$ (including a steep decline, $\alpha=-5$, or an equal density at all radii $\alpha=0$) as long as the electron-density near the photosphere is not $\ll 6\times10^6\,{\rm cm^{-3}}$. In Fig.~\ref{fig:Xshooter_helium_3panel} (right panel), we show the strength of the P Cygni for various assumptions of the electron temperature used in the recombination-rate coefficient, $T_e$. Notably, the feature is weaker in the limit of low electron densities and high electron temperatures, as this produces the weakest recombination rates. We note the 2800\,K assumed in the fiducial model is observationally motivated by the empirical goodness-of-fit for a simple blackbody model \citep{Sneppen2023_bb} and the 5-10\,\% consistency between electron and radiation temperature seen near ionization transitions \citep{Sneppen2023c,Sneppen2024}, but nearly the same results are obtained across a broad temperature range. 

In Fig.~\ref{fig:Xhe_verus}, we highlight the robustness of our $X_{\text{He}}$ constraint on physical properties of the ejecta in another way. Here, we show the optical depth at the photosphere of the He\,{\sc i}\,$\lambda 1083.3$\,nm line given various helium mass fractions, total ejecta-mass, radioactive heating per unit time per ion and typical ionization degree of the \rprocess elements. We assume for these plots the power-law slope $\alpha=-5$ (from 0.1-0.5c), that the average nucleon-number is $A\sim100$ and we only vary one parameter per panel and keep the other fixed. A low ionization state and lower total ejecta mass implies a smaller electron density, which in turn allows more helium to be hidden in the ejecta. Conversely, a lower deposition rate decreases the ionization rate and thus suggest stronger bounds. 
$\dot{q}$ and $Z_{\mathrm{ion}}$ should not be thought of as independent with high deposition rates also increasing the ionization degree, but we note this is a relatively subtle effect with orders-of-magnitude increase in $\dot{q}$ resulting in at most a few more ionized electrons per nuclei. In the low $\dot{q}$-limit thermal ionization will maintain typically single ionized ejecta at these temperatures, while for very large $\dot{q}$ (1-10 eV/s) particles will only be a few times ionized. Regardless, across a broad range of physical parameters we robustly obtain $\tau_{1083\,\mathrm{nm}}>1$ if $X_{\mathrm{He}}\gtrsim0.05$. 

\subsection{Work per ion dependence on composition}\label{sec:work_per_ion}

In Fig.~\ref{fig:Work_per_ion_versus_Xhe}, we also show the work per ion dependence on the composition used for computing $\tau_{1083\,\mathrm{nm}}(X_{\mathrm{He}})$.
These evaluations of the Spencer-Fano solver, were computed while iteratively solving for the ionization-balance of both He (including \hei, \heii, \heiii) and Sr (including \sri, \srii, \sriii given recombination rates in \cite{Banerjee2025,Singh2025} and consistent with independent calculations (C. Ballance, priv comm.)) as a function of the total density and deposition rate. The electron density is in the motivated range ($6\cdot 10^{6} \,\mathrm{cm}^{-3}<n_e<10^{8} \,\mathrm{cm}^{-3}$) for expected ejecta masses. We explored varying the deposition rate from 0.01 to 10 times the fiducial value (see Fig.~\ref{fig:Work_per_ion_versus_Xhe}). 
While we consider Sr a reasonable representative element for evaluating the \rprocess stopping power, we note if one instead considered Fe as representative of the heavy-elements one would obtain characteristically 20\% higher $w_i$.
When helium is the dominant source of free electrons, the composition starts to matter more because the growing number of electrons increases the amount of non-thermal energy lost to heating them. For compositions with $X_\mathrm{He} \leq 0.05$, this changes $w_i$ by up to 25\%, but for $X_\mathrm{He} = 0.4$ the change can be as large as 50\% (see Fig.~\ref{fig:Work_per_ion_versus_Xhe}).

We conclude that, under our assumptions, the variation of the work per ion for helium for different compositions is modest (factor of two to three, at most). Since, in our calculations, this quantity is degenerate with the energy deposition ($\dot{q}$), to which we have argued (Sect.\ref{sec:helium_uncertainties}) that our conclusions are insensitive over more than an order-of-magnitude variation, this composition dependence does not significantly impact our findings.

\section{Hydrodynamic models used in Sect.~\ref{sec:HMNS_lifetime}}\label{sec:hydrodynamic-models}

\begin{figure}
  \includegraphics[width=\linewidth]{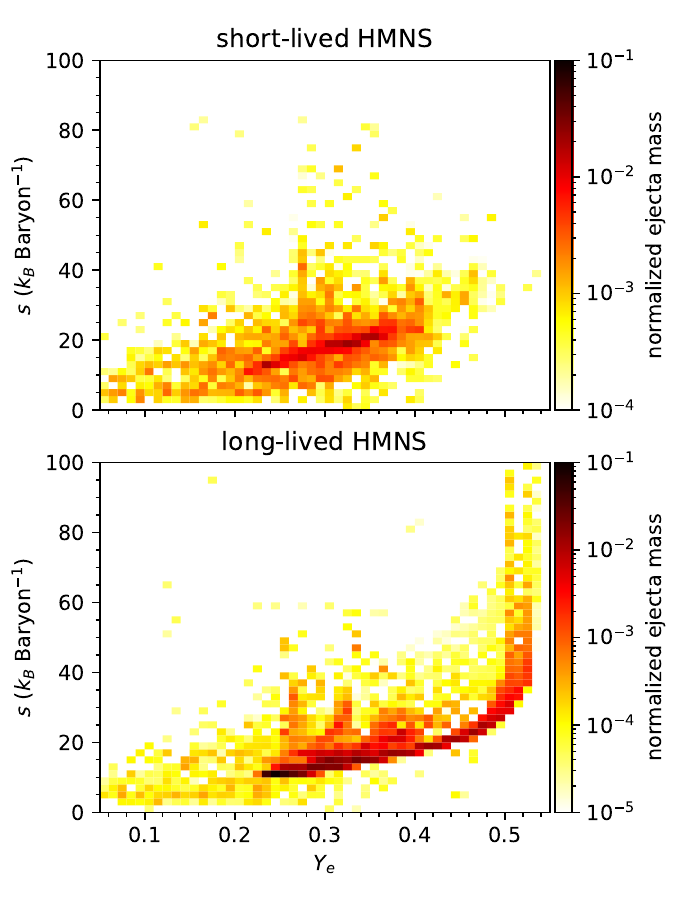}
  \caption{Mass-distribution histograms in the space of electron fraction and entropy per baryon (both measured when the temperature drops below 5\,GK) for ejecta material resulting in two numerical simulations in which the HMNS remnant is short-lived (model ``sym-n1-a6-short'' with $\tau_{\mathrm{BH}}=10$\,ms; left panel) and long-lived (model ``sym-n1-a6'' with $\tau_{\mathrm{BH}}=122$\,ms; right panel). The neutrino wind in the long-lived model produces a substantial amount of ejecta with $Y_e\gtrsim 0.45$.} 
    \label{fig:Y-s-histograms}
\end{figure}
\begin{figure}
  \includegraphics[width=\linewidth]{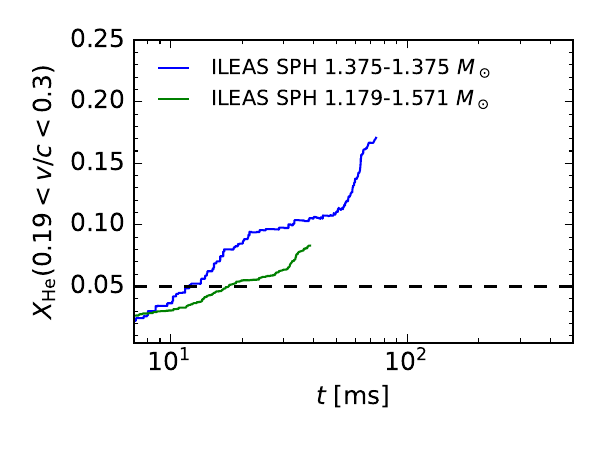}
  \caption{Same as Fig.~\ref{fig:He_HMNS_time} but for the two merger models with $M_1+M_2=2.75$ and $q=1$ and 0.75 evolved only with the 3D GR SPH code ILEAS, i.e., without mapping at $t=10\,$ms to the 2D SR grid code as for the long-term evolution models. The similarly steep increase of $X_{\mathrm{He}}(t)$ in these models supports the robustness of this behavior with regard to the mapping, axisymmetry, and more approximate treatment of GR in the post-merger models.} 
    \label{fig:vimal_xhe}
\end{figure}

In this appendix, we provide supplementary information about the hydrodynamic merger simulations used to connect the helium mass fraction with the HMNS lifetime, assess the sensitivity of helium production across certain numerical approximations, and summarize the nucleosynthesis post-processing.

Most of the merger models used in Sect.~\ref{sec:HMNS_lifetime} to discuss the dependence of the helium mass fraction $X_{\mathrm{He}}$ on the HMNS-remnant lifetime $\tau_{\mathrm{BH}}$ are taken directly from Ref.~\citep{Just2023}, to which we refer for a detailed description of the model setup. The models are built from three-dimensional merger simulations (until 10\,ms after merger), performed with a general-relativistic (GR) SPH code \citep{Oechslin2002} using a leakage-plus-absorption (``ILEAS'') neutrino treatment \citep{Ardevol-Pulpillo2019a} and assuming the conformal flatness condition, and two-dimensional (axisymmetric) simulations covering the post-merger (i.e., $>10\,$ms) evolution, conducted with a special-relativistic (SR) finite-volume code \citep{Obergaulinger2008a} adopting spectral M1 neutrino transport \citep{Just2015b} and a gravitational potential that approximately reproduces general-relativistic effects \citep{Marek2006}. Angular-momentum transport due to small-scale turbulence is described using a recently developed two-parameter viscosity prescription inspired by the $\alpha$-viscosity scheme of Ref.~\citep{Shakura1973}.

With respect to Ref.~\citep{Just2023} only the short-lived ($\tau_{\mathrm{BH}}=10\,$ms) models as well as the model without viscosity are new models. Each of the models listed in the legend of Fig.~\ref{fig:He_HMNS_time} is characterized through the model name as follows: Short-lived (``short'') models have a total mass of both NS progenitors of $M_1+M_2=2.8\,M_\odot$, while all others have $2.75\,M_\odot$; ``sym'' (``asy'') models have a mass ratio of $q=M_1/M_2=1$ ($q=0.75$); and the parameters used to describe the turbulent viscosity are indicated by the values after ``n'' and ``a'' (broadly speaking, the impact of viscosity is smaller for models with higher ``n''- and lower ``a''-values; cf. \citep{Just2023}). The suffix ``novis'' denotes the model with vanishing viscosity. In contrast to the long-lived models from Ref.~\citep{Just2023}, in which the time of BH formation is self-consistently obtained by the point when the HMNS becomes gravitationally unstable, in the ``short'' models we trigger an early collapse by hand by removing at $t=\tau_{\mathrm{BH}}=10\,$ms a small amount of material in the innermost part of the HMNS, which leads to an immediate collapse. Although this treatment is not fully hydrodynamically consistent, the impact on the resulting ejecta properties should be minor, because the total mass of the system that would actually collapse at $\tau_{\mathrm{BH}}=10\,$ms will only be a few percent higher than the $2.8\,M_\odot$ used for our current short-lived models.

In Fig.~\ref{fig:Y-s-histograms} we provide mass-distribution histograms for a short- and a long-lived model, illustrating that the long-lived case is characterized by a distinct neutrino wind component at $Y_e\gtrsim 0.45$.

To test the impact of mapping at $t=10\,$ms, axisymmetry, and the more approximate treatment of general relativity in the long-term models, we continued the 3D GR SPH simulations of the two merger models with $M_1+M_2=2.75$ and $q=1$ and 0.75 for a few tens of milliseconds longer than the mapping time of 10\,ms. The $X_{\mathrm{He}}(t)$ lines shown in Fig.~\ref{fig:vimal_xhe} closely resemble the steep increase of $X_{\mathrm{He}}(t)$ of the 2D SR grid simulations (for any viscosity) displayed in Fig.~\ref{fig:He_HMNS_time}, suggesting robustness of this behavior with respect to the aforementioned technical differences between both types of simulations.

The abundances of elements synthesized in the ejected material in the above hydrodynamical models are obtained in a post-processing step using the same methods as in Refs.~\citep{Just2023} (called network B therein) and \citep{xiong2024production, Fernandez2024a}. To this end, a nuclear network calculation is run for about $\sim$5000{--}10000 tracer particles (per model) sampling the ejecta. For each trajectory, the full network is started at 8~GK to account for quasi-statistical equilibrium \citep{Meyer1998,Hix.Thielemann:1999a}. The network assumes neutron-capture and photodissociation rates based on the HFB21 mass model~\cite{Goriely2010} as described in \cite{mendoza2015nuclear}, as well as the $\beta$-decay rates from \cite{marketin2016largescale}.

\begin{figure}
  \centering
    \includegraphics[width=0.99\columnwidth,viewport=22 20 355 345 ,clip=]{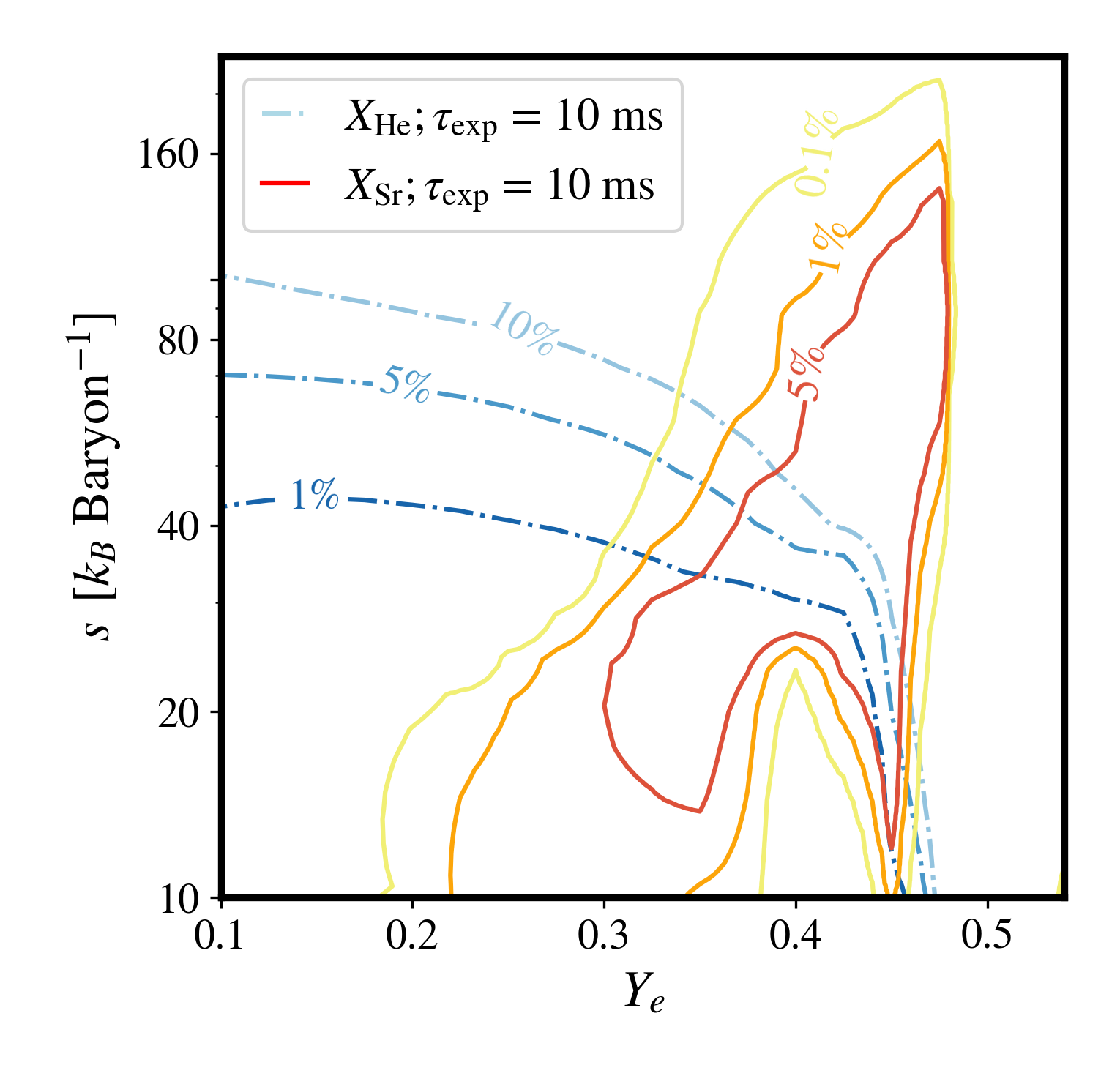}
  \caption{  
  Same as Fig.~\ref{fig:He_nucleosynthesis} but showing (selected contours of) the mass fractions of both strontium (solid lines) and helium (dot-dashed lines). In contrast to helium, strontium production becomes inefficient for $Y_e\gtrsim 0.5$ but instead is more efficient in a broader range of $0.2\lesssim Y_e\lesssim 0.45$ at lower entropies $s/k_B\lesssim 30$.
  }
    \label{fig:HeSr_nucleosynthesis}
\end{figure}

\begin{figure}
    \includegraphics[width=0.95\linewidth,viewport=2 5 275 305 ,clip=]{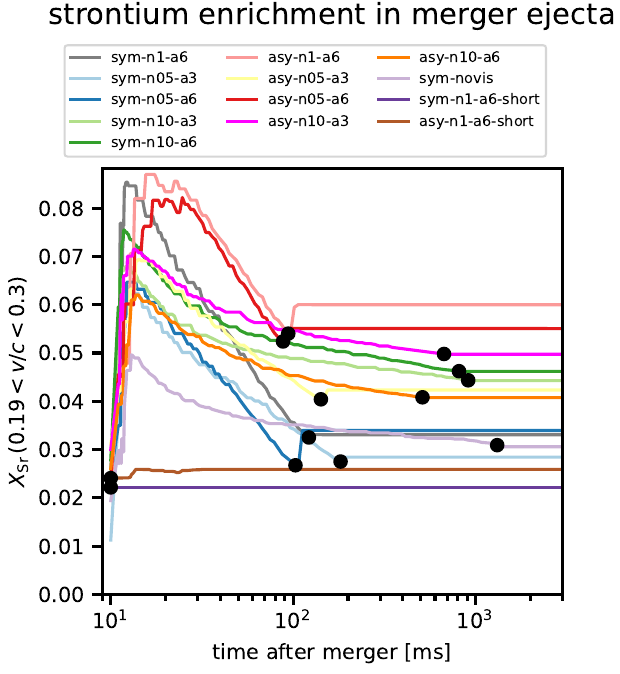}
    \caption{Same as Fig.~\ref{fig:He_HMNS_time} but for the mass fraction of strontium instead of helium. Due to the high $Y_e$ in the polar neutrino-driven wind, strontium enrichment into the considered velocity range ceases after 10{--}20\,ms. Since the total ejecta mass keeps growing with time, the relative mass fraction declines. The ejecta from the BH-torus system (formed at times indicated by the black circles) carry strontium as well but are too slow to contribute, resulting in a flat line. Unlike helium, for strontium the final mass fraction is only poorly correlated with the time of BH formation $\tau_{\mathrm{BH}}$.
     }
    \label{fig:Sr_HMNS_time}
\end{figure}


\section{Strontium as tracer of black-hole formation time}\label{app:strontium} 

In this appendix we examine the possibility of using the abundance of strontium, instead of helium, to formulate constraints on the NS-remnant lifetime. This idea, briefly considered in \citet{Perego2022}, is motivated by the strong evidence that strontium is related to the spectral feature of AT2017gfo at early epochs \cite{Watson2019, Sneppen2024_helium}. As we argue in this section, a lifetime constraint based on the spectral strontium feature remains, however, highly uncertain, mainly for two reasons.

The first reason, which was already acknowledged by Ref.~\cite{Perego2022}, is that the strontium-mass estimate in Ref.~\cite{Watson2019} was derived from a spectral model that adopts the LTE approximation. The strontium mass inferred from LTE is best viewed as a lower limit on the true strontium mass estimated from NLTE, because \srii (the relevant ionization state for observational inference) is the dominant ionization state under LTE and the motivated KN temperatures. A reliable measure of the strontium mass of AT2017gfo must be inferred from an NLTE model (as was done for helium in the present work but for strontium is not available so far), which can introduce orders-of-magnitude variations from the LTE-based value (see also Sect.~\ref{sec:strontium}).

The second reason strontium is problematic is that it is not likely a unique signature of NS-remnant outflows. To understand this argument, we first compare the nucleosynthesis conditions under which strontium is synthesized with the corresponding ones for helium (see Fig.~\ref{fig:HeSr_nucleosynthesis}). While helium is created most efficiently for high values of $Y_e\gtrsim 0.45-0.5$, significant strontium yields are instead obtained for a more broad range at lower values $0.2\lesssim Y_e\lesssim 0.45$ (for typical ejecta entropies of a few tens of $k_B$). Looking at the histograms in Fig.~\ref{fig:Y-s-histograms}, the most striking difference in $Y_e$ and $s$ between cases of long-lived and short-lived NS remnants appears to be given by the polar-wind component with a $Y_e$ close to 0.5 and entropies $30\lesssim s/k_B\lesssim 100$. These considerations suggest helium to be a more suitable element to unambiguously distinguish long-lived from short-lived models because, first, strontium may not be efficiently produced in the high-$Y_e$ component (because of the sharp drop of $X_{\rm Sr}$ around $Y_e\sim 0.45-0.48$ in Fig.~\ref{fig:HeSr_nucleosynthesis}) and, second, it may additionally be co-produced in low-$Y_e$ material that is ejected in both long-lived and short-lived models.

This assessment is supported by our hydrodynamical models (cf. Sect.~\ref{sec:hydrodynamic-models}). In Fig.~\ref{fig:Sr_HMNS_time} we see that strontium is enriched in the relevant velocity layers only during the first 10{--}20\,ms of NS-remnant evolution and effectively shuts off afterwards. This is because $Y_e^{\rm eq,abs}$ (and therefore the actual $Y_e$) in the polar neutrino wind (right panel of Fig.~\ref{fig:yeeq}) lies in a favorable range for strontium production only briefly. After a few tens of milliseconds, strontium production is quenched in the high-velocity polar wind due to the rise of $Y_e^{\rm eq,abs}$ towards $\gtrsim 0.45$. Strontium keeps being produced in low-velocity outflows launched from the NS remnant and subsequently from the BH-torus remnant \cite[e.g.,][]{Just2023}, but these ejecta are too slow to be observable through the P~Cygni feature at times $t\lesssim 5\,$days and therefore are not accounted for in Fig.~\ref{fig:Sr_HMNS_time}. As a result, the final amount of strontium does not correlate reliably with the lifetime, e.g., a long-lived NS remnant does not necessarily imply large strontium abundances. Obviously, a poor correlation would make it difficult to constrain the lifetime from a potentially observed strontium abundance. We note that the correlation is unlikely to improve even if, hypothetically, the strontium abundance could be observationally determined in the more slowly expanding ejecta ($v<0.2\,c$), because in these layers strontium can originate from both NS remnants and BH-torus remnants.

Of course, Fig.~\ref{fig:Sr_HMNS_time} is subject to uncertainties carried by our hydrodynamic models, which have to be assessed by future, more sophisticated models. A stronger correlation between the strontium mass (fraction) and \taubh may emerge if more accurate hydrodynamic models find that strontium is synthesized (also) in high-velocity ejecta from the NS remnant. The NS remnants in our viscous 2D models can release material only in the form of neutrino-driven winds and viscous ejecta, whereas recent 3D simulations additionally found spiral winds \cite[][]{Nedora2019y, Nedora2021} or magneto-hydrodynamic outflows \cite[e.g.,][]{Combi2023a, Curtis2024a, Aguilera-Miret2023b, Kiuchi2024a, Musolino2024c, Kalinani2025a} to be operating. Spiral winds, which have been considered as the source of strontium by ~\citet{Perego2022}, tend to be too slow to fulfil the condition $v> 0.2\,c$ \cite[][]{Nedora2021}, and the same is true for viscous NS-remnant winds \cite{Metzger2014, Fujibayashi2018, Just2023}. Magnetically-driven ejecta can be very fast in principle, but their properties are less well explored so far, given the challenging computational requirements of resolving magneto-hydrodynamic effects. However, due to their polar nature, $Y_e$ is likely to be close to $Y_e^{\rm eq,abs}$ in magnetically-driven ejecta, which would imply their nucleosynthesis conditions to be similar to polar neutrino-driven winds in non-magnetized models (i.e.,\ show inefficient strontium production).

In summary, we argue for helium over strontium for constraining the BH-formation time, because the mass of strontium in AT2017gfo is still very uncertain and because helium is presumably a better tracer of fast, high-$Y_e$ winds characteristic of long-lived merger remnants.

\begin{figure*}
    \includegraphics[width=5.7cm]{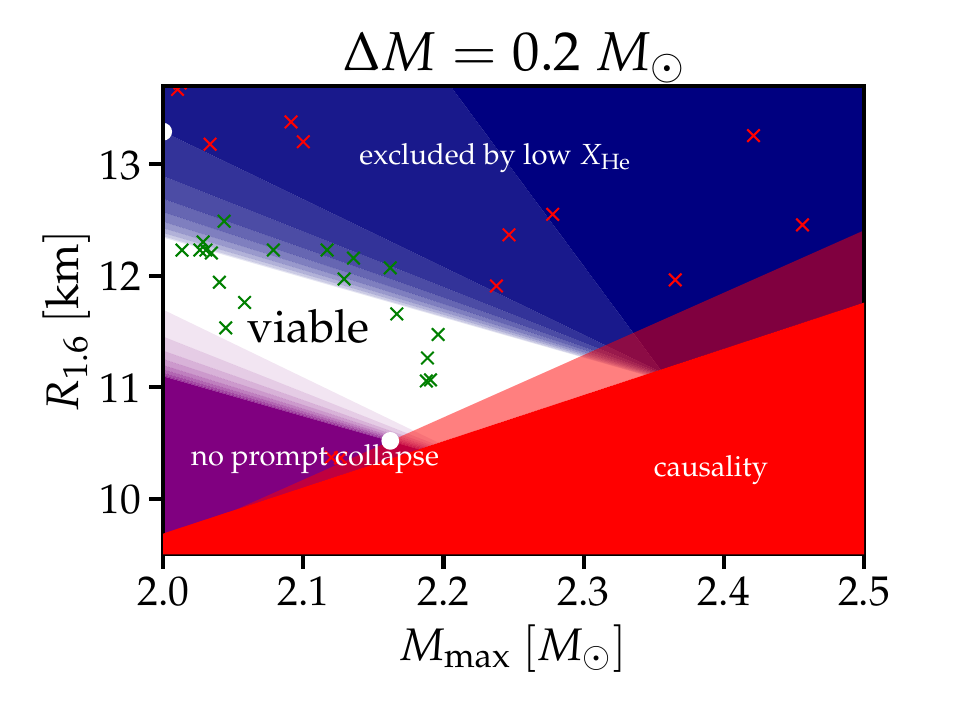}
  \includegraphics[width=5.7cm]{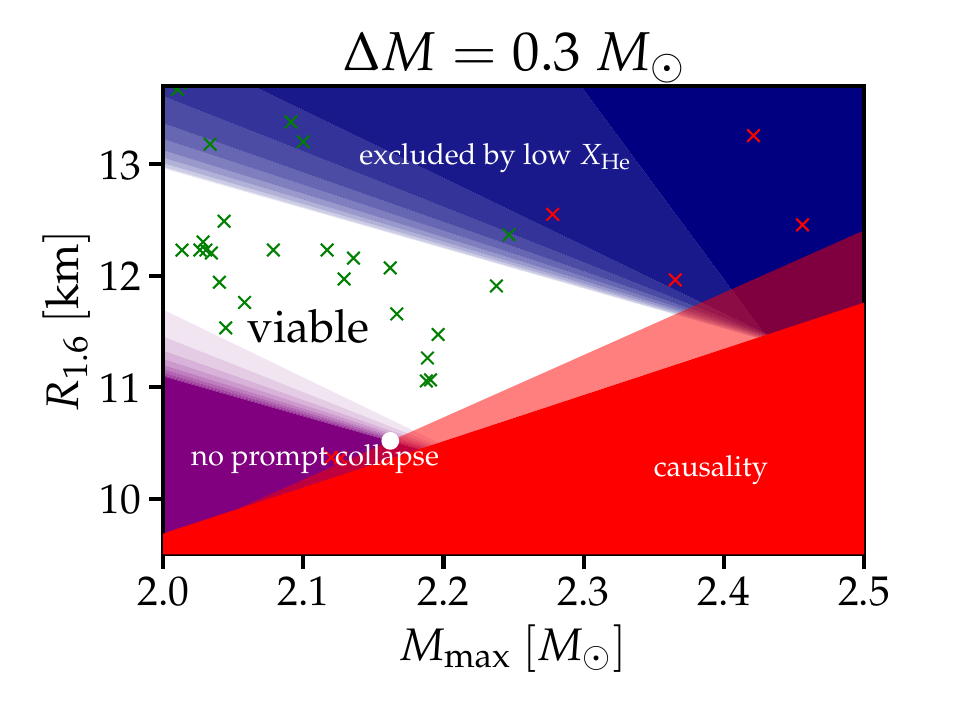}
  \includegraphics[width=5.7cm]{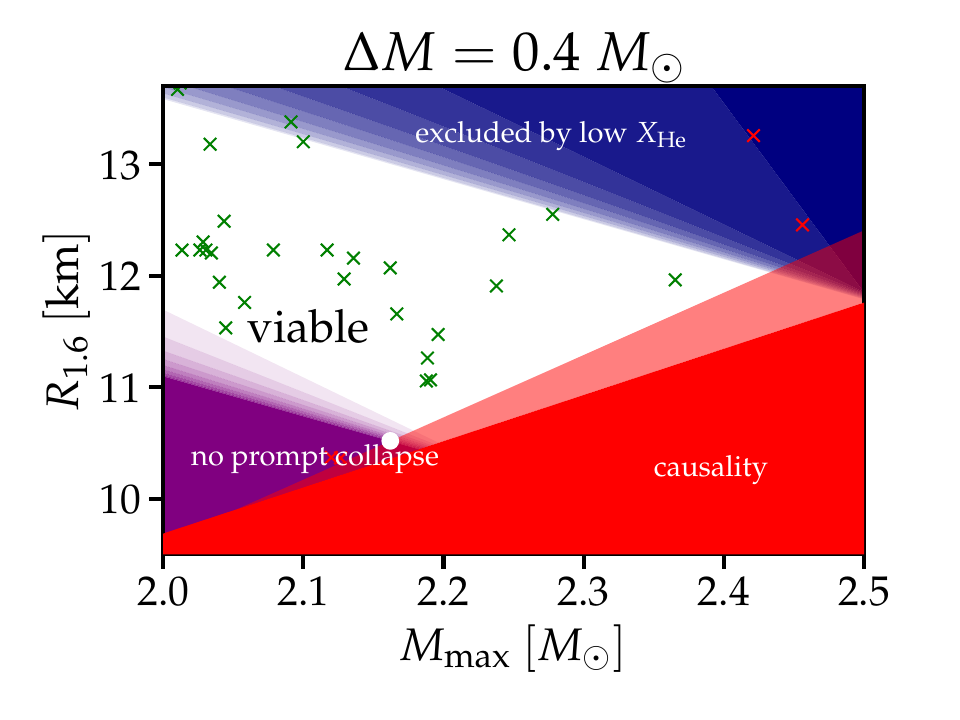}\\
      \includegraphics[width=5.7cm]{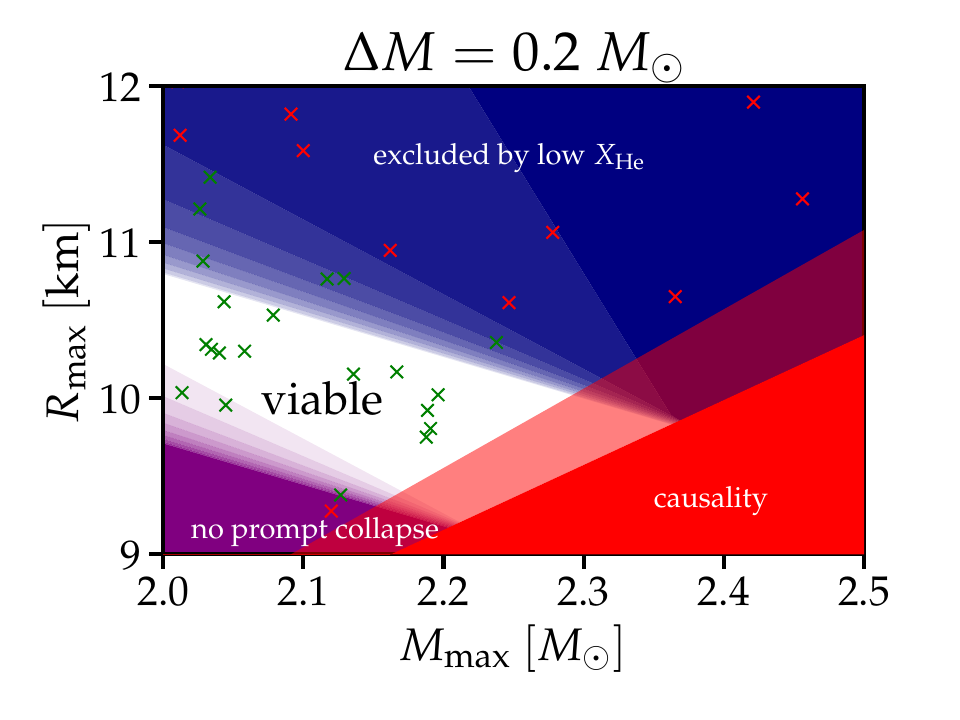}
  \includegraphics[width=5.7cm]{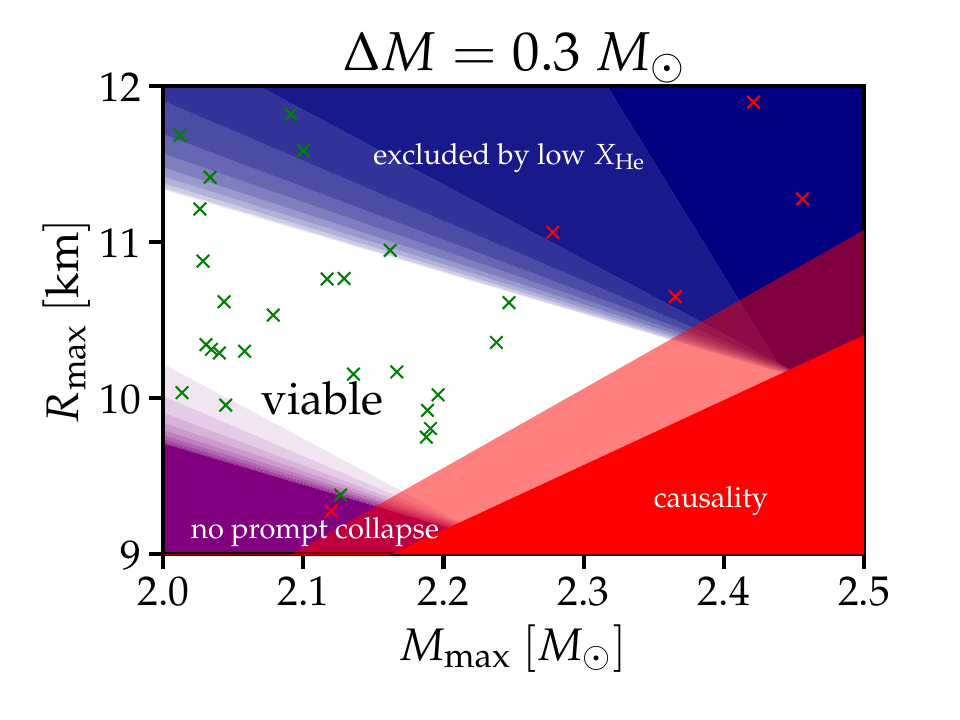}
  \includegraphics[width=5.7cm]{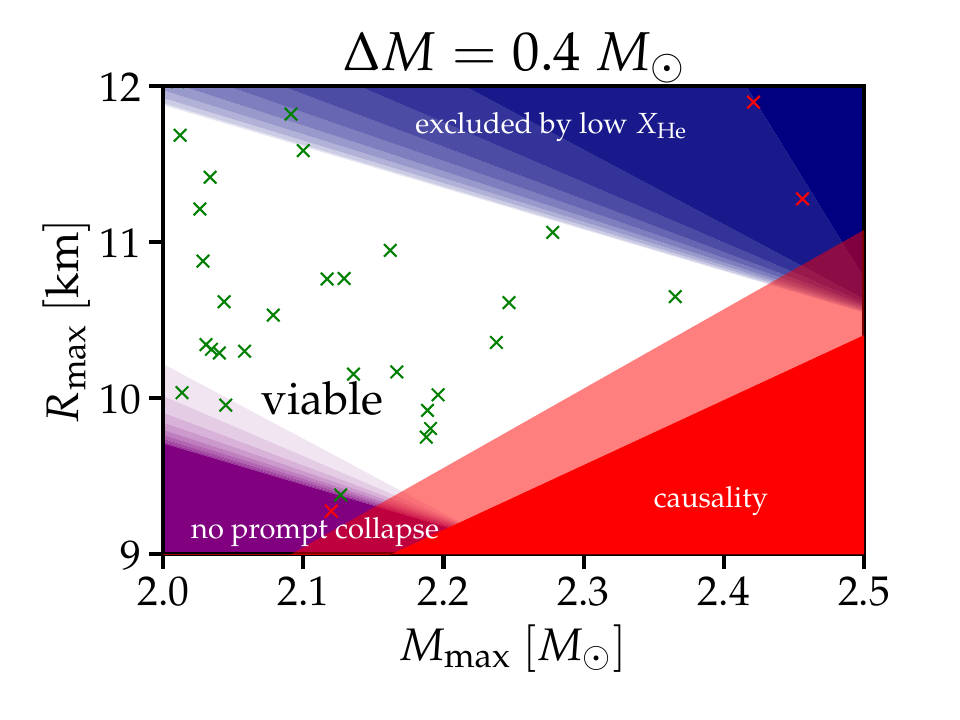}\\
      \includegraphics[width=5.7cm]{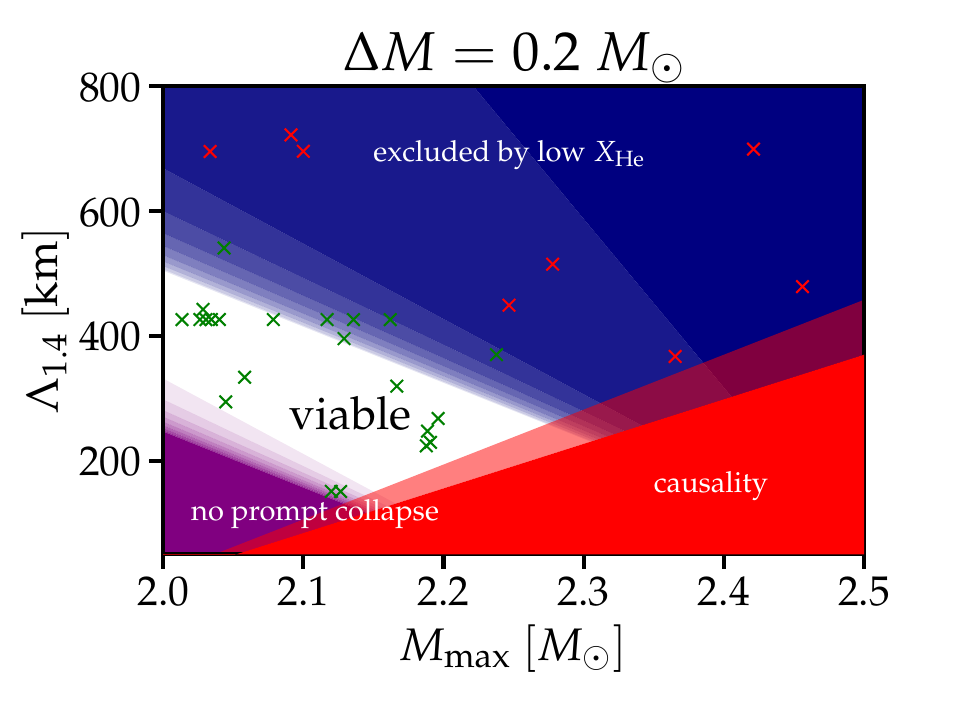}
  \includegraphics[width=5.7cm]{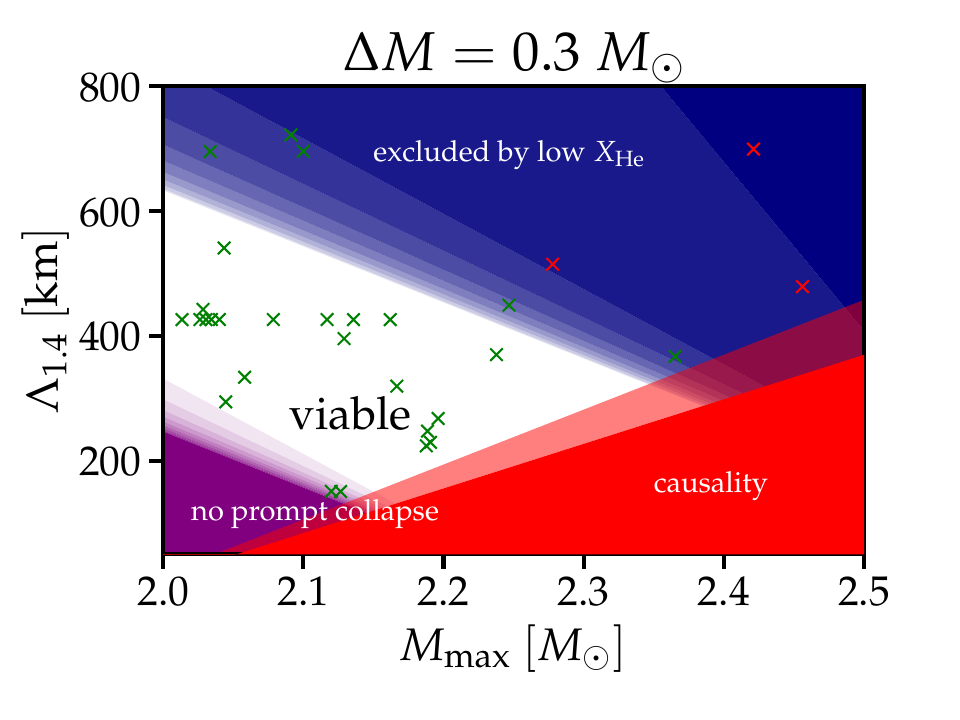}
  \includegraphics[width=5.7cm]{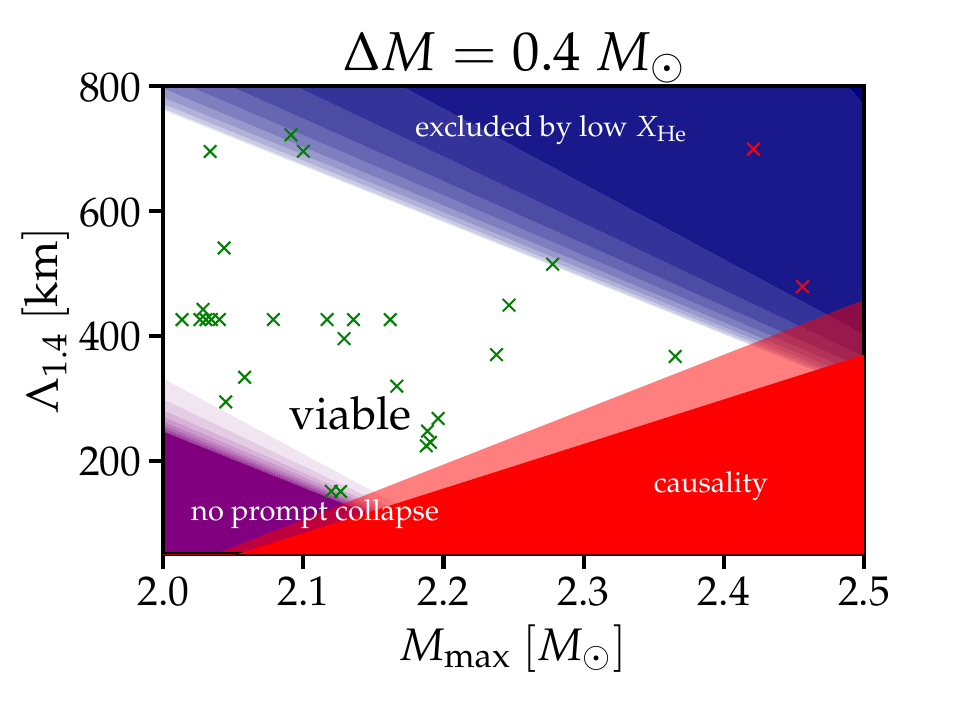}
  \caption{Constraints on stellar parameters similar to Figs.~\ref{fig:r16limit} and~\ref{fig:lam14limit} adopting $\Delta M=0.2\,$M$_\odot$ (left panels), $\Delta M=0.3\,$M$_\odot$ (middle panels) and $\Delta M=0.4\,$M$_\odot$ (right panels). Color shading of the blue and purple area indicates confidence levels of exclusion (in 10\% steps) from considering the posterior distribution of the binary mass ratio of GW170817. Note the initial steep increase with the 50\%~level very close to the 10\% level for the upper limit and the steep decrease between the 90\%~level and the 50\%~level for the lower limit. Green and red data points are stellar parameters from a large set of microphysical EoS models considered in~\citep{Bauswein2021}. Red crosses refer to EoS models which are excluded by our constraints. Dots in the upper row show which constraints are visualized in Figs.~\ref{fig:tovlimit} and~\ref{fig:tovlimit0304} (see main text).}
  \label{fig:sensedm}
\end{figure*}

\begin{figure*}
    \includegraphics[width=8.7cm]{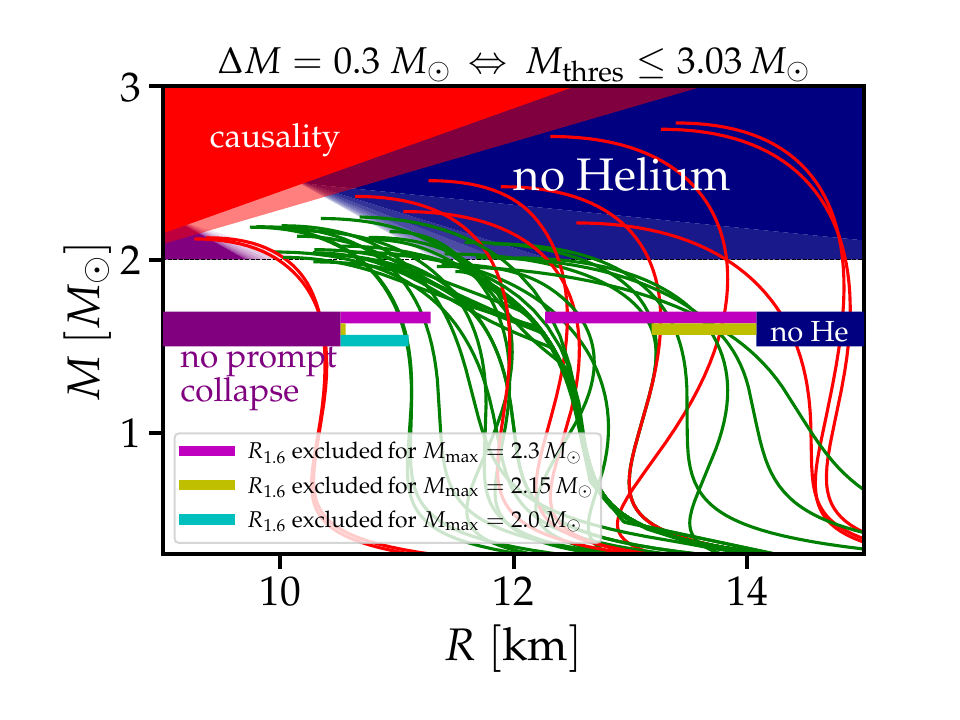}
  \includegraphics[width=8.7cm]{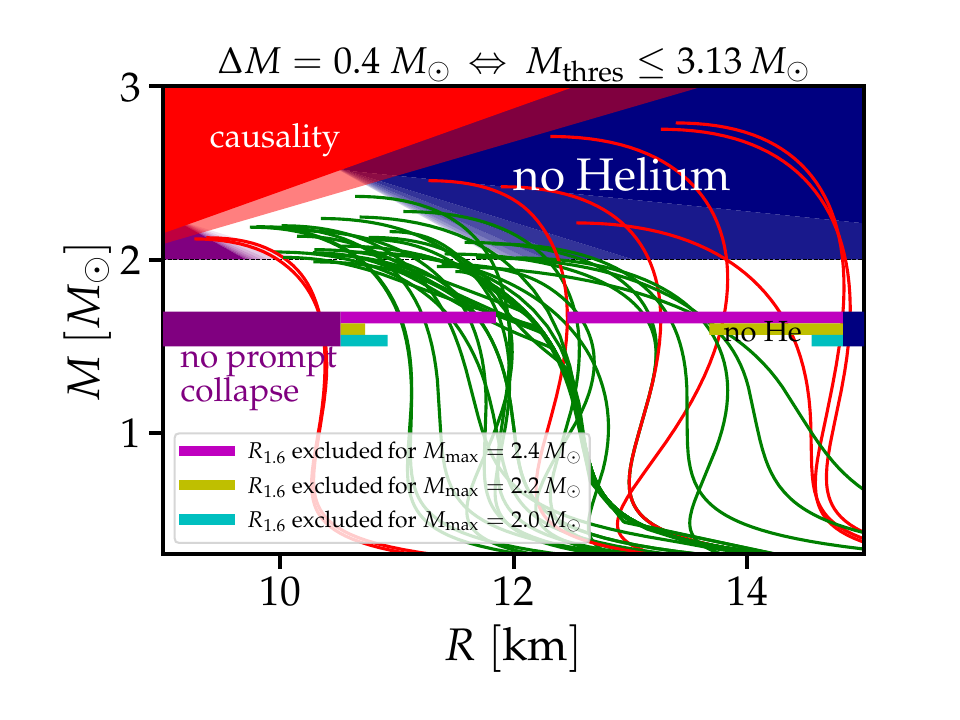}
  
    \caption{Same as Fig.~\ref{fig:tovlimit} but for $\Delta M=0.3\,$M$_\odot$ (left) and $\Delta M=0.4\,$M$_\odot$ (right). The adopted absolute $R_{1.6}$ constraints are displayed in Fig.~\ref{fig:sensedm} by white dots (see main text for explanations). Note the different choices for $M_\mathrm{max}$ compared to Fig.~\ref{fig:tovlimit} for which $R_{1.6}$ limits are displayed by the horizontal bars at $\approx1.6~M_\odot$.}
    \label{fig:tovlimit0304}
\end{figure*}

\begin{figure*}
       \includegraphics[width=5.7cm]{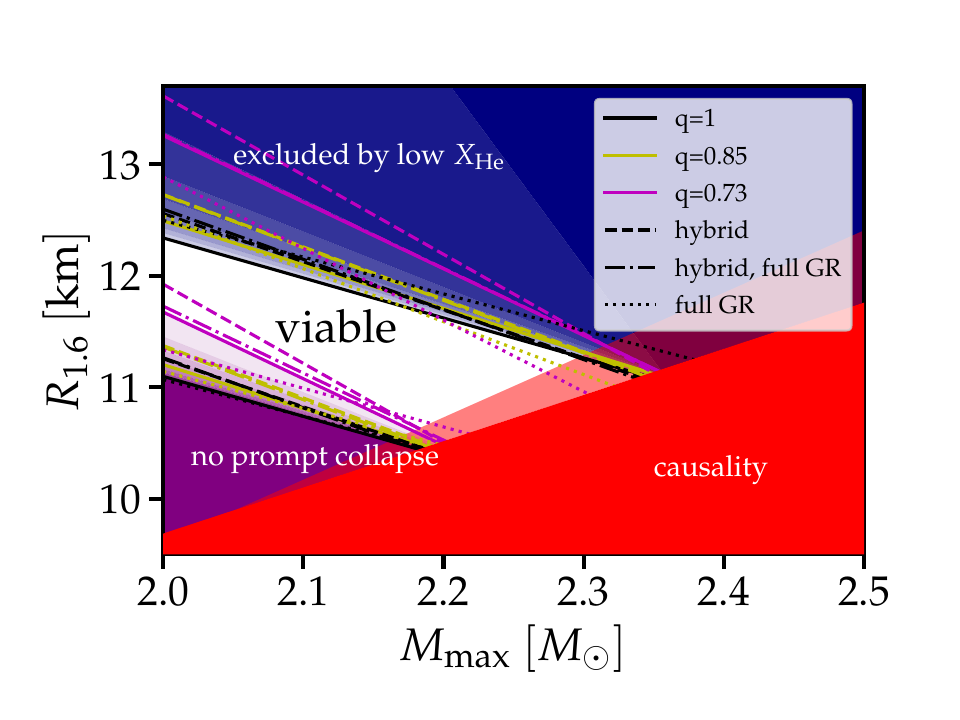}
  \includegraphics[width=5.7cm]{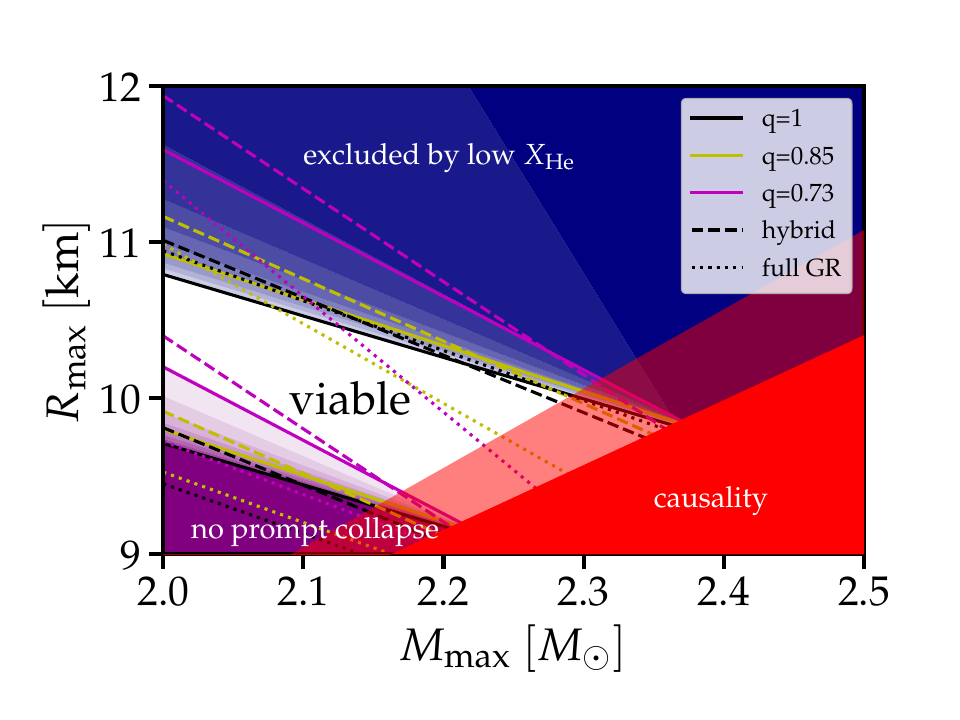}
  \includegraphics[width=5.7cm]{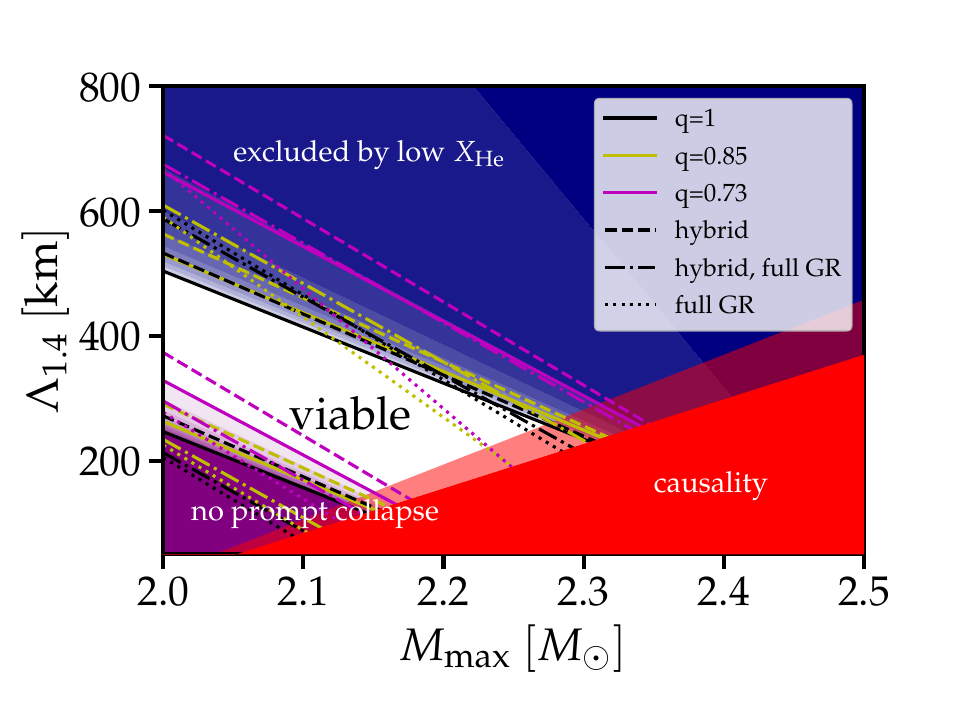}

    \caption{Same as Figs.~\ref{fig:r16limit} and~\ref{fig:lam14limit} but including alternative fit formulae for $M_\mathrm{thres}$ (assuming $\Delta M=0.2\,$M$_\odot$). Colors refer to fixed binary mass ratios of $q=1$ (black), $q=0.85$ (yellow) and $q=0.73$ (magenta). Solid lines adopt the same fit formula as in the main text (set `b' in Tab.~VI of~\citep{Bauswein2021}). For the dashed curves a fit formula is employ, which is based on data including nine EoS models with strong first-order phase transitions (set `b+h' in Tab.~VI of~\citep{Bauswein2021}). Dashed-dotted lines are obtained from the fit formula of~\citep{Koelsch2022} using the set `K+b+e+h' (Tab.~VIII). Note that most of the underlying data for this fit stems from~\citep{Bauswein2021}. Dotted curve results from a fit of~\citep{Koelsch2022} (set `K+P', Tab.~X).}
    \label{fig:difffit}
\end{figure*}

\section{Sensitivity of EoS constraints}\label{app:sense}
We provide additional plots exemplifying the sensitivity of our EoS constraints including the results for different choices of $\Delta M$, a description of how to obtain an upper limit on $M_\mathrm{max}$ for given $R/\Lambda$ and the resulting EoS constraints, and a discussion of the impact of different fit formulae for $M_\mathrm{thres}$.

In Fig.~\ref{fig:sensedm} we show the limits on $R_{1.6}$, $R_\mathrm{max}$ and $\Lambda_{1.4}$ adopting different values for $\Delta M$, i.e.,\ the proximity of the measured total binary mass of GW170817 to the threshold mass for prompt black hole formation. As expected from Eq.~\eqref{eq:upper}, the upper limits become less constraining if $\Delta M$ increases although even for the very conservative case of $\Delta M=0.4\,$M$_\odot$ the absence of significant amounts of helium in the ejecta still excludes some current microphysical EoS models which are compatible with pulsar observations~\citep{Antoniadis2013,Fonseca2021,Romani2022} and the tidal deformability inference from GW170817~\citep{Abbott2019}. For better illustration in Fig.~\ref{fig:sensedm} we overplot stellar parameters from a set of microphysical EoSs (same EoS sample as in~\citep{Bauswein2021}, i.e.,\ the identical set that was employed to construct fit formulae for $M_\mathrm{thres}$ in~\citep{Bauswein2021}). These panels also demonstrate the potential of our limits, which place a combined constraint on $R/\Lambda$ and $M_\mathrm{max}$. For instance, some models with $R_{1.6}\sim 12$~km are excluded because their $M_\mathrm{max}$ is too large. Table~\ref{tab:eostab} lists the overplotted EoS models and shows which ones are excluded for a given $\Delta M$. The table also specifies whether an EoS model is excluded by the upper or the lower limit.

The lower limits on $R/\Lambda$ are independent of $\Delta M$ because they are solely based on the argument that GW170817 did not undergo a prompt collapse. Imposing additionally that the EoS cannot become arbitrarily stiff, which is ultimately limited by causality, rules out stellar parameters in the lower right corner of the panels. We distinguish between a more conservative ``causality limit'' adopting $v_s=c$ above a certain density 
and a more empirical relation which simply reproduces the stellar parameter found in a large set of microphysical candidate EoSs (light shading). See~\citep{Bauswein2019,Bauswein2021} for more detailed explanations. 

In brief, we obtain the ``causality'' limit by the following consideration. For a given $R_{1.6}$, the maximum mass cannot become too large with the maximum $M_\mathrm{max}^\mathrm{up}(R_{1.6})$ given by a fiducial stellar configuration that has maximal stiffness, i.e.,\ $v_\mathrm{s}=c$, beyond the central density of the 1.6~M$_\odot$ NS. For a large set of microphysical EoSs (the sample used in~\citep{Bauswein2021}), we adopt the EoS up to the central density of the 1.6\,M$_\odot$ NS and extend the EoS to higher densities imposing $v_\mathrm{s}=c$. From this modified EoS we obtain the Tolman-Oppenheimer-Volkoff (TOV) solution with the highest possible $M_\mathrm{max}^\mathrm{up}$. Considering the pairs $\{R_{1.6},M_\mathrm{max}^\mathrm{up}\}$ for the full set of modified EoSs, we determine a line with coefficients $w_1$ and $w_2$ such that $M_\mathrm{max}^\mathrm{up}(R)<w_1 R +w_2$. Being less conservative, we obtain a second set of $w_1$ and $w_2$ as an ``empirical'' limit by considering the original pairs $\{R_{1.6},M_\mathrm{max}\}$ with the actual $M_\mathrm{max}$ from the unmodified, microphysical EoS. None of the actual models (green crosses) lies beyond the light red area in Fig.~\ref{fig:sensedm}. The constraints are quantified as $M_\mathrm{max}(R)\leq w_1 R+w_2$ (and equivalently for $\Lambda$) with $w_1$ and $w_2$ provided in Tab.~\ref{tab:nopc}; for $R_\mathrm{max}$ we adopt the result from~\cite{Koranda1997,Lattimer2016}. The exact limits obviously depend on the chosen set of EoS models and for consistency we here repeat the procedure for the EoS sample from~\citep{Bauswein2021}. Determining the lines to embrace all models for the conservative and the empirical limit is obviously ambiguous. Finally, for a given EoS model with a given $R_{1.6}$, the computation of $M_\mathrm{max}^\mathrm{up}(R_{1.6})$ as described may not exactly yield the highest possible $M_\mathrm{max}$ because a different EoS at lower densities with the same $R_{1.6}$ may yield a slightly larger limit. However, all these effects are relatively small in practice and the displayed areas thus provide very reasonable limits.

The intersections of these limits with the bound from the no-prompt-collapse constraint provide absolute lower limits on $R/\Lambda$ (see Fig.~\ref{fig:sensedm}). We list these limits in Tab.~\ref{tab:nopc} together with the absolute upper limits, i.e.,\ the constraints for $M_\mathrm{max}=2.0\,$M$_\odot$ for different values of $\Delta M$. All limits adopt the 90\% confidence level resulting form the posterior distribution of the binary mass ratio of GW170817. The table also provides the absolute upper limits on the maximum mass $M_\mathrm{max}$, which is given by the intersection of the upper limit $R/\Lambda$ limit with the causality (or empirical) constraint  (see Fig.~\ref{fig:sensedm}). For a given $\Delta M$, these limits slightly depend on whether one determines the intersection with the upper bound on $R_{1.6}$, $R_\mathrm{max}$ or $\Lambda_{1.4}$ (with the corresponding $w_1$ and $w_2$). However, comparing the constraints on $M_\mathrm{max}$ for the respective causal or empirical limit, one finds a very good agreement between these numbers. For instance, we find $M_\mathrm{max}\lesssim2.31$, $M_\mathrm{max}\lesssim2.38$ and $M_\mathrm{max}\lesssim2.45$ for $\Delta M=0.2\,$M$_\odot$, $\Delta M=0.3\,$M$_\odot$ and $\Delta M=0.4\,$M$_\odot$, respectively, adopting the empirical relations. 

For the cases $\Delta M=0.3\,$M$_\odot$ and $\Delta M=0.4\,$M$_\odot$ we visualize the resulting constraints in comparison to mass-radius relations from existing microphysical EoSs in Fig.~\ref{fig:tovlimit0304}. These plots are equivalent to Fig.~\ref{fig:tovlimit} with the same color scheme as in the main text, to which we refer for further explanations (see also Tab.~\ref{tab:eostab}).

Figure~\ref{fig:difffit} exemplifies the impact of other fit formulae for $M_\mathrm{thres}$ on the different constraints for $R_{1.6}$, $R_\mathrm{max}$ and $\Lambda_{1.4}$ adopting $\Delta M=0.2\,$M$_\odot$. See Tab.~\ref{tab:fits} for fits and coefficients. Different colors of the lines again refer to fixed binary mass ratios and the solid curves display the limits resulting from the fit formulae used in the main text and Figs.~\ref{fig:sensedm} and~\ref{fig:tovlimit0304}. For the dashed lines we employ fit formulae based on a dataset which includes EoS models undergoing a strong phase transition to deconfined quark matter (set `b+h' in Tab.~VI of~\citep{Bauswein2021}). The resulting constraints become slightly weaker in particular for relatively small $M_\mathrm{max}\approx2.0\,$M$_\odot$. We do not adopt these fit formulae as our standard choice because the additional hybrid models considered in these fits are on the rather extreme side with regard to the stiffness of the quark matter EoS (see~\citep{Blacker2024}) and thus possibly represent an unlikely range of models.
The dashed-dotted lines in Fig.~\ref{fig:difffit} display the results from using the fit formulae (set `K+b+e+h') from Ref.~\citep{Koelsch2022}. These fits are based on data, which include $M_\mathrm{thres}$ determinations from different simulation tools. Most models are taken from~\citep{Bauswein2021} but also fully relativistic models are computed. Note that the set includes hybrid models as well. The dotted lines correspond to fits to a dataset which only comprises fully relativistic calculations (set `K=P' from~\citep{Koelsch2022}) as opposed to the approximate solution of the Einstein equations within the conformal flatness approximation in~\citep{Bauswein2021}. We do not consider these fits as default choice because they are based on a significantly smaller number of EoS models and calculations for a wider range of binary mass ratios, which is not required for the range inferred from GW170817. 
Thus, the coverage of the range in $q$ and EoSs models applicable to our constraints is not as comprehensive as the on in~\citep{Bauswein2021}. Note that the comparisons in~\citep{Koelsch2022,Kashyap2022} show a generally very good agreement between the calculations in full general relativity and the ones within the conformal flatness approximation~\cite{Kashyap2022}. At any rate our plots (Fig.~\ref{fig:difffit}) show that the main uncertainties originate form the unknown $M_\mathrm{max}$ and binary mass ratio $q$.

\begin{table*}
\begin{tabular}{|l|c|c|c|c|c|c|c|c|c|}  \hline

$X$  & $c_1$ & $c_2$ & $c_3$ & $c_4$ & $c_5$ & $c_6$ & $c_7$ & dev. & Ref.  \\ \hline

\multicolumn{10}{|c|}{$M_\mathrm{thres}(q,M_\mathrm{max},X)=c_1 M_\mathrm{max} + c_2 X +c_3 + c_4 \delta q^3 M_\mathrm{max}+ c_5 \delta q^3 X$}  \\ \hline

$R_{1.6}$  & 0.578  & $1.610\times10^{-1}$ & -0.218  & 8.987  & -1.767  & - & -  & 0.017 & \cite{Bauswein2021}\\ \hline

$R_\mathrm{max}$ & 0.491  & $1.847\times10^{-1}$  & -0.051  & 9.382  & -2.088  & - & - & 0.021 & \cite{Bauswein2021} \\ \hline

$\Lambda_{1.4}$ & 0.698  & $7.772\times10^{-4}$ & 1.137  & 0.900  & $-9.050\times10^{-3}$  & - & - &  0.029 & \cite{Bauswein2021} \\ \hline

$R_{1.6}$ hybrid & 0.663 & $1.535\times10^{-1}$ & -0.329 & 9.058 & -1.784 & - & - & 0.035 & \cite{Bauswein2021}  \\ \hline

$R_\mathrm{max}$ hybrid & 0.613 & $1.665\times10^{-1}$ & -0.135 & 8.209 & -1.859 & - & - & 0.032 & \cite{Bauswein2021}  \\ \hline

$\Lambda_{1.4}$ hybrid & 0.750 & $7.670\times10^{-4}$ & 1.016 & 0.878 & $-9.763\times10^{-3}$ & - & - &  0.040 &  \cite{Bauswein2021} \\ \hline

$R_{1.6}$ hybrid, full GR & 0.675 & $1.500\times10^{-1}$ & -0.315 & 5.313 & -1.031 &- & - & 0.0365 & \cite{Koelsch2022} \\ \hline

$\Lambda_{1.4}$  hybrid, full GR & 0.67 & $5.338\times10^{-4}$ & 1.271 & -0.042 & $-3.347\times10^{-4}$ &- & -  & 0.0456 & \cite{Koelsch2022} \\ \hline

\multicolumn{10}{|c|}{$M_\mathrm{thres}(q,M_\mathrm{max},X)=c_1 M_\mathrm{max} + c_2 X +c_3 + c_4 \delta q M_\mathrm{max}+ c_5 \delta q X + c_6 \delta q^3 M_\mathrm{max}+ c_7 \delta q^3 X$}  \\ \hline

$R_{1.6}$ full GR & 0.462 & 0.14 & 0.251 & 0.76 & -0.12 & 0.00817 & -0.188 &  0.0287 & \cite{Koelsch2022} \\ \hline

$R_\mathrm{max}$ full GR & 0.428 & 0.134 & 0.602 & 0.886 & -0.16 & -0.539 & -0.112  & 0.0341 & \cite{Koelsch2022} \\ \hline

$\Lambda_{1.4}$ full GR & 0.686 & $5.049\times10^{-4}$ & 1.249 & 0.162 & $-3.921\times10^{-4}$ & -0.808 & $-1.406\times10^{-3}$  & 0.0303 & \cite{Koelsch2022} \\ \hline

\hline

\end{tabular}
\caption{Different fits describing the EoS dependence of the threshold binary mass $M_\mathrm{thres}$ for prompt BH formation including an explicit dependence on the binary mass ratio $q$ through $\delta q=1-q$ (see main text). First column specifies the stellar parameter $X$, which may either be $R_{1.6}$, $R_\mathrm{max}$ or $\Lambda_{1.4}$ and the underlying set of data (see caption of Fig.~\ref{fig:difffit}). Fit parameters $c_i$ are given in second to eighth columns. The units of the fit parameters $c_i$ are such that masses are in M$_\odot$, radii in km and tidal deformabilities dimensionless. Next column specifies the average deviation between fit and the underlying data. Last column provides the reference from which the fit is taken and where further details can be found.}
\label{tab:fits}
\end{table*}

\begin{table}
\caption{Absolute lower and upper limits on different stellar parameters $X$ independent of $M_\mathrm{max}$ for different $\Delta M$ (see main text). Radii are given in km. $w_1$ and $w_2$ describe an upper limit on the maximum mass via $M_\mathrm{max}\leq w_1 X + w_2$. `causal' and `empirical' in the first column indicate how $w_1$ and $w_2$ were obtained, where `causal' implies a more conservative approach and limit on $X$ assuming maximum stiffness of the EoS at higher densities and `empirical' only embraces a large but incomplete set of currently available microphysical models. Every combination $\Delta M$, $w_1$ and $w_2$ determines an absolute upper limit on $M_\mathrm{max}$ (in solar masses) listed after the rows with the corresponding limits on $R_{1.6}$, $R_\mathrm{max}$ and $\Lambda_{1.4}$ for the same set of $\Delta M$, $w_1$ and $w_2$. All limits are determined adopting the 90\% confidence level resulting from the distribution of the binary mass ratio of GW170817.}
\begin{tabular}{|l |c| |c| c|c |c|}
\hline\hline
Parameter (limit) & $\Delta M/M_\odot$ & lower limit & upper limit &   $w_1$ & $w_2$  \\
\hline

$R_{1.6}$ (causal)           & 0.2  &  10.44 &  13.29  &   0.241  &  -0.336  \\
$R_{1.6}$ (empirical)        & 0.2  &  10.52 &  13.29  &   0.179  &  0.282  \\
$M_\mathrm{max}$ (causal)    & 0.2  &  -     &  2.353  &   0.241  &  -0.336  \\
$M_\mathrm{max}$ (empirical) & 0.2  &  -     &  2.316  &   0.179  &  0.282  \\
$R_\mathrm{max}$ (causal)    & 0.2  &  9.17  &  11.63  &   0.240  &  0.0  \\
$R_\mathrm{max}$ (empirical) & 0.2  &  9.31  &  11.63  &   0.197  &  0.321  \\
$M_\mathrm{max}$ (causal)    & 0.2  &  -     &  2.369  &   0.240  &  0.0  \\
$M_\mathrm{max}$ (empirical) & 0.2  &  -     &  2.313  &   0.197  &  0.321  \\
$\Lambda_{1.4}$ (causal)     & 0.2  &  117.0 &  668.7  &   0.0014 &  1.982  \\
$\Lambda_{1.4}$ (empirical)  & 0.2  &  131.4 &  668.7  &   0.0011 &  1.980  \\
$M_\mathrm{max}$ (causal)    & 0.2  &  -     &  2.341  &   0.0014 &  1.982  \\
$M_\mathrm{max}$ (empirical) & 0.2  &  -     &  2.312  &   0.0011 &  1.980  \\
\hline

$R_{1.6}$ (causal)           & 0.3  &  10.44 & 14.08  &   0.241  &  -0.336  \\
$R_{1.6}$ (empirical)        & 0.3  &  10.52 & 14.08  &   0.179  &  0.282  \\
$R_{1.6}$ (causal)           & 0.3  &  -     & 2.431  &   0.241  &  -0.336  \\
$R_{1.6}$ (empirical)        & 0.3  &  -     & 2.384  &   0.179  &  0.282  \\
$R_\mathrm{max}$ (causal)    & 0.3  &   9.17 & 12.33  &   0.240  &  0.0  \\
$R_\mathrm{max}$ (empirical) & 0.3  &   9.31 & 12.33  &   0.197  &  0.321  \\
$R_\mathrm{max}$ (causal)    & 0.3  &   -    & 2.447  &   0.240  &  0.0  \\
$R_\mathrm{max}$ (empirical) & 0.3  &  -     & 2.384  &   0.197  &  0.321  \\
$\Lambda_{1.4}$ (causal)     & 0.3  &  117.0 &  837.2 &   0.0014 &  1.982  \\
$\Lambda_{1.4}$ (empirical)  & 0.3  &  131.4 &  837.2 &   0.0011 &  1.980  \\
$\Lambda_{1.4}$ (causal)     & 0.3  &  -     & 2.429 &   0.0014 &  1.982  \\
$\Lambda_{1.4}$ (empirical)  & 0.3  &  -     & 2.393 &   0.0011 &  1.980  \\
\hline

$R_{1.6}$ (causal)           & 0.4  &  10.44 &  14.82 &   0.241  &  -0.336  \\
$R_{1.6}$ (empirical)        & 0.4  &  10.52 &  14.82 &   0.179  &  0.282  \\
$R_{1.6}$ (causal)           & 0.4  &  -     &  2.509 &   0.241  &  -0.336  \\
$R_{1.6}$ (empirical)        & 0.4  &  -     &  2.453 &   0.179  &  0.282  \\
$R_\mathrm{max}$ (causal)    & 0.4  &   9.17 &  13.03 &   0.240  &  0.0  \\
$R_\mathrm{max}$ (empirical) & 0.4  &   9.31 &  13.03 &   0.197  &  0.321  \\
$R_\mathrm{max}$ (causal)    & 0.4  &  -     &  2.526 &   0.240  &  0.0  \\
$R_\mathrm{max}$ (empirical) & 0.4  &  -     &  2.455 &   0.197  &  0.321  \\
$\Lambda_{1.4}$ (causal)     & 0.4  &  117.0 &   1005.5 &   0.0014 &  1.982  \\
$\Lambda_{1.4}$ (empirical)  & 0.4  &  131.4 &   1005.5 &   0.0011 &  1.980  \\
$\Lambda_{1.4}$ (causal)     & 0.4  &  -     &  2.517 &   0.0014 &  1.982  \\
$\Lambda_{1.4}$ (empirical)  & 0.4  &  -     &  2.473 &   0.0011 &  1.980  \\

\hline
\hline
\end{tabular}
\label{tab:nopc}
\end{table}

\begin{table*}
\begin{tabular}{|l|c|c|c|c||c|c|c|c||c|c|c|c||c|c|}\hline

EoS & $R_{1.6}$ (km) & 0.2 & 0.3 & 0.4 & $R_\mathrm{max}$ (km) & 0.2 & 0.3 & 0.4 & $\Lambda_{1.4}$ & 0.2 & 0.3 & 0.4 & $M_\mathrm{max}~(M_\odot)$ & Ref. \\ \hline

BHBLP\footnote{Figure~\ref{fig:lifetime} suggests $\Delta M\approx 0.3~M_\odot$ for this EoS implying that this model is actually not ruled out for $\Delta M=0.2~M_\odot$. But we note the coarse sampling of the sequence of merger models and that the data point at $M_\mathrm{tot}=2.7~M_\odot$ only provides a lower bound on $\tau_\mathrm{BH}$ (see Fig.~\ref{fig:lifetime}).}      & 13.203 & (u) &   &   & 11.588 & (u) &   &   & 696.1  & (u) &   &   & 2.100 &  \cite{Banik2014}     \\ \hline
DD2Y       & 13.182 & u &   &   & 11.418 &   &   &   & 695.7  & u &   &   & 2.033 &  \cite{Fortin2018,Marques2017}     \\ \hline
DD2        & 13.258 & u & u & u & 11.899 & u & u & u & 699.5  & u & u & u & 2.421 &  \cite{Hempel2010,Typel2010}     \\ \hline
DD2F       & 12.232 &   &   &   & 10.534 &   &   &   & 426.3  &   &   &   & 2.079 &  \cite{Typel2005,Typel2010,Alvarez-Castillo2016}     \\ \hline
APR        & 11.263 &   &   &   & 9.923  &   &   &   & 247.5  &   &   &   & 2.189 &  \cite{Akmal1998}     \\ \hline
BSK20      & 11.658 &   &   &   & 10.170 &   &   &   & 319.6  &   &   &   & 2.167 &  \cite{Goriely2010}     \\ \hline
eosUU      & 11.066 &   &   &   & 9.807  &   &   &   & 229.5  &   &   &   & 2.191 &  \cite{Wiringa1988}     \\ \hline
LS220      & 12.491 &   &   &   & 10.619 &   &   &   & 541.1  &   &   &   & 2.043 &  \cite{Lattimer1991}     \\ \hline
LS375      & 13.776 & u & u & u & 12.320 & u & u & u & 957.4  & u & u & u & 2.711 &  \cite{Lattimer1991}     \\ \hline
GS2        & 13.381 & u &   &   & 11.823 & u &   &   & 722.3  & u &   &   & 2.091 &  \cite{Shen2011}     \\ \hline
NL3        & 14.807 & u & u & u & 13.394 & u & u & u & 1369.4 & u & u & u & 2.789 &  \cite{Hempel2010,Lalazissis1997a}     \\ \hline
Sly4       & 11.533 &   &   &   & 9.957  &   &   &   & 294.6  &   &   &   & 2.045 &  \cite{Douchin2001}     \\ \hline
SFHO       & 11.761 &   &   &   & 10.303 &   &   &   & 333.9  &   &   &   & 2.058 &  \cite{Steiner2013}     \\ \hline
SFHOY      & 11.758 &   &   &   & 10.295 &   &   &   & 333.9  &   &   &   & 1.988 &  \cite{Fortin2018,Marques2017}     \\ \hline
SFHX       & 11.972 &   &   &   & 10.769 &   &   &   & 395.8  &   &   &   & 2.129 &  \cite{Steiner2013}     \\ \hline
TM1        & 14.361 & u & u & u & 12.544 & u & u & u & 1150.8 & u & u & u & 2.212 &  \cite{Sugahara1994a,Hempel2012}     \\ \hline
TMA        & 13.673 & u &   &   & 12.030 & u &   &   & 935.3  & u & u &   & 2.010 &  \cite{Toki1995,Hempel2012}     \\ \hline
BSK21      & 12.552 & u & u &   & 11.064 & u & u &   & 515.0  & u & u &   & 2.278 &  \cite{Goriely2010}     \\ \hline
GS1        & 14.877 & u & u & u & 13.268 & u & u & u & 1402.3 & u & u & u & 2.753 &  \cite{Shen2011}     \\ \hline
eosAU      & 10.365 & l & l & l & 9.381  &   &   &   & 150.9  &   &   &   & 2.127 &  \cite{Wiringa1988}     \\ \hline
WFF1       & 10.370 & l & l & l & 9.277  & l & l & l & 151.0  &   &   &   & 2.120 &  \cite{Wiringa1988,Read2009a}     \\ \hline
WFF2       & 11.057 &   &   &   & 9.752  &   &   &   & 223.9  &   &   &   & 2.188 &  \cite{Wiringa1988,Read2009a}     \\ \hline
MPA1       & 12.458 & u & u & u & 11.280 & u & u & u & 479.1  & u & u & u & 2.456 &  \cite{Muther1987,Read2009a}     \\ \hline
ALF2       & 12.628 &   &   &   & 11.393 &   &   &   & 569.4  &   &   &   & 1.975 &  \cite{Alford2005,Read2009a}     \\ \hline
H4         & 13.731 & u &   &   & 11.687 & u &   &   & 853.2  & u & u &   & 2.012 &  \cite{Lackey2006,Read2009a}     \\ \hline
DD2F-SF-1  & 12.158 &   &   &   & 10.156 &   &   &   & 426.3  &   &   &   & 2.136 &  \cite{Kaltenborn2017,Bastian2018,Fischer2018,Bauswein2019,Bastian2020}    \\ \hline
DD2F-SF-2  & 12.071 &   &   &   & 10.949 & u &   &   & 426.2  &   &   &   & 2.162 &  \cite{Kaltenborn2017,Bastian2018,Fischer2018,Bauswein2019,Bastian2020}    \\ \hline
DD2F-SF-3  & 12.205 &   &   &   & 10.315 &   &   &   & 426.3  &   &   &   & 2.034 &  \cite{Kaltenborn2017,Bastian2018,Fischer2018,Bauswein2019,Bastian2020}    \\ \hline
DD2F-SF-4  & 12.232 &   &   &   & 10.345 &   &   &   & 426.3  &   &   &   & 2.031 &  \cite{Kaltenborn2017,Bastian2018,Fischer2018,Bauswein2019,Bastian2020}    \\ \hline
DD2F-SF-5  & 11.943 &   &   &   & 10.291 &   &   &   & 426.3  &   &   &   & 2.040 &  \cite{Kaltenborn2017,Bastian2018,Fischer2018,Bauswein2019,Bastian2020}    \\ \hline
DD2F-SF-6  & 12.231 &   &   &   & 10.036 &   &   &   & 426.3  &   &   &   & 2.013 &  \cite{Kaltenborn2017,Bastian2018,Fischer2018,Bauswein2019,Bastian2020}    \\ \hline
DD2F-SF-7  & 12.232 &   &   &   & 10.766 &   &   &   & 426.3  &   &   &   & 2.117 &  \cite{Kaltenborn2017,Bastian2018,Fischer2018,Bauswein2019,Bastian2020}    \\ \hline
DD2F-SF-8  & 12.232 &   &   &   & 11.214 &   &   &   & 426.3  &   &   &   & 2.026 &  \cite{Kaltenborn2017,Bastian2018,Fischer2018,Bauswein2019,Bastian2020}    \\ \hline
VBAG       & 12.214 &   &   &   & 11.496 &   &   &   & 422.3  &   &   &   & 1.932 &  \cite{Cierniak2018}     \\ \hline
ENG        & 11.909 &   &   &   & 10.358 & u &   &   & 370.0  &   &   &   & 2.238 &  \cite{Engvik1996,Read2009a}     \\ \hline
APR3       & 11.963 & u & u &   & 10.653 & u & u &   & 367.2  & u &   &   & 2.365 &  \cite{Akmal1998,Read2009a}     \\ \hline
GNH3       & 13.772 & u &   &   & 11.350 &   &   &   & 857.8  & u & u &   & 1.961 &  \cite{Glendenning1985,Read2009a}     \\ \hline
SAPR       & 11.474 &   &   &   & 10.023 &   &   &   & 267.9  &   &   &   & 2.196 &  \cite{Schneider2019}     \\ \hline
SAPRLDP    & 12.369 & u &   &   & 10.614 & u &   &   & 449.3  & u &   &   & 2.247 &  \cite{Schneider2019}     \\ \hline
SSkAPR     & 12.304 &   &   &   & 10.880 &   &   &   & 442.6  &   &   &   & 2.028 &  \cite{Schneider2019}     \\ \hline
                        
\end{tabular}

\caption{EoS models shown in Figs.~\ref{fig:tovlimit},~\ref{fig:sensedm} and~\ref{fig:tovlimit0304} with the radius $R_{1.6}$ of a nonrotating NS with 1.6~$M_\odot$, the $R_\mathrm{max}$ as the radius of the maximum-mass NS with $M_\mathrm{max}$ and the tidal deformability $\Lambda_{1.4}$ of a 1.4~$M_\odot$ NS. The columns to the right of the $R_{1.6}$, $R_\mathrm{max}$ and $\Lambda_{1.4}$ entries, respectively, indicate by a `u' or `l' which EoSs are excluded by our upper or lower limit for $\Delta M=0.2~M_\odot$, $\Delta M=0.3~M_\odot$ and $\Delta M=0.4~M_\odot$ for the given $M_\mathrm{max}$ and $R_{1.6}$, $R_\mathrm{max}$ or $\Lambda_{1.4}$ employing the 90\% confidence level of the posterior distribution of the binary mass ratio of GW170817. The last column provides the references for the respective EoS.}
\label{tab:eostab}
\end{table*}

\end{document}